%% file: PRD_verfin.tex
\documentclass[preprint,floatfix,aps,prd,showpacs,amsmath,amssymb,amsfonts,superscriptaddress,nofootinbib]{revtex4-2}
\usepackage{graphicx}
\usepackage{color}
\usepackage{enumitem}
\usepackage{dsfont}
\usepackage[normalem]{ulem}
\usepackage{hyperref}
\usepackage{mathtools}
\usepackage{comment}
\usepackage{tensor}
\usepackage[dvipsnames]{xcolor}
\usepackage{soul}
\usepackage{mathtools}
\usepackage[utf8]{inputenc}
\usepackage{multirow} 
\usepackage{subfigure}
\usepackage{varioref}
\usepackage{adjustbox}
\usepackage{rotating}
\usepackage{tensor}
\usepackage{multirow}
\usepackage[active]{srcltx}

\expandafter\ifx\csname package@font\endcsname\relax\else
\expandafter\expandafter
\expandafter\usepackage
\expandafter\expandafter
\expandafter{\csname package@font\endcsname}%
\fi
\allowdisplaybreaks
\hypersetup{
	colorlinks=true,       
	linkcolor=blue,          
	citecolor=blue,        
}

\usepackage{array}
\newcolumntype{P}[1]{>{\centering\arraybackslash}p{#1}}

\usepackage[section]{placeins}

\usepackage{tikz}
\newcommand{\circled}[1]{%
  \tikz[baseline=(char.base)]%
    \node[draw,circle,inner sep=0pt,minimum size=1.5em,line width=0.4pt](char){\scriptsize #1};%
}
\usetikzlibrary{positioning}

\usepackage{fancybox}

\newcommand{\bea}{\begin{aligned}}
\newcommand{\eea}{\end{aligned}}
\def\bea{\begin{eqnarray}}
\def\eea{\end{eqnarray}}
\def\beq{\begin{equation}}
\def\eeq{\end{equation}}
\def\bse{\begin{subequations}}
\def\ese{\end{subequations}}
\def\d{\mathrm{d}}

\newcommand{\upchi}{\protect\raisebox{2.5pt}{$\chi$}}

\begin{document}
\title{\emph{Disformal} interactions in the Dark Sector: From driving Early Dark Energy to confronting cosmological tensions}
\author{Pulkit Bansal} 
\email{pulkit.l.bansal@gmail.com}
\affiliation{Department of Physics, Indian Institute of Technology Bombay, Mumbai 400076, India}
\author{Joseph P. Johnson}
\email{josephpj@iisermohali.ac.in}
\affiliation{Department of Physical Sciences, Indian Institute of Science Education and Research Mohali, SAS Nagar, Punjab 140306, India}
\author{S. Shankaranarayanan}
\email{shanki@iitb.ac.in}
\affiliation{Department of Physics, Indian Institute of Technology Bombay, Mumbai 400076, India}
%
\begin{abstract}

The $\Lambda$CDM model faces significant challenges, including an incomplete understanding of the dark sector and persistent tensions in the Hubble constant and the clustering amplitude. To address these issues, we propose a \emph{general disformal coupling} between dark energy (DE) and dark matter from a field-theoretic action which can generate a rich variety of interactions including conformal and pure-momentum coupling scenarios. Our analysis reveals that a pure disformal coupling naturally produces a unique \emph{interacting Early Dark Sector}, wherein the interactions with dark matter suppress the Hubble friction on the DE scalar field leading to a kinetic-driven cosmological constant-like behavior at early times followed by its dilution as $a^{-6}$ and eventually leading to a potential-driven epoch characteristic of late-time dark energy. In contrast to existing Early Dark Energy (EDE) models that rely on finely-tuned potentials, the EDE-like behavior, in our framework, is purely a consequence of the disformal coupling paired with the dilution of dark matter, offering a more fundamental and less \emph{ad hoc} solution to cosmological tensions. This framework also predicts a \emph{suppression of power} in the CMB temperature spectrum on large angular scales, offering a potential physical explanation for the observed low-$\ell$ anomaly. By deriving these effects from a fundamental action, our work provides a unified, testable alternative to $\Lambda$CDM that can be constrained by next-generation cosmological surveys and gravitational wave observations.

\end{abstract}
\maketitle
\newpage
\section{Introduction}

The standard cosmological model, $\Lambda$-Cold Dark Matter ($\Lambda$CDM), has been remarkably successful in describing a wide array of astronomical observations \cite{2000-Padmanabhan-TheoreticalAstrophysicsVolume,2005-Mukhanov-PhysicalFoundationsCosmology,2008-Weinberg-Cosmology,Peebles:2020bdl}. Despite its successes, the $\Lambda$CDM model faces significant challenges, both theoretical and observational \cite{Perivolaropoulos:2021jda}. The fundamental nature of its two main components, dark energy (DE) and dark matter (DM), remains unknown~\cite{Scott:2018adl,Peebles:2025lmy}. Furthermore, persistent tensions have emerged between different cosmological datasets \cite{Abdalla:2022yfr}. The most prominent of these is the \emph{Hubble tension} \cite{2021-DiValentino.etal-Class.Quant.Grav.,Schoneberg:2021qvd,Kamionkowski:2022pkx}: a statistically significant discrepancy between the value of the Hubble constant ($H_0$) measured from the local distance ladder \cite{2016-Riess.others-Astrophys.J.} and the value inferred from Cosmic Microwave Background (CMB) anisotropies \cite{2018-Planck-AA}. A related tension for the matter clustering amplitude, $\sigma_8$\,, between CMB predictions and measurements from weak lensing surveys, which has existed for years, recently reduced to less than 1$\sigma$ with the KiDS final data release \cite{2025-KiDS}. These tensions, coupled with recent findings from the Dark Energy Spectroscopic Instrument (DESI) that hint at a dynamically evolving dark energy rather than a constant $\Lambda$ \cite{2025-Gu.others-}, suggest that the standard model may be incomplete and that new physics is required.

A central issue in moving beyond $\Lambda$CDM lies in how the dark sector's components are modeled. In standard analyses, dark matter and dark energy are treated as perfect fluids~\cite{2000-Padmanabhan-TheoreticalAstrophysicsVolume,2005-Mukhanov-PhysicalFoundationsCosmology,2008-Weinberg-Cosmology}. While computationally convenient and effective for describing background evolution and large-scale phenomena, the fluid description represents a low-energy, phenomenological approximation of the underlying physics~\cite{Prigogine:1989zz,CosmoVerseNetwork:2025alb}. A significant limitation is its inherent \emph{degeneracy}: a single fluid equation of state can be realized by multiple, distinct field theories~\cite{Padmanabhan:2002cp}. For instance, both a canonical scalar field and a $k-$essence scalar field, despite their fundamentally different kinetic terms and dynamical properties, can yield a perfect fluid with an identical equation of state~\cite{Copeland:2006wr, Bamba:2012cp}. This degeneracy makes it impossible to discern the true nature of the cosmic components from fluid models alone.

This limitation becomes especially critical when considering interactions between dark energy and dark matter {whose understanding is crucial for potentially addressing the shortcomings of the $\Lambda$CDM model \cite{2021-DiValentino.etal-Class.Quant.Grav., Wang:2016lxa, Wang:2024vmw, Wang:2025znm}}. In the fluid picture, such interactions are typically introduced \emph{ad hoc} as source terms in the conservation equations, typically dependent on the energy densities and possibly the Hubble rate \cite{Yang:2016evp, Marcondes:2016reb, Feng:2016djj, Costa:2016tpb}. However, without a foundational field-theoretic underpinning, the true dynamics of such an interaction remain obscure. An interaction term introduced at the fluid level does not reveal whether it arises from a direct coupling between the fundamental fields, a transformation of the gravitational sector, or an emergent property of the system.

A field-theoretic description, in contrast, provides a more fundamental and robust framework for investigating DE-DM interactions~\cite{2021-Johnson.Shankaranarayanan-Phys.Rev.D, 2022-Johnson.etal-JCAP, Bansal:PRD:2025}. By starting from an action principle involving specific scalar fields (or other fundamental fields) for dark energy and dark matter, interactions are naturally encoded in the Lagrangian. This approach not only provides a concrete physical origin for the interaction but also allows for a detailed investigation of the dynamics of these interacting fields. Crucially, a field-theoretic framework can reveal momentum-dependent interactions or other complex behaviors that are inherently averaged out or unresolvable in a simple fluid description. Furthermore, it offers the potential to connect cosmological interactions to fundamental particle physics or high-energy gravitational theories, thereby shedding light on the microphysical origin of dark energy and dark matter and their observed interplay. Thus, while fluid descriptions are practical, a deeper, field-theoretic investigation is indispensable for truly understanding the nature and dynamics of dark sector interactions~\cite{2021-Johnson.Shankaranarayanan-Phys.Rev.D,Bansal:PRD:2025}.

This need for a more fundamental description becomes particularly urgent when exploring solutions to the aforementioned cosmological tensions. Among the theoretical models proposed to address these discrepancies, \emph{Early Dark Energy (EDE)} models have garnered significant attention, primarily for their potential to alleviate the Hubble tension \cite{2016-Karwal.Kamionkowski-Phys.Rev.D, 2018-Poulin.etal-Phys.Rev.Lett}. Following its phenomenological success, several field-theoretic models of EDE have been proposed~\cite{Alexander:2019rsc, Smith:2019ihp, Agrawal:2019lmo, Niedermann:2019olb, Sakstein:2019fmf, Karwal:2021vpk, McDonough:2021pdg, Lin:2022phm, Ye:2020btb} (see Ref.~\cite{2023-Poulin.etal-Phys.DarkUniv.} for a recent review). In majority of these models, the typical EDE scalar field is initially {frozen in its potential due to Hubble friction} followed by rolling down and subsequent decay of its energy density faster than matter. This transient dark energy component in the pre-recombination era alters the expansion history to reconcile conflicting observations. However, such models rely on finely-tuned, specific forms of scalar field potentials and are susceptible to radiative corrections that can disrupt the desired evolution. Moreover, these approaches often treat early- and late-time dark energy as separate entities, failing to explore potential cosmological connections between the two. This highlights the need for a robust and well-motivated, first-principles framework to connect promising solutions like EDE to the underlying physics of the dark sector.

Motivated by these limitations, in this work, we propose a general action based on \emph{disformal couplings} that provides a fundamental field-theoretic framework for DE-DM interactions. Unlike previous models, our approach is capable of generating a rich variety of interaction scenarios, including the widely studied \emph{conformal couplings}~\cite{Amendola:1999er, Boehmer:2008av, Beyer:2010mt, CarrilloGonzalez:2017cll, 2021-Johnson.Shankaranarayanan-Phys.Rev.D, 2022-Johnson.etal-JCAP, Bansal:PRD:2025}, disformal couplings \cite{VanDeBruck:2017mua, Benisty:2021cmq, Benisty:2022lox, Benisty:2023dkn} as well as a unique \emph{pure momentum coupling} at the linear perturbation level.

Importantly, our framework includes a class of pure disformal couplings that naturally predicts an \emph{interacting Early Dark Sector}. In this model, the dynamics of dark energy resembles that of the phenomenological fluid-EDE models \cite{2016-Karwal.Kamionkowski-Phys.Rev.D}, but it fundamentally differs from typical scalar-field-EDE models \cite{Smith:2019ihp, Agrawal:2019lmo}. Specifically, the disformal interactions cause a suppression of \emph{Hubble friction}, leading to a constant kinetic energy for the DE scalar field at early times. As the energy density of dark matter dilutes, the coupling weakens, triggering the dilution of the DE kinetic energy (at a rate of $a^{-6}$), which eventually leads to a potential-dominated phase characteristic of late-time dark energy.

Compared to existing EDE scenarios, our model does not rely on a finely-tuned potential. The EDE-like behavior (the three-stage evolution) emerges naturally as a consequence of the \emph{interaction dynamics} between the dark energy and dark matter fluids coupled with the large and diluting dark matter energy density at early times, not the shape of the potential.
Additionally, this model leads to a suppression of power on large angular scales in the CMB temperature spectrum compared to the $\Lambda$CDM model. By deriving these interactions from a first-principles action, our work provides a robust theoretical foundation for exploring the new physics required to move beyond $\Lambda$CDM and resolve existing tensions.

This work is organized as follows. In Sec.~\ref{Sec:Framework}, we consider a general action comprised of arbitrary scalar field Lagrangians describing dark matter and dark energy including a non-minimal coupling of the DE field to gravitation and perform a general disformal transformation. We then establish a field-to-fluid mapping of the dark matter field for the resulting DE-DM interacting theory in the Einstein frame, which is useful for relating to cosmological observations. In Sec.~\ref{Sec:Cosmology}\,, we discuss the cosmology of the general interacting DE-DM framework and a subclass of the interactions which lead to pure-momentum coupling scenarios. In Sec.~\ref{Sec:Numerics}, we explore the interaction landscape within our framework by identifying seven different subclasses of DE-DM interactions (labeled as Models M1-M7 in Table \ref{Table:iDEDM}). We then perform a detailed investigation of the observational consequences for the background cosmology, matter and CMB power spectra for three specific cases of interest: pure-momentum coupling models (Models M6 and M7) and pure disformal interactions (Model M3). We then summarize our findings for all models in Sec.~\ref{Subsec:Summary} (Table \ref{Table:Summary}). We conclude and discuss future directions in Sec.~\ref{Sec:Conclusions}. The appendices contain the following details: 
Appendix \ref{Appendix:DT-DHOST-Quintessence} provides the details supporting the derivation of our interacting framework in Sec.~\ref{Subsec:Disformal-Framework}. 
In Appendix \ref{Appendix:Summary-Previous-Works}, we summarize the interacting frameworks derived from $f(R)$ and Horndeski theories in our previous works \cite{2021-Johnson.Shankaranarayanan-Phys.Rev.D, Bansal:PRD:2025}. 
In Appendix \ref{Appendix:Conformal} and \ref{Appendix:M4,M5}, we investigate the cosmological implications of conformal couplings (Models M1, M2) and other disformal couplings (Models M4, M5), respectively. 
Appendix \ref{Appendix:CLASS-Comparison} details the implementation of our general interacting DE-DM framework using a custom Python code and testing its accuracy against the Boltzmann code \texttt{CLASS} \cite{Lesgourgues:2011re} for the $\Lambda$CDM model.

In this work, we use the the metric signature $(-,+,+,+)$ and natural units where $c=1, M_{\mathrm{Pl}}^2 = (8 \pi G)^{-1}$. Greek alphabets denote the 4-dimensional space-time coordinates, and Latin alphabets denote the 3-dimensional spatial coordinates. Overbarred quantities (like $\overline{\rho}(t), \overline{p}(t)$) are evaluated for the FLRW background, a \textit{overdot} represents the derivative with respect to cosmic time $t$, and $H$ denotes the cosmic Hubble parameter. Unless otherwise specified, subscript `$\upchi$', `$\phi$' and `$X$' denote partial derivatives with respect to $\upchi$, $\phi$, and the kinetic term of the scalar field $\phi$, respectively. Variables with a \emph{Tilde} ($\Tilde{x}$) correspond to quantities in the original frame; otherwise, they correspond to the Einstein frame.

\section{Field-theoretic description and fluid mapping of a disformally interacting dark sector}
\label{Sec:Framework}

Building on the systematic field-theoretic approach to interacting dark energy-dark matter developed in our previous works \cite{2021-Johnson.Shankaranarayanan-Phys.Rev.D,Bansal:PRD:2025}, we now generalize the analysis. 
A summary of the interacting framework developed in these works is presented in Appendix \ref{Appendix:Summary-Previous-Works}. 
While our earlier studies focused exclusively on interactions generated by conformal transformations from quadratic-Horndeski theories, this work extends that framework to \emph{disformal transformations}~\cite{Bekenstein:1992pj}. Disformal transformations offer a wider range of metric possibilities than their conformal counterparts, primarily because they do not generally preserve the angles between spacetime curves. This can alter the shape of light cones, depending on the form of the disformal factor $D(\phi, X)$. It is important to note, however, that some of these transformations may lead to unphysical results, such as a change in the metric's signature~\cite{Bekenstein:1992pj}. In this work, we choose the disformal couplings that are physical at all times and length scales. 

\subsection{Scalar field interactions from a disformal origin}
\label{Subsec:Disformal-Framework}

We consider the following action:
\bea \label{Eq:NMC Action}
S = \int \mathrm{d}^4x\sqrt{-\Tilde{g}}\left[ \frac{M_{\mathrm{Pl}}^2}{2}\Tilde{f}(\phi, \Tilde{X})\Tilde{R} + \Tilde{\mathcal{L}}_{\phi}(\Tilde{g}_{\mu\nu}, \phi) + \Tilde{L}_{\chi}(\Tilde{g}_{\mu\nu}, \chi) \right] + \int\mathrm{d}^4x\sqrt{-g}\mathcal{L}_{_{\mathrm{M}}}(g_{\mu\nu}, \Theta_{_{\mathrm{M}}}) \,.
\eea
Here, $\Tilde{g}_{\mu\nu}$, $\Tilde{\mathcal{L}}_{\phi}$ and $\Tilde{\mathcal{L}}_{\chi}$ are in the \emph{Jordan frame}, while $g_{\mu\nu}$ and $\mathcal{L}_{_{\mathrm{M}}}$ (visible matter) are in the \emph{Einstein frame}.
This action describes two gravitationally interacting scalar fields ($\phi$ and $\upchi$). 
A non-minimal coupling exists between the field $\phi$ and the Ricci scalar described by the arbitrary function $\Tilde{f}(\phi, \Tilde{X})$. $\Tilde{X} = -\Tilde{g}^{\mu\nu}\Tilde{\nabla}_{\mu}\phi\Tilde{\nabla}_{\nu}\phi/2$ is the kinetic term of $\phi$. 
The non-minimal coupling function $\Tilde{f}(\phi, \Tilde{X})$ is the most general form and encompasses all the models considered in Ref.~\cite{Bansal:PRD:2025}, as well as the new cases we will explore in this work.
Note that the scalar field Lagrangians, $\Tilde{\mathcal{L}}_{\phi}$ and $\Tilde{\mathcal{L}}_{\chi}$, must be chosen to avoid ghosts or instabilites. Examples include Horndeski theories and certain classes of DHOST theories \cite{BenAchour:2016cay}. For details, please refer to Appendix \ref{Appendix:DT-DHOST-Quintessence}.

Our first step is to remove this non-minimal coupling term via a metric transformation. For this purpose, we perform a general disformal transformation given by~\cite{Bekenstein:1992pj}:
\begin{equation} \label{Eq:GDT}
\Tilde{g}_{\mu\nu} = C(\phi, X)g_{\mu\nu} + D(\phi, X)\phi_{\mu}\phi_{\nu}\, ,
\end{equation}
where $C$ and $D$ are arbitrary functions of $\phi$ and $X$, and $X \equiv -g^{\mu\nu}\nabla_{\mu}\phi\nabla_{\nu}\phi/2 \equiv -\phi_{\mu}\phi^{\mu}/2$. 
Under the above disformal transformation, the action \eqref{Eq:NMC Action} transforms to the following form:
\bea \label{Eq:Transformed NMC action}
S = \int \d^4x\sqrt{-g}
& &
\left[ \frac{M_{\mathrm{Pl}}^2}{2}\Tilde{f}(\phi, \Tilde{X}) \sqrt{C(C - 2XD)}R + \mathcal{L}_{\phi}(g_{\mu\nu}, \phi)
\nonumber \right. \\ 
& & \left.
+ \, C^{3/2}\sqrt{C - 2DX}\Tilde{L}_{\chi}\left(C(\phi, X)g_{\mu\nu} + D(\phi, X)\phi_{\mu}\phi_{\nu}, \chi \right) 
\nonumber \right. \\ 
& & \left.
 + \mathcal{L}_{_{\mathrm{M}}}(g_{\mu\nu}, \Theta_{_{\mathrm{M}}}) 
\right]\,.
\eea 
 Let us now examine the transformation of each term in action \eqref{Eq:NMC Action} to those in action \eqref{Eq:Transformed NMC action}. Under the disformal transformation \eqref{Eq:GDT}, the Ricci scalar $\Tilde{R}$ transforms into a combination of  $R$ and terms explicitly related to the scalar field $\phi$ and its derivatives. 
Consequently, the transformed Lagrangian for the scalar field $\phi$, denoted as $\mathcal{L}_{\phi}$, will include these additional contributions from the metric transformation of the Ricci scalar and the original Lagrangian $\Tilde{L}_{\phi}$, along with the Jacobian factor from the metric transformation of $\sqrt{-\Tilde{g}}$.
The Lagrangian $\Tilde{L}_{\chi}$ remains invariant in form, as the disformal transformation depends only on $\phi$, and we do not perform any redefinition of the scalar field $\upchi$. 

For the disformal transformation to be invertible and to preserve the metric signature and causality, the conformal and disformal factors $C(\phi, X)$ and $D(\phi, X)$, respectively, must satisfy the following conditions:
\begin{equation} \label{Eq:Disformal invertibility condition}
C > 0 \, ,~~C(C - XC_X + 2X^2D_X) \neq 0\, ,~~ C - 2DX > 0\,. 
\end{equation}
The non-minimal coupling between the field $\phi$ and gravitation is effectively removed if the non-minimal coupling function $\Tilde{f}(\phi, \Tilde{X})$ satisfies the following condition: 
\bea \label{Eq:Remove NMC}
\Tilde{f}(\phi, \Tilde{X}) \sqrt{C(C - 2DX)} = 1\,.
\eea 
This transforms the action into the following form in the Einstein frame:
{\small
\bea \label{Eq:EinsteinF action}
S = \int \d^4x\sqrt{-g}
& &
\left[ \frac{M_{\mathrm{Pl}}^2}{2}R + \mathcal{L}_{_{\mathrm{M}}}(g_{\mu\nu}, \Theta_{_{\mathrm{M}}})
+ \mathcal{L}_{\phi}(g_{\mu\nu}, \phi)
\nonumber \right. \\ 
& & \left.
+ C^{3/2}\sqrt{C - 2DX}\Tilde{L}_{\chi}\left(C(\phi, X)g_{\mu\nu} + D(\phi, X)\phi_{\mu}\phi_{\nu}, \chi \right) \right].
\eea 
}
This action demonstrates a key result: the non-minimal gravitational interaction in the original frame is mapped to a non-gravitational interaction between the fields $\phi$ and $\upchi$ in the Einstein frame, where the gravitational sector is standard. 
This structure means \cite{Amendola:2003eq, Bean:2008ac, Amendola_2012, Miranda:2017rdk, VanDeBruck:2017mua, Karwal:2021vpk}:
\begin{enumerate}
\item {\bf In the Einstein Frame:} Visible matter evolves independently (following geodesics), while dark matter ($\upchi$) and dark energy ($\phi$) are directly coupled to each other.
\item {\bf In the Jordan Frame:} Dark matter and dark energy are uncoupled, and visible matter is coupled to the scalar field $\phi$.
\end{enumerate}
This systematic process allows for the generation of a wide class of non-gravitational interactions in the dark sector. The precise nature of these interactions is determined by the initial non-minimal coupling $\Tilde{f}(\phi, \Tilde{X})$ and the chosen disformal transformation functions, $C$ and $D$. This highlights that the final Lagrangian describing the interacting scalar fields is not unique.
Therefore, through disformal transformations, one can \emph{systematically generate non-gravitational interactions} between the scalar fields $\phi$ and $\upchi$ with arbitrary field descriptions.

Our next objective is to explore the DE-DM interactions generated by a general disformal coupling. To this end, we extend the field-fluid mapping of our earlier works, summarized in Appendix \ref{Appendix:Summary-Previous-Works}, by describing the $\upchi$-field as a $k-$essence scalar field~\cite{2000-Armendariz-Picon.etal-Phys.Rev.Lett.,2001-Armendariz-Picon.etal-Phys.Rev.D}. Additionally, in Appendix \ref{Appendix:DT-DHOST-Quintessence}, we provide a detailed discussion of how one can obtain a $k$-essence description of the scalar field $\phi$ in the Einstein frame by starting with a ghost-free class of DHOST theories in the Jordan frame. This is important as we shall consider quintessence descriptions of the field $\phi$ when we conduct a numerical investigation of the cosmological implications of the interacting framework later.

\subsection{Dark sector interactions with a generalized disformal coupling}
\label{Subsec:DE-DM:GDC}

For the purposes of this study, we shall assume a $k-$essence description for the scalar field $\upchi$ in the original frame~\cite{2000-Armendariz-Picon.etal-Phys.Rev.Lett.,2001-Armendariz-Picon.etal-Phys.Rev.D}. This is motivated by the desire to maintain transparency and understand the effects of specific interaction types. The $k-$essence Lagrangian is defined as:
\bea
\Tilde{L}_{\chi}(\Tilde{g}_{\mu\nu}, \upchi) = \Tilde{P}_1\left( \upchi, \Tilde{Y} \right) \, ,
\eea
where $\Tilde{P}_1$ is an arbitrary function of the field $\upchi$ and its kinetic term $\Tilde{Y} = -\Tilde{g}^{\mu\nu}\Tilde{\nabla}_{\mu}\upchi\Tilde{\nabla}_{\nu}\upchi/2$. The action~\eqref{Eq:EinsteinF action} then transforms to the following form in the Einstein frame:
\begin{equation} \label{Eq:GDT:Final action}
S = \int \d^4x\sqrt{-g}
\left[ \frac{M_{\mathrm{Pl}}^2}{2}R + \mathcal{L}_{\phi}(g_{\mu\nu}, \phi)
+ C^{3/2}\sqrt{C - 2XD}\times P_1\left( \chi, Z \right)
\right]\, ,
\end{equation}
where
\begin{equation}
\begin{aligned}
&
Y \equiv -\frac{1}{2}g^{\mu\nu}\nabla_{\mu}\chi\nabla_{\nu}\chi 
= - \frac{1}{2} \chi_{\mu}\chi^{\mu}\,,
\\
&
Z \equiv \frac{Y}{C} + \frac{D}{2C}\frac{(\chi^{\mu}\phi_{\mu})^2}{(C - 2XD)} \,\,.
\end{aligned}
\end{equation}
Varying the action with respect to the metric $g_{\mu\nu}$ will yield the Einstein equations, 
\begin{equation} \label{Eq:Einstein equations}
    G_{\mu\nu} = \frac{1}{M_{\mathrm{Pl}}^2}T_{\mu\nu}\, ,
\end{equation}
where, the energy-momentum tensor is given by:
{\small
\begin{eqnarray} \label{Eq:GDT:Total EM-tensor}
T_{\mu\nu} = 
& &
T_{\mu\nu}^{(\mathcal{L}_\phi)} + g_{\mu\nu}C^{3/2}\sqrt{C - 2DX}\,P_1 + \sqrt{C(C - 2DX)}\chi_{\mu}\chi_{\nu}\,P_{1Z}
+ \frac{\sqrt{C}D^2(\phi_{\alpha}\chi^{\alpha})^2}{(C - 2DX)^{3/2}}\phi_{\mu}\phi_{\nu}\,P_{1Z}
\nonumber \\
& - &
\sqrt{\frac{C}{C - 2DX}}D(\phi_{\alpha}\chi^{\alpha})(\chi_{\mu}\phi_{\nu} + \phi_{\mu}\chi_{\nu})\,P_{1Z} - \frac{C^{3/2}D}{\sqrt{C - 2DX}}\phi_{\mu}\phi_{\nu}\,P_1
\nonumber \\
& + &
C_{X}\phi_{\mu}\phi_{\nu}\left[ \sqrt{\frac{C}{C - 2DX}}(2C - 3DX)\,P_1 - \sqrt{\frac{C - 2DX}{C}}\left( Y + \frac{D(C - XD)(\phi_{\alpha}\chi^{\alpha})^2}{(C - 2XD)^2} \right)\,P_{1Z} \right]
\nonumber \\
& + &
D_X\phi_{\mu}\phi_{\nu}\left[ -\frac{C^{3/2}}{\sqrt{C - 2DX}}X\,P_1 + \frac{1}{2}\left( \frac{C}{C - 2DX} \right)^{3/2}(\phi_{\alpha}\chi^{\alpha})^2\,P_{1Z}  \right]\,.
\end{eqnarray}
}
The specific form of the energy-momentum tensor $T_{\mu\nu}^{(\mathcal{L}_\phi)}$ is directly determined by the Lagrangian $\mathcal{L}_{\phi}$ which describes the dynamics of the scalar field $\phi$.
However, the non-gravitational interactions between $\phi$ and $\upchi$ in the Einstein frame leads to violation of the 
energy conservation of the individual components. Hence, there is no unique decomposition of the stress-energy tensor. 
To enable the mapping of the $\upchi$ field to a perfect fluid and define its corresponding four-velocity, we split the total energy-momentum tensor in the following manner:
{\small
\begin{subequations}
\begin{eqnarray}
T^{(\phi)}_{\mu\nu} &= & 
T_{\mu\nu}^{(\mathcal{L}_\phi)}
+ D_X\phi_{\mu}\phi_{\nu}\left[ -\frac{C^{3/2}}{\sqrt{C - 2DX}}X\,P_1 + \frac{1}{2}\left( \frac{C}{C - 2DX} \right)^{3/2}(\phi_{\alpha}\chi^{\alpha})^2\,P_{1Z}  \right] \nonumber \\
&+ & 
C_{X}\phi_{\mu}\phi_{\nu}\left[ \sqrt{\frac{C}{C - 2DX}}(2C - 3DX)\,P_1 - \sqrt{\frac{C - 2DX}{C}}\left( Y + \frac{D(C - XD)(\phi_{\alpha}\chi^{\alpha})^2}{(C - 2XD)^2} \right)\,P_{1Z} \right] \nonumber \\
\label{SubEq:GDT:DE EM-tensor} 
& - & 
\frac{C^{3/2}D}{\sqrt{C - 2DX}}\phi_{\mu}\phi_{\nu}\,P_1\,, \\
T^{(\chi)}_{\mu\nu} & = & 
g_{\mu\nu}C^{3/2}\sqrt{C - 2DX}\,P_1 + \sqrt{C(C - 2DX)}\chi_{\mu}\chi_{\nu}\,P_{1Z} + \frac{\sqrt{C}D^2(\phi_{\alpha}\chi^{\alpha})^2}{(C - 2DX)^{3/2}}\phi_{\mu}\phi_{\nu}\,P_{1Z} \nonumber \\
& -& \sqrt{\frac{C}{C - 2DX}}D(\phi_{\alpha}\chi^{\alpha})(\chi_{\mu}\phi_{\nu} + \phi_{\mu}\chi_{\nu})\,P_{1Z} \, .
\label{SubEq:GDT:DM EM-tensor}
\end{eqnarray}
\end{subequations}
}
The interaction between the two scalar fields can be described as~\cite{2021-Johnson.Shankaranarayanan-Phys.Rev.D}:
\begin{equation} \label{Eq:Interaction strength definition}
    Q_{\nu} = \nabla^{\mu}T^{(\chi)}_{\mu\nu} = -\nabla^{\mu}T^{(\phi)}_{\mu\nu}\,.
\end{equation}
The field representation yields a detailed, though complex, expression for the interaction strength $Q_{\nu}$. To establish a more direct connection with physical observables, we present $Q_{\nu}$  
using quantities derived from the energy-momentum tensor of the $\upchi$ field 
($T_{\mu\nu}^{(\chi)}$): 
\begin{eqnarray} \label{Eq:GDT:IS:TensorF}
Q_{\nu} = 
& &
\frac{T^{{(\chi)}}}{2C}\left[ C_{\nu} - \frac{D(C_{\alpha}\phi^{\alpha})\phi_{\nu}}{(C - 2DX)} \right]  
- \frac{D}{(C - 2DX)}T^{{(\chi)}}_{\alpha\beta}\phi^{\alpha\beta}
\\ 
& + &
\frac{D}{C(C - 2DX)}\left[ T^{{(\chi)}}_{\alpha\beta}C^{\alpha}\phi^{\beta}\phi_{\nu} + \frac{T^{{(\chi)}}_{\alpha\beta}q^{\alpha\beta}}{6}\left( C_{\mu}\phi^{\mu}\phi_{\nu} + 2XC_{\nu} \right) \right]
\nonumber 
\\
& + &
\frac{1}{(C - 2DX)}\left[ - \frac{D}{2C}\left( T^{{(\chi)}}_{\alpha\beta}\phi^{\alpha}\phi^{\beta} + \frac{2X}{3}T^{{(\chi)}}_{\alpha\beta}q^{\alpha\beta} \right)\left( 2XD_{\nu} + D_{\mu}\phi^{\mu}\phi_{\nu} \right) \right. 
\nonumber 
\\ 
& - & \left. 
T^{{(\chi)}}_{\alpha\beta}D^{\alpha}\phi^{\beta}\phi_{\nu} + \frac{D_{\nu}}{2}T_{\alpha\beta}\phi^{\alpha}\phi^{\beta}
\right] + \frac{C}{\sqrt{C - 2DX}}\nabla_{\mu}\left[ \frac{D}{3C\sqrt{C - 2DX}}T^{(\chi)i}_{i}\phi_{\nu} \right] \nonumber
\, . 
\nonumber
\end{eqnarray} 
where $T^{(\chi)}$ corresponds to the trace of $T_{\mu\nu}^{(\chi)}$. 
The analysis presented here represents the most general form of DE-DM interactions arising from a disformal transformation of the field theoretic action \eqref{Eq:NMC Action}, assuming a $k-$essence description for $\upchi$. This is because setting the disformal factor $D=0$ recovers the extended conformal coupling interaction strength derived in our previous work~\cite{Bansal:PRD:2025}. The close correspondence of this conformal coupling limit to the findings in Refs.~\cite{2021-Johnson.Shankaranarayanan-Phys.Rev.D,Bansal:PRD:2025} allows us to establish a physical mapping of the two scalar fields to the dark sector. Consequently, we will identify $\phi$ as the dark energy (DE) field and $\upchi$ as the dark matter (DM) fluid throughout the rest of this work.

\subsection{Field-Fluid mapping of dark matter for the generalized disformal coupling}
\label{Section:GDT:FFMapping}

As mentioned in the introduction, matter fields in cosmology are often described as a perfect fluid. Although field theory provides a fundamental description, cosmological observations are often more effectively analyzed using this fluid framework~\cite{Calabrese:2009zza,Kopp:2018zxp}. To facilitate this connection, we translate the field-theoretic representation of dark matter into a perfect fluid. The energy-momentum tensor for such a perfect fluid takes the following form:
\begin{eqnarray} \label{Eq:Perfect fluid:EM-tensor}
T_{\mu\nu} = p_{\mathrm{DM}}g_{\mu\nu} + (\rho_{\mathrm{DM}} + p_{\mathrm{DM}})u_{\mu}u_{\nu}\, .
\end{eqnarray}
To map $T_{\mu\nu}^{(DM)}$ ($\equiv T_{\mu\nu}^{(\chi)}$) to the perfect fluid tensor, we must consistently define the DM fluid four-velocity, energy density and pressure. 
We begin this process by defining the DM fluid four-velocity in the following manner:
\begin{eqnarray} \label{Eq:GDT:DM fluid:four-velocity}
    u_{\mu} = -\left( \chi_{\mu} - \frac{D(\phi_{\alpha}\chi^{\alpha})}{(C - 2DX)}\phi_{\mu} \right)\times \left( - g^{\alpha\beta}\chi_{\alpha}\chi_{\beta} + \frac{2D(C - DX)(\phi_{\alpha}\chi^{\alpha})^2}{(C - 2DX)^2} \right)^{-1/2}\, .
\end{eqnarray}
This corresponds to mapping the DM fluid energy density $\rho_{\mathrm{DM}}$ and pressure $p_{\mathrm{DM}}$, via the following relations:
\begin{subequations} \label{Eq:GDT:DM fluid:rho & p}
\begin{align}
\rho_{\mathrm{DM}} = 
& \
\sqrt{C(C - 2DX)}\left[ 2\left(Y + \frac{D(C - DX)(\phi_{\alpha}\chi^{\alpha})^2}{(C - 2DX)^2} \right)\,P_{1Z} - C\,P_1 \right]\, ,
\\
p_{\mathrm{DM}} = 
& \
C^{3/2}\sqrt{C - 2DX}\,P_1 \, .
\end{align}
\end{subequations}
With these definitions, the DM perfect fluid is then described by the energy-momentum tensor in Eq.~\eqref{Eq:Perfect fluid:EM-tensor}. Additionally, we can rewrite $T_{\mu\nu}^{(DE)}$ in the following form:
\bea \label{Eq:GDT:DE-tensor:Fluid form}
T_{\mu\nu}^{(DE)} = 
& & 
T_{\mu\nu}^{(\mathcal{L}_\phi)} - \frac{D}{(C - 2DX)}p_{\mathrm{DM}}\phi_{\mu}\phi_{\nu} + \frac{C_X}{2C}\left[ - \rho_{\mathrm{DM}} + \frac{(3C - 4DX)}{(C - 2DX)}p_{\mathrm{DM}} \right]\phi_{\mu}\phi_{\nu}
\nonumber \\ 
& + & 
\frac{D_{X}}{2C}\left[ (\rho_{\mathrm{DM}} + p_{\mathrm{DM}})(u_{\alpha}\phi^{\alpha})^2 - \frac{2XC}{(C - 2DX)}p_{\mathrm{DM}} \right]\phi_{\mu}\phi_{\nu}\,.
\eea 
Consequently, the interaction strength $Q_{\nu}$ \eqref{Eq:GDT:IS:TensorF} can be rewritten as:
\bea \label{Eq:GDT:IS:Fluid form}
Q_{\nu} = 
& & 
\frac{(3p_{\mathrm{DM}} - \rho_{\mathrm{DM}})}{2C}\left[ C_{\nu} - \frac{D(C_{\alpha}\phi^{\alpha})\phi_{\nu}}{(C - 2DX)}  \right]
\nonumber \\
& + &
 \frac{D}{C(C - 2DX)}\left[ (\rho_{\mathrm{DM}} + p_{\mathrm{DM}})(\phi_{\alpha}u^{\alpha})(C_{\beta}u^{\beta})\phi_{\nu} + \frac{p_{\mathrm{DM}}}{2}\left( C_{\alpha}\phi^{\alpha}\phi_{\nu} + 2XC_{\nu} \right) \right]
\nonumber \\ 
& + &
\frac{1}{(C - 2DX)}\left[ - p_{\mathrm{DM}}XD_{\nu} - (\rho_{\mathrm{DM}} + p_{\mathrm{DM}})(\phi_{\alpha}u^{\alpha})(D_{\beta}u^{\beta})\phi_{\nu} 
\right. \\ 
& + & \left.
\frac{1}{2C}(\rho_{\mathrm{DM}} + p_{\mathrm{DM}})(\phi_{\alpha}u^{\alpha})^2\left( D_{\nu}(C - 2DX) - DD_{\beta}\phi^{\beta}\phi_{\nu} \right) \right]
\nonumber \\
& - &
\frac{D}{(C - 2DX)}(\rho_{\mathrm{DM}} + p_{\mathrm{DM}})\phi_{\alpha\beta}u^{\alpha}u^{\beta}\phi_{\nu} - \frac{Dp_{\mathrm{DM}}\phi_{\nu}}{2(C - 2DX)^2}\phi^{\mu}\left[ C_{\mu} - 2XD_{\mu} + 2D\phi_{\mu\sigma}\phi^{\sigma} \right]
\nonumber \\ 
& + &
\frac{D}{(C - 2DX)}\left[ p_{\mathrm{DM}}\phi_{\mu\nu}\phi^{\mu} + \phi_{\nu}\phi^{\mu}\nabla_{\mu}p_{\mathrm{DM}} \right]\, .
\nonumber
\eea 
For a pressureless DM fluid, $T_{\mu\nu}^{(DE)}$ and $Q_{\nu}$ reduce to the following forms:
\bea 
\label{Eq:GDT:DEtensor:Pressureless}
T_{\mu\nu}^{(DE)} &=& T_{\mu\nu}^{(\mathcal{L}_\phi)} + \rho_{DM}\left[ \frac{-C_X + D_X(u_{\alpha}\phi^{\alpha})^2}{2C} \right]\phi_{\mu}\phi_{\nu}\,, \\
\label{Eq:GDT:IS:Pressureless}
Q_{\nu} = 
& & 
\frac{\rho_{\mathrm{DM}}}{2C(C - 2DX)}\left[ D(C_{\alpha}\phi^{\alpha})\phi_{\nu} - (C - 2DX)C_{\nu} + 2D(\phi_{\alpha}u^{\alpha})(C_{\beta}u^{\beta})\phi_{\nu}
\right. \nonumber \\ 
& - & \left.
2C(\phi_{\alpha}u^{\alpha})(D_{\beta}u^{\beta})\phi_{\nu} + (u_{\alpha}\phi^{\alpha})^2\left( (C - 2DX)D_{\nu} - DD_{\beta}\phi^{\beta}\phi_{\nu} \right)
\right.
\nonumber \\
& - & \left.
2CD\phi_{\alpha\beta}u^{\alpha}u^{\beta}\phi_{\nu} \right]\,.
\eea
This is the first key result of this work, regarding which, we want to discuss the following points. First, in the special case where $D(\phi, X) = 0$, the above expression reduces to the conformal coupling results~\cite{Bansal:PRD:2025}. 
Furthermore, assuming an equation of state $p = \rho/3$, we find that while the interaction strength $Q_{\nu}$ vanishes for pure conformal couplings corresponding to $C(\phi, X) \neq 0, D(\phi, X) = 0$ \cite{Bansal:PRD:2025}, it remains non-zero in the presence of disformal couplings.
This implies that a dark energy-dark radiation type model can exhibit non-gravitational interactions of disformal origin~\cite{Bekenstein:1992pj}, a feature absent in conformal and extended conformal couplings. 
Consequently, a dark sector one field-one fluid interacting theory with interactions of the form \eqref{Eq:GDT:IS:Fluid form} can be mapped to a classical field theoretic description with DE-DM interactions. We also note from Eq.~\eqref{Eq:GDT:DM fluid:rho & p} that the DM fluid description cannot distinguish between quintessence and $k-$essence scalar fields $\upchi$, indicating the non-uniqueness of the interaction term \eqref{Eq:GDT:IS:Fluid form} for a field-fluid mapping of dark matter.

\section{Cosmology of the disformally interacting dark sector}
\label{Sec:Cosmology}

As discussed in Sec.~\ref{Sec:Framework}, the general disformal transformations describe a wide range of DE-DM interactions. We, now, look at the dynamics of these interactions in a spatially flat  Friedmann-Lemaître-Robertson-Walker (FLRW) cosmology. In a particular subclass of these models, we find a unique interaction signature at the linear perturbation level. Specifically, while the equation of motion for the dark matter density contrast, $\delta_{\mathrm{DM}}$, simplifies to a form with no coupling, the velocity perturbation equation contains non-zero terms resulting directly from the kinetic coupling. This leads to a framework with a \emph{pure momentum coupling at linear order in perturbation theory}.

To go about this, we work in the Newtonian gauge, described by the following metric in cosmic time~\cite{1992-Mukhanov.etal-PRep}: 
\bea \label{Eq:FLRW-LE}
{\rm d}s^2 = -\left[  1 + 2\Psi \right]{\rm d}t^2 + a^2(t)\left[  1 + 2\Phi \right]{\rm d\textbf{x}}^2\,,
\eea 
where $t$ denotes cosmic time, $a(t)$ denotes the scale factor and $\Psi \equiv \Psi(t, {\bf x})$ and $\Phi \equiv \Phi(t, {\bf x})$ denote the Newtonian potentials. At the background level in this spatially flat FLRW universe, a pressureless DM perfect fluid obeys the equation of motion:
\begin{equation}
\Dot{\rho}_{\mathrm{DM}} + 3H\rho_{\mathrm{DM}} = -Q_{0}\,,    
\end{equation} 
where $Q_{0}$ denotes the zeroth component of the interaction strength $Q_{\nu}$ in the FLRW background. $Q_{0} > 0$ corresponds to an energy transfer from DM to DE and vice versa. 
Within the homogeneous FLRW background, the spatial components of $Q_{\nu}$ vanish, leading to \emph{zero} net momentum coupling at this level.
For general disformal interactions, $Q_{0}$ takes the following form:
{\small
\bea  \label{Eq:GDT:IS:FLRW:Pressureless}
Q_{0} = 
\frac{\rho_{\mathrm{DM}}\Dot{\phi}}{2C[C - 2DX]}\left[ \left[ C_X(4DX - C) - 2CD_XX - 2CD \right] \Ddot{\phi} + C_{\phi}[4DX - C] - 2CD_{\phi}X \right],
\eea 
}
where $X = \Dot{\phi}^2/2$ is the DE kinetic term in the FLRW background. We assume a quintessence DE scalar field in the rest of this work, for which the energy density and pressure in the FLRW background are given by:
\bea
&&
\rho_{\mathrm{DE}} = X + V(\phi) + \rho_{\mathrm{DM}}\,X\,\left[ \frac{2XD_X - C_X}{C} \right]\,,
\nonumber \\ 
&&
p_{\mathrm{DE}} = X - V(\phi)\,,
\eea
where $V(\phi)$ represents the scalar field potential. It is clear that the explicit kinetic term dependence of the couplings $C$ and $D$ directly contributes to the DE energy density while leaving its pressure unchanged. This imposes a constraint on the allowed forms of these kinetic couplings, as they must be chosen to avoid negative energy density states for DE. The DE scalar field is governed by the following equation of motion:
\begin{align}
\Ddot{\phi}
& 
\left[ 1 
+ \frac{\rho_{\mathrm{DM}}}{C^2} \left( 
   -CC_X 
   + XC_X^2 
   - XCC_{XX} 
   + 4XCD_X 
   - 2X^2C_XD_X 
   + 2X^2CD_{XX} 
\right) \right. \nonumber \\
& 
\left.
\quad {} + \frac{\rho_{\mathrm{DM}}}{2(C - 2DX)}
\left( 
    1 - \frac{X}{C}(C_X - 2XD_X) 
\right)
\left( 
    2D + 2XD_X + C_X\left( 1 - \frac{4DX}{C} \right) 
\right)
\right]
\nonumber \\ 
&
= - 3H\Dot{\phi} - V_{\phi} - \frac{\rho_{\mathrm{DM}}X}{C^2}\left[ C_{\phi}C_X - CC_{\phi X} - 2XC_{\phi}D_X + 2XCD_{\phi X} \right]
\nonumber \\ 
&
~~~~
+ \frac{\rho_{\mathrm{DM}}}{2(C - 2DX)}\left[ 1 - \frac{X}{C}(C_X - 2XD_X) \right]\left[ C_{\phi}\left( \frac{4DX}{C} - 1 \right) - 2XD_{\phi} \right]\,.
\end{align}
Note that in the limit $C\rightarrow1, D\rightarrow0$, the above equation reduces down to the standard Klein-Gordan equation.

We now examine the modifications to linear perturbations in the matter fields introduced by the coupling. The perturbed energy density and pressure of the DE field are given by:
\bea
\delta\rho_{\mathrm{DE}} =
& & 
+ ~ \Dot{\overline{\phi}}\Dot{\delta\phi} - \Dot{\overline{\phi}}^2\Psi + \overline{V}_{\phi}\delta\phi - \overline{\rho}_{\mathrm{DM}}\left[ \overline{X}\delta_{\mathrm{DM}} + \Dot{\overline{\phi}}\Dot{\delta\phi} - \Dot{\overline{\phi}}^2\Psi \right]\left[ \frac{\overline{C}_X - 2\overline{X}\overline{D}_X}{\overline{C}} \right]
\nonumber \\ 
& &
- ~ \frac{\overline{X}\overline{\rho}_{\mathrm{DM}}}{\overline{C}^2}\left[ \delta\phi\left( \overline{C}(\overline{C}_{\phi X} - 2\overline{X}\overline{D}_{\phi X}) - \overline{C}_{\phi}(\overline{C}_X - 2\overline{X}\overline{D}_X) \right)
\right. \nonumber \\ 
& & \left. 
\hspace{1.75cm}
+ (\Dot{\overline{\phi}}\Dot{\delta\phi} - \Dot{\overline{\phi}}^2\Psi)\left( \overline{C}( \overline{C}_{XX} - 2\overline{D}_X - 2\overline{X}\overline{D}_{XX} ) - \overline{C}_X( \overline{C}_X - 2\overline{X}\overline{D}_X ) \right) \right]
\nonumber \\
\delta p_{\mathrm{DE}} =
& & 
+ \Dot{\overline{\phi}}\Dot{\delta\phi} - \Dot{\overline{\phi}}^2\Psi - \overline{V}_{\phi}\delta\phi\,.
\eea
Consistent with our background analysis, the kinetic dependence of these interactions modifies the perturbed energy density of the DE field while its pressure remains unchanged.
The equations of motion governing the DM density contrast $\delta_{\mathrm{DM}}$ and velocity perturbation $v_{\mathrm{DM}}$ are modified to the following form:
\bea \label{Eq:pt:DM}
\Dot{\delta}_{\mathrm{DM}} + 3\Dot{\Phi} - \frac{k}{\overline{a}}v_{\mathrm{DM}} =
& & 
\left[ \frac{\overline{Q}_{0}\delta_{\mathrm{DM}} - \delta Q_{0}}{\overline{\rho}_{\mathrm{DM}}} \right]
\,,
\\ 
\Dot{v}_{\mathrm{DM}} + \overline{H}v_{\mathrm{DM}} + \frac{k}{\overline{a}}\Psi = 
& &  
\frac{\overline{Q}_{0}}{\overline{\rho}_{\mathrm{DM}}}\left[ v_{\mathrm{DM}} + \frac{k}{\overline{a}}\frac{\delta\phi}{\Dot{\overline{\phi}}} \right] + \frac{k}{\overline{a}}\left[ \frac{\overline{C}_X - 2\overline{X}\overline{D}_X}{2\overline{C}} \right]\left[ \Ddot{\overline{\phi}}\delta\phi - \Dot{\overline{\phi}}\Dot{\delta\phi} + \Dot{\overline{\phi}}^2\Psi \right]\,,
\nonumber 
\eea
where $\delta Q_{0}$ denotes the time (zeroth) component of the perturbed interaction term, representing energy transfer at the linear level \footnote{Explicit form of this term as well as the perturbed DE scalar field equation of motion is complex and not particularly insightful, so it is not provided here.}.

{The disformal coupling framework allows for a specific subclass of models where the interaction between dark energy and dark matter vanishes in the background.} This occurs if the coupling functions
$C(\phi, X)$ and $D(\phi, X)$ in Eq.~\eqref{Eq:GDT:IS:FLRW:Pressureless} satisfy the following relations:
\bea \label{Eq:Zero IS:Conditions} 
C_{\phi}\left[ \frac{2DX}{C} - \frac{1}{2} \right] =
&&
XD_{\phi}\,, 
\nonumber \\ 
C_{X}\left[ \frac{2DX}{C} - \frac{1}{2} \right] =
&&
D + XD_{X}\,,
\eea
the interaction terms $\overline{Q}_{0}$ as well as $\delta Q_{0}$ vanish, which corresponds to no energy exchange between the DE and DM fluids at the background as well as the perturbed level. From Eq.~\eqref{Eq:pt:DM}, we note that the equation of motion governing the DM density contrast $\delta_{\mathrm{DM}}$ reduces to the form with no coupling. Interestingly, the velocity perturbation equation contains non-zero terms resulting from the kinetic coupling. Therefore, this subclass of the general interacting framework leads to \emph{pure momentum coupling between the DE and DM fluids at linear order in perturbation theory}.

Within this subclass of models, we can define a momentum coupling function of the form:
\bea
\label{Eq:MT}
MT(\phi, X) = \frac{2XD_X - C_X}{2C}\,.
\eea
The coupled equations of motion governing the DM fluid, then reduce to the following:
\bea
&&
\Dot{\rho}_{\mathrm{DM}} + 3H\rho_{\mathrm{DM}} = 0\,,
\nonumber \\
&&
\Dot{\delta}_{\mathrm{DM}} + 3\Dot{\Phi} - \frac{k}{\overline{a}}v_{\mathrm{DM}} = 0\,,
\nonumber \\
&&
\Dot{v}_{\mathrm{DM}} + \overline{H}v_{\mathrm{DM}} + \frac{k}{\overline{a}}\Psi = 
- \frac{k}{\overline{a}}\overline{MT}(\phi, \overline{X})\left[ \Ddot{\overline{\phi}}\delta\phi - \Dot{\overline{\phi}}\Dot{\delta\phi} + \Dot{\overline{\phi}}^2\Psi \right]\,,
\eea
At the background level, the DE field $\phi$ is governed by the following equation:
\bea
\Ddot{\phi}\left[ 1 + 2\rho_{\mathrm{DM}}MT + 2\rho_{\mathrm{DM}}XMT_X \right] + 3H\Dot{\phi} + V_{\phi} + 2\rho_{\mathrm{DM}}XMT_{\phi} = 0\,.
\eea
Additionally, the DE field energy density and pressure at the background and linear level are given by:
\bea
\rho_{\mathrm{DE}} = 
&&
X + V(\phi) + 2\,\rho_{\mathrm{DM}}\,X\,MT(\phi, X)\,,
\nonumber \\ 
p_{\mathrm{DE}} = 
&&
X - V(\phi)\,,
\nonumber \\
\delta\rho_{\mathrm{DE}} =
& & 
\left[ \Dot{\overline{\phi}}\Dot{\delta\phi} - \Dot{\overline{\phi}}^2\Psi \right]\left[ 1 + 2\overline{\rho}_{\mathrm{DM}}\overline{MT} + 2\overline{\rho}_{\mathrm{DM}}\overline{X}\overline{MT}_X \right] + 2\overline{X}\overline{\rho}_{\mathrm{DM}}\left[ \overline{MT}_{\phi}\delta\phi + \overline{MT}\delta_{\mathrm{DM}} \right] + \overline{V}_{\phi}\delta\phi\,,
\nonumber \\
\delta p_{\mathrm{DE}} =
& & 
\Dot{\overline{\phi}}\Dot{\delta\phi} - \Dot{\overline{\phi}}^2\Psi - \overline{V}_{\phi}\delta\phi\,.
\eea
The unique dynamics of this subclass of models, where energy exchange vanishes while a non-trivial momentum coupling persists, showcases the rich phenomenology enabled by the general disformal transformations. As we will demonstrate in the rest of this work, the general disformal framework provides a powerful tool for exploring new scenarios that can potentially address current cosmological tensions. Interestingly, the separation of energy and momentum coupling offers a promising way to resolving discrepancies, such as those related to the Hubble constant and the growth of large-scale structure, by providing new dynamics that depart from standard interacting dark energy models.

\section{Cosmological Implications of the disformal dark sector}
\label{Sec:Numerics}

\begin{table}[!htb]
\hspace*{-0.6cm}
    \centering
\begin{tabular}{||c|l|l||} 
\hline \hline  
Model 
& Form of coupling 
& Modified scalar field equations
\\
\hline \hline 
M1 & $\begin{array}{l}
    C(\phi) = c_0\,\phi^{cf_0}\,e^{cf_1\,\phi} \\
    \displaystyle D = 0
  \end{array}$
  & \rule[8pt]{0pt}{20pt} 
  $ 
  \begin{array}{l}
       \displaystyle \Ddot{\phi} + 3H\Dot{\phi} + V_{\phi} + \frac{C_{\phi}}{2C}\rho_{\mathrm{DM}} = 0\,,
       \\
       \rho_{\mathrm{DE}} = X + V
  \end{array}
  $ \rule[-16pt]{0pt}{20pt}
\\
\hline
M2 & $\begin{array}{l}
    C(X) = c_0\,e^{ck_1\,X} \\
    \displaystyle D = 0
  \end{array}$ 
  & \rule[20pt]{0pt}{20pt} 
  $
  \begin{array}{l}
       \Ddot{\phi} + \frac{\displaystyle 3H\Dot{\phi} + V_{\phi}}{\displaystyle 1 + \rho_{\mathrm{DM}}\left[ -\frac{C_X}{2C} + \frac{XC_X^2}{2C^2} - \frac{XC_{XX}}{C} \right]} = 0\,,
       \\
       \rho_{\mathrm{DE}} = X + V - \displaystyle \rho_{\mathrm{DM}}\left[\frac{XC_X}{C}\right]        
  \end{array}
  $ \rule[-36pt]{0pt}{20pt}
\\
\hline
M3 & $\begin{array}{l}
    C = c_0 \\
    \displaystyle D = d_0
  \end{array}$ 
  & \rule[20pt]{0pt}{20pt} 
  $
  \begin{array}{l}
    \displaystyle \Ddot{\phi} + \frac{3H\Dot{\phi} + V_{\phi}}{1 + \displaystyle \frac{\rho_{\mathrm{DM}}d_0}{\left[ c_0 - 2d_0X \right] }} = 0\,,
    \\
    \rho_{\mathrm{DE}} = X + V
  \end{array}
  $ \rule[-30pt]{0pt}{20pt}
\\
\hline
M4 & $\begin{array}{l}
    C = c_0 \\
    \displaystyle D(\phi) = d_0\,\phi^{df_0}\,e^{df_1\,\phi}
  \end{array}$ 
  & \rule[24pt]{0pt}{20pt} 
  $
  \begin{array}{l}
    \displaystyle \Ddot{\phi} + \frac{3H\Dot{\phi} + V_{\phi} + \displaystyle \frac{D_{\phi}X\rho_{\mathrm{DM}}}{c_0 - 2DX} }{1 + \displaystyle \frac{\rho_{\mathrm{DM}}D}{\left[ c_0 - 2DX \right] }} = 0\,,
    \\
    \rho_{\mathrm{DE}} = X + V
  \end{array}
  $ \rule[-32pt]{0pt}{20pt}
\\
\hline
M5 & $\begin{array}{l}
    C = c_0 \\
    \displaystyle D(X) = d_0\,e^{dk_1\,X}
  \end{array}$ 
  & \rule[32pt]{0pt}{20pt} 
  $ 
  \begin{array}{l}
       \displaystyle \Ddot{\phi} + \frac{3H\Dot{\phi} + V_{\phi}}{1 + \rho_{\mathrm{DM}}
  \left[
  \begin{array}{l}
   \frac{4XD_X + 2X^2D_{XX}}{c_0} 
   \\
   + \frac{(c_0 + 2X^2D_X)(D+XD_X)}{c_0(c_0 - 2DX)}
  \end{array} 
  \right]}
   = 0\,,
   \vspace{2mm}\\
   \rho_{\mathrm{DE}} = X + V + \displaystyle \rho_{\mathrm{DM}}\left[\frac{2X^2D_X}{C}\right]
  \end{array}
  $ \rule[-50pt]{0pt}{20pt}
\\
\hline
M6 & $MT(X)=m_0\,X^{mk_0}\,e^{mk_1\,X}$
  & \rule[16pt]{0pt}{20pt} 
  $
  \begin{array}{l}
    \displaystyle \Ddot{\phi} + \frac{3H\Dot{\phi} + V_{\phi}}{1 + 2\rho_{\mathrm{DM}}(MT + XMT_X)} = 0\,,
    \\
   \rho_{\mathrm{DE}} = X + V + \displaystyle 2\rho_{\mathrm{DM}}X\,MT(X)
  \end{array}
  $ \rule[-16pt]{0pt}{20pt}
\\
\hline
M7 & $MT(X)=\displaystyle \frac{m_1}{X}$ 
  & \rule{0pt}{20pt} 
  $
  \begin{array}{l}
    \Ddot{\phi} + 3H\Dot{\phi} + V_{\phi} = 0\,,
    \\
    \rho_{\mathrm{DE}} = X + V + 2\rho_{\mathrm{DM}}\,m_1
  \end{array}
  $ \rule[-16pt]{0pt}{20pt}
\\
\hline \hline
\end{tabular}
    \caption{Modified cosmological background evolution of the DE scalar field resulting from various sub-classes of our general interacting DE-DM framework assuming a pressureless DM fluid.}
    \label{Table:iDEDM}
\end{table}

We now turn to a detailed assessment of the cosmological implications of the DE-DM interactions arising from our general disformal coupling framework. Our analysis will focus on the effects of these interactions on the background cosmology, large-scale structure (LSS) formation, and the cosmic microwave background (CMB) observations. Specifically, we will examine the evolution of the DE field, the Hubble expansion rate, the linear matter power spectrum, and the unlensed CMB temperature power spectrum.

For our numerical analysis, we fix the standard cosmological parameters $\{\Omega_{\mathrm{b}}h^2, \Omega_{\mathrm{cdm}}h^2,$ $\theta_\mathrm{s}, \tau_{\mathrm{reio}}, n_{\mathrm{s}}, A_{\mathrm{s}}\}$ to the Planck 2018 constraints on the $\Lambda$CDM model \cite{2018-Planck-AA}. The normalization of the DE scalar field potential, $V_0$\,, is computed to match the present-day value of $\Omega_{\mathrm{DE}}$ derived from these fixed parameters. The results presented in this work are valid for the following DE potential:
\bea \label{Eq:DE Potential}
V(\phi) = V_0~\phi^n~e^{-\lambda\phi}\,,
\eea
with $n\leq 1$, which encompasses linear, inverse, constant, and exponential potentials. 
Our choice of the scalar potential is motivated solely by the requirement for \emph{late-time cosmic acceleration} imposing no extreme fine-tuning. We have adopted widely used combinations of simple exponential and power-law potentials common in the quintessence literature \cite{delaMacorra:2004da, Tsujikawa:2013fta}.
We present the numerical results for inverse DE scalar field potential ($n=-1, \lambda=0$) and set the initial field value to $\phi_{i} = 10$. This ensures that the dynamics of the uncoupled model remain close to the $\Lambda$CDM model. {For all models, we have evolved the system over the redshift range $z \in [10^8, 0]$\,.}

The perturbed equations of motion in the presence of conformal and disformal couplings are highly complex. 
To provide a general overview of the interaction landscape within our framework, we have classified different models based on power-law and exponential dependencies of the coupling functions $C$ and $D$ on the DE field $\phi$ and its kinetic term $X$. 
This allows us to keep the analysis tractable and capture the essential physics of these interactions \cite{Amendola:2003eq, 2022-Johnson.etal-JCAP, Pourtsidou:2013nha, VanDeBruck:2017mua}.
These are presented in Table \ref{Table:iDEDM}. Models M1 and M2 correspond to purely conformal couplings; M3, M4, and M5 correspond to disformal couplings; while M6 and M7 are examples of the unique pure-momentum coupling scenario.

We will now discuss these models in detail in the subsequent subsections. Our primary interest lies in the disformally interacting model M3 and the pure-momentum coupling models (M6, M7) due to their unique features. A brief discussion of the cosmological implications of purely conformal interactions, {which have been investigated by the current authors in Refs.~\cite{2021-Johnson.Shankaranarayanan-Phys.Rev.D, 2022-Johnson.etal-JCAP, Bansal:PRD:2025}}, is provided in Appendix \ref{Appendix:Conformal}. In Appendix \ref{Appendix:M4,M5}, we provide a detailed discussion of the cosmology of scalar-field-dependent and kinetic-term-dependent pure disformal coupling models (M4, M5).

\subsection{Pure-Momentum Coupling between DE and DM fluids}
\label{Subsec:PMC-Numerics}

\begin{figure}[!htb]
\minipage[b]{0.5\textwidth}
  \includegraphics[width=1\linewidth]{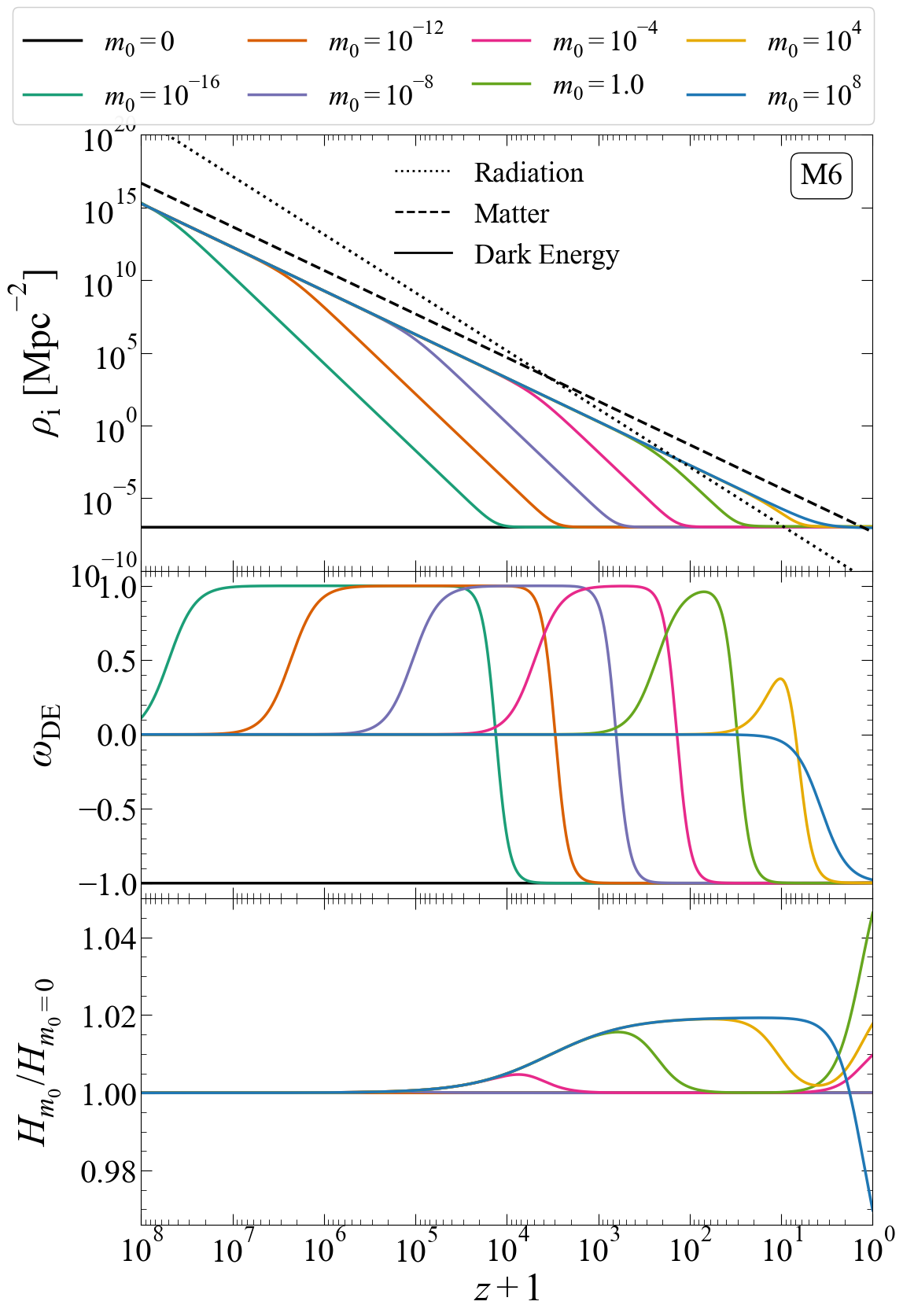}
\endminipage\hfill
\minipage[b]{0.5\textwidth}
  \includegraphics[width=1\linewidth]{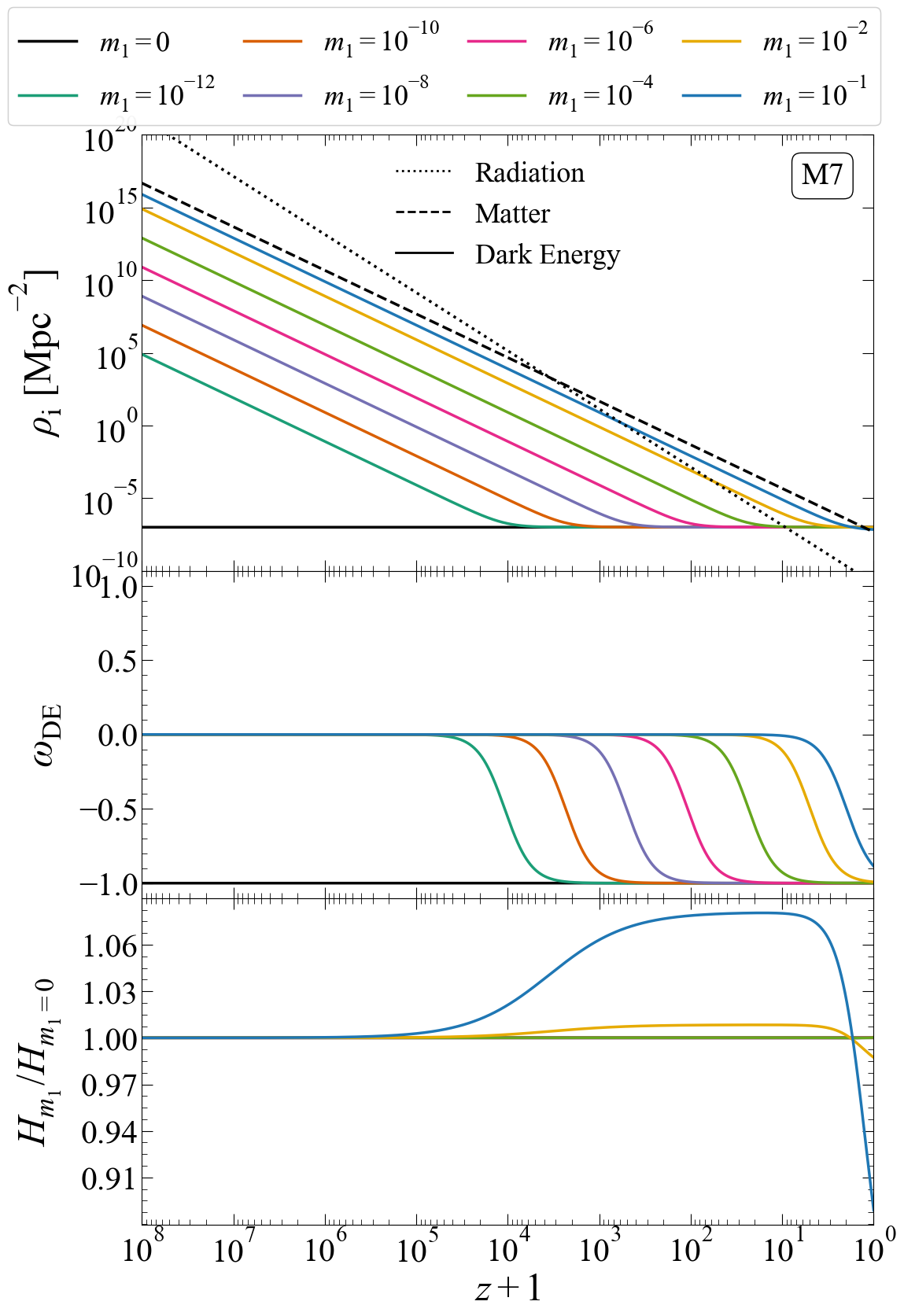}
\endminipage
\caption{Modified evolution of the energy densities of different components (top), DE equation of state (center) and the Hubble parameter (bottom) due to pure-momentum DE-DM interactions. \emph{Left}: Model M6 and \emph{Right}: Model M7 from Table \ref{Table:iDEDM}\,.}
\label{Fig:M6,M7:Bg}
\end{figure}

We first turn our attention to Models M6 and M7, that exhibit a pure-momentum coupling between the dark matter and dark energy fluids. While it is not straightforward to find analytical forms for the coupling functions $C(\phi, X)$ and $D(\phi, X)$ that satisfy the conditions in Eq.~\eqref{Eq:Zero IS:Conditions} and simultaneously yield a simple form for the momentum coupling function $MT(\phi, X)$ as defined in Eq.~\eqref{Eq:MT}, we consider a few specific examples. A detailed analysis of these coupling functions lies beyond the scope of this work and is deferred to a future investigation.

For our analysis, we treat the power-law and exponential forms of the coupling functions assumed for Model M6 in Table \ref{Table:iDEDM} as a \emph{toy model}. In contrast, Model M7 can be constructed from a simpler form of $C$ and $D$, where $C(\phi, X)$ is a constant and $D(\phi, X) \propto 1/X$\,.

At the background level, the absence of an energy exchange means the DM fluid evolves identically to its uncoupled counterpart. The modified equations of motion for the DE scalar field and the form of its energy density $\rho_{\mathrm{DE}}$ for Models M6 and M7 are provided in Table \ref{Table:iDEDM}\,.

\subsubsection{Cosmological evolution within Model M6}

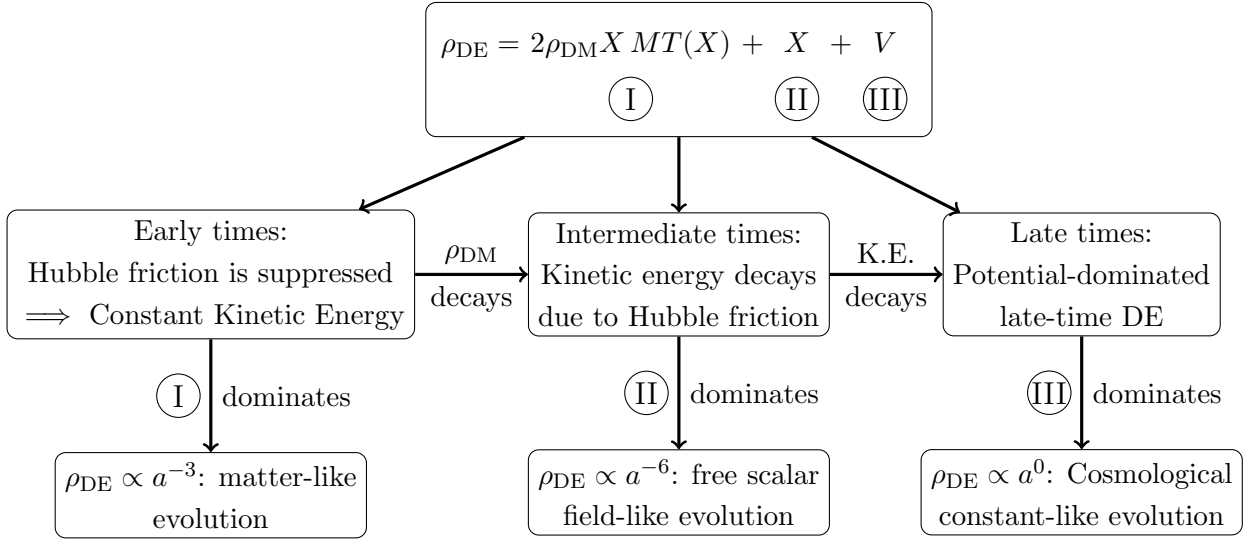
\begin{figure}[h]
    \centering
    \begin{tikzpicture}[
        node distance=1cm,
        box/.style={draw, rectangle, rounded corners, minimum width=3.5cm, minimum height=1cm, align=center}
    ]
    \node[box] (Step) {
    \begin{tabular}{c c c c c c}
        $\rho_{\mathrm{DE}} = $ & $2\rho_{\mathrm{DM}}X\,MT(X)$ & $+$ & $X$ & $+$ & $V$ \\
        & \circled{\normalsize{I}} & & \circled{\normalsize{II}} & & \circled{\normalsize{III}}
    \end{tabular}
    };
        \node[box, below=1cm of Step] (Branch1B) {Intermediate times: \\[-1.25ex] Kinetic energy decays \\[-1.25ex] due to Hubble friction};
        \node[box, left=1.5cm of Branch1B] (Branch1A) {Early times: \\[-1.25ex] Hubble friction is suppressed \\[-1.25ex] $\implies$ Constant Kinetic Energy};
        \node[box, right=1.5cm of Branch1B] (Branch1C) {Late times: \\[-1.25ex] Potential-dominated \\[-1.25ex] late-time DE};
        \node[box, below=1.5cm of Branch1A] (Branch2A) {$\rho_{\mathrm{DE}}\propto a^{-3}$: matter-like \\[-1.25ex] evolution};
        \node[box, below=1.5cm of Branch1B] (Branch2B) {$\rho_{\mathrm{DE}}\propto a^{-6}$: free scalar \\[-1.25ex] field-like evolution};
        \node[box, below=1.5cm of Branch1C] (Branch2C) {$\rho_{\mathrm{DE}} \propto a^0$: Cosmological \\[-1.25ex] constant-like evolution};
        \draw[->, very thick] (Step) -- node[above]{} node[below]{} (Branch1A);
        \draw[->, very thick] (Step) -- node[above]{} node[below]{} (Branch1B);
        \draw[->, very thick] (Step) -- node[above]{} node[below]{} (Branch1C);
        \draw[->, very thick] (Branch1A) -- node[above]{$\rho_{\mathrm{DM}}$} node[below]{decays} (Branch1B);
        \draw[->, very thick] (Branch1B) -- node[above]{K.E.} node[below]{decays} (Branch1C);
        \draw[->, very thick] (Branch1A) -- node[left]{\circled{\normalsize{I}}} node[right]{dominates} (Branch2A);
        \draw[->, very thick] (Branch1B) -- node[left]{\circled{\normalsize{II}}} node[right]{dominates} (Branch2B);
        \draw[->, very thick] (Branch1C) -- node[left]{\circled{\normalsize{III}}} node[right]{dominates} (Branch2C);
    \end{tikzpicture}
    \caption{Flow chart detailing the evolution of DE scalar field $\phi$ and its energy density $\rho_{\mathrm{DE}}$ from early till late times due to Model M6-like pure-momentum coupling between DE and DM fluids. Note that one obtains a functionally identical evolution of the DE field for kinetic-energy (K.E.) dependent conformal and disformal coupling models M2 and M5 from Table \ref{Table:iDEDM}\,.}
    \label{Schematic:M6}
\end{figure}
We start by discussing the cosmological evolution in Model M6. For numerical analysis, we assume a constant momentum coupling function, $MT(\phi, X) = m_0$\,. It is important to note that similar results are obtained for more general power-law or exponential couplings.
As shown in the left panels of Fig.~\ref{Fig:M6,M7:Bg}\,, the background evolution of the DE energy density ($\rho_{\mathrm{DE}}$), its equation of state ($\omega_{\mathrm{DE}}$), and the Hubble parameter ($H$) are significantly modified. 

As discussed in Sec.~\ref{Sec:Cosmology}\,, the non-zero momentum coupling modifies the description of the DE scalar field by suppressing Hubble friction at early times. This suppression results in a constant kinetic energy for the DE field. The DE energy density is given by 
\bea
\rho_{\mathrm{DE}} = X + V + 2\rho_{\mathrm{DM}}X\,MT(\phi, X)\,.
\eea
\begin{figure}[!htb]
\minipage{1\textwidth}
  \includegraphics[width=1\linewidth]{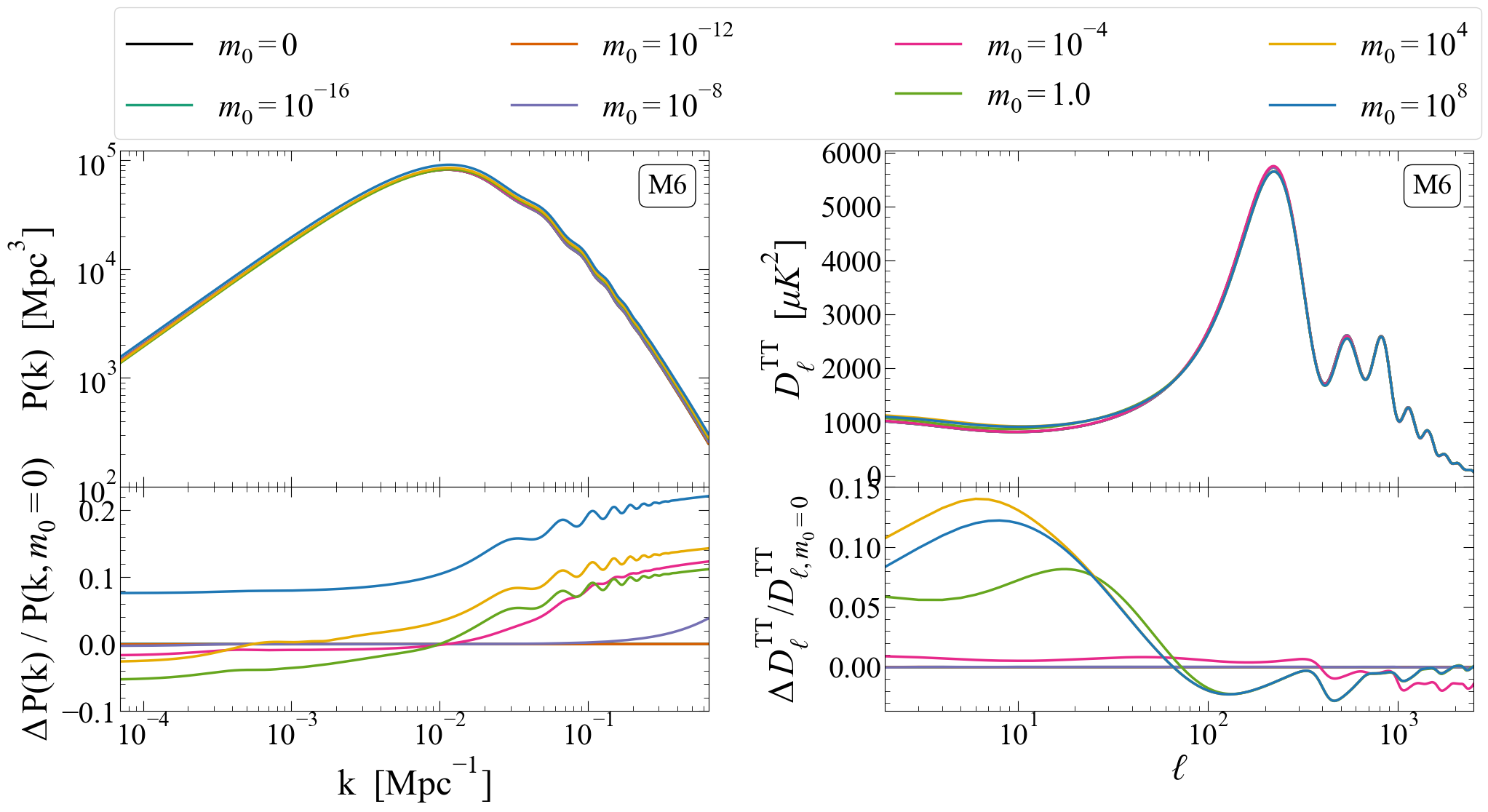}
\endminipage
\caption{Modified linear matter power spectrum (left), CMB temperature power spectrum (right) and relative deviation from their uncoupled counterparts due to pure-momentum exchange, corresponding to Model M6 from Table \ref{Table:iDEDM}\,, between the DE and DM fluids.}
\label{Fig:M6:mPk-DlTT}
\end{figure}
Consequently, the DE field behaves like a dark matter fluid at early times. As the DM energy density decays, the Hubble friction becomes dominant causing the DE kinetic energy to dilute at a rate proportional to $a^{-6}$. Eventually, the decaying kinetic energy of the DE field becomes subdominant to its potential energy, which drives the late-time cosmological constant-like evolution of dark energy. This evolution is clearly seen in the DE equation of state, $\omega_{\mathrm{DE}}$\,, shown in the left-center panel of Fig.~\ref{Fig:M6,M7:Bg}\,. Fig.~\ref{Schematic:M6} provides a schematic of this cosmological evolution. A similar background evolution is observed in other models with kinetic couplings, such as the conformal and disformal models M2 and M5, respectively.

The transition redshift, $z_t$\,, from matter-like to the free scalar field-like evolution of dark energy is determined by the coupling strength, $m_0$\,. A larger $m_0$ shifts $z_t$ to later times. 
We find that a transition redshift in the range of roughly $10^2 < z_t < 10^4$ leads to a larger value for the Hubble constant, $H_0$\,, as illustrated in the bottom-right panel of Fig.~\ref{Fig:M6,M7:Bg}\,.
The degree of modification to the cosmological evolution in this model is determined by both $m_0$ and the initial kinetic energy, $X_i$\,, of the DE scalar field. The value of $X_i$ is crucial because it sets the amplitude of the DE's contribution to the total energy budget from the onset of the matter-dominated epoch, which significantly affects $H_0$\,. In Fig.~\ref{Fig:M6,M7:Bg}\,, we have dynamically set $X_i$ for different values of $m_0$ such that the evolution of the DE field starts with the same initial energy density, $\rho_{\mathrm{DE}}(z=z_i)$\,.

\begin{figure}[!htb]
\minipage{1\textwidth}
  \includegraphics[width=1\linewidth]{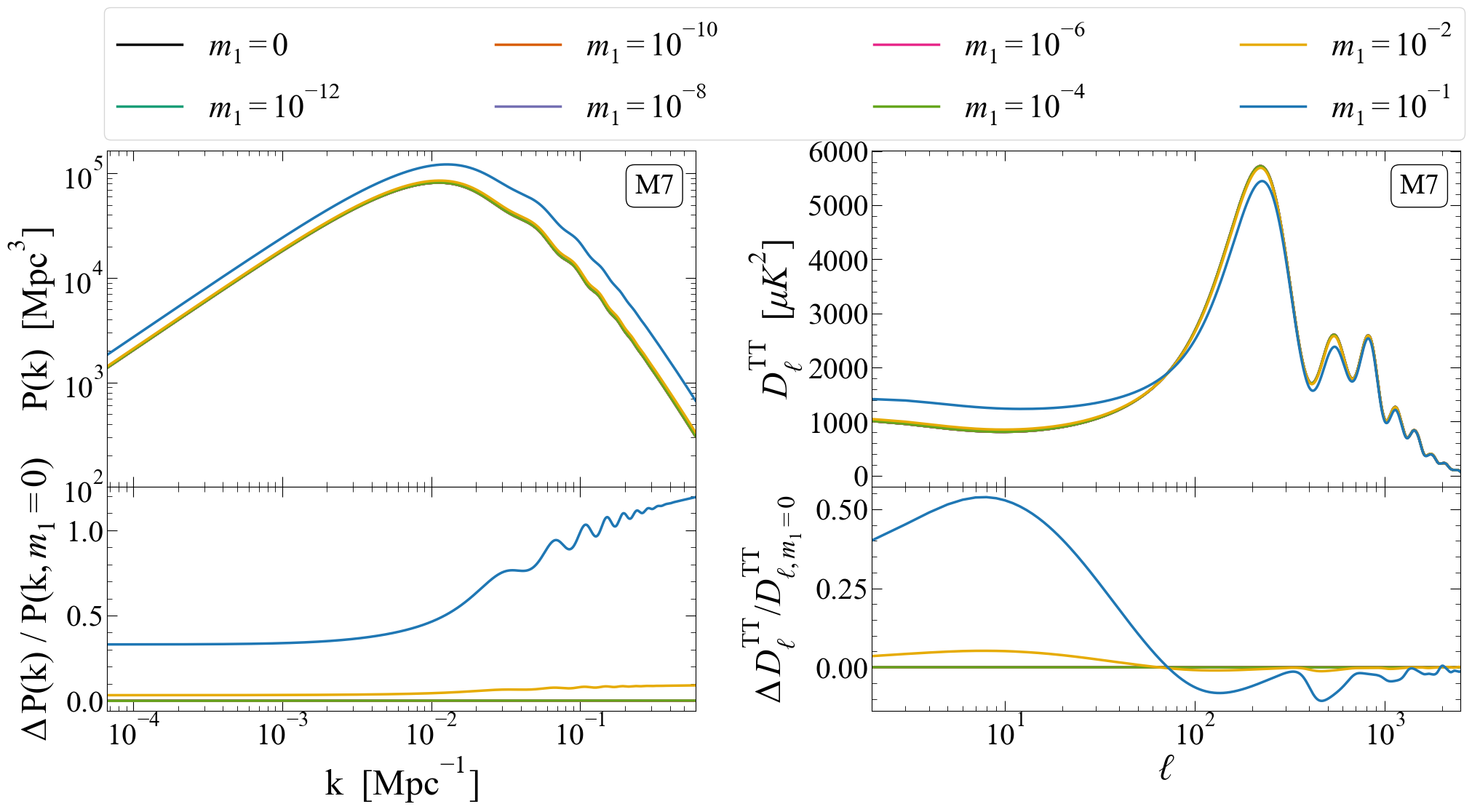}
\endminipage
\caption{Modified linear matter power spectrum (left), CMB temperature power spectrum (right) and relative deviation from their uncoupled counterparts due to pure-momentum exchange, corresponding to Model M7 from Table \ref{Table:iDEDM}\,, between the DE and DM fluids.}
\label{Fig:M7:mPk-DlTT}
\end{figure}

We now look at the impact on linear perturbations. The linear matter and CMB temperature power spectra for this model are plotted in Fig.~\ref{Fig:M6:mPk-DlTT}\,. We find that the model predicts a power suppression over large scales in the matter power spectrum, but an enhanced structure growth over small scales, leading to a larger value of $\sigma_8$\,. On the other hand, the CMB temperature spectrum shows a power enhancement on large angular scales (low multipoles, $\ell$), while power is suppressed on small angular scales (high multipoles). For large coupling strengths, the low-$\ell$ power enhancement in the CMB spectrum changes sign around the multipole range of $\ell = 65-72$\,. As we will discuss in detail in Sec.~\ref{SubSec:Disformal-Numerics}, the Planck 2018 results demonstrated that the best-fit amplitude for the low-$\ell$ spectrum ($\ell < 30$) is approximately $10\%$ lower than for the high-$\ell$ spectrum ($\ell \ge 30$)~\cite{2018-Planck-AA}.

\subsubsection{Cosmological evolution within Model M7}

We now examine Model M7, which assumes a specific form for the momentum coupling function: $MT(\phi, X) = m_1/X$\,. The cosmological background evolution for this model is shown in the right panels of Fig.~\ref{Fig:M6,M7:Bg}\,.

Unlike Model M6, this specific form of coupling does not modify the DE scalar field's equation of motion from its standard Klein-Gordan form (cf. Table \ref{Table:iDEDM}).
As a result, kinetic energy of the the DE field simply redshifts away due to Hubble friction. Consequently, for a non-zero value of the constant $m_1$, the DE field initially evolves like a DM fluid and then transitions to a potential-dominated, cosmological constant-like behavior. This is evident from the evolution of the equation of state, $\omega_{\mathrm{DE}}$\,, which transitions from $0$ to $-1$, as plotted in the right-center panel of Fig.~\ref{Fig:M6,M7:Bg}\,.

The value of $m_1$ determines the transition redshift, $z_t$\,, and the impact on the Hubble constant. We find that smaller values of $m_1$\,, which cause an earlier transition (at high redshift), do not produce a visible modification of the background dynamics. Only values of $m_1$ large enough to cause a transition at $z_t < 10$ lead to a noticeable deviation. Interestingly, unlike M6, this model leads to \emph{smaller values of $H_0$}, which further exacerbates the existing tension between the CMB-inferred value of the present-day expansion rate and local measurements.

The linear matter and CMB TT power spectra for this model are plotted in Fig.~\ref{Fig:M7:mPk-DlTT}\,. Similar to Model M6, we observe an \emph{enhanced growth of structures} on small scales, leading to a larger value of $\sigma_8$\,. However, we do not find any power suppression on large scales in the matter power spectrum, unlike M6. The CMB temperature power spectrum shows similar characteristic deviations to Model M6: enhanced power on large angular scales and suppressed power on small angular scales, with a change in the sign around $\ell=70$\,.

\subsection{Pure disformal coupling, and a unified Early and late dark energy}
\label{SubSec:Disformal-Numerics}

\begin{figure}[!htb]
\minipage{1\textwidth}
  \includegraphics[width=1\linewidth]{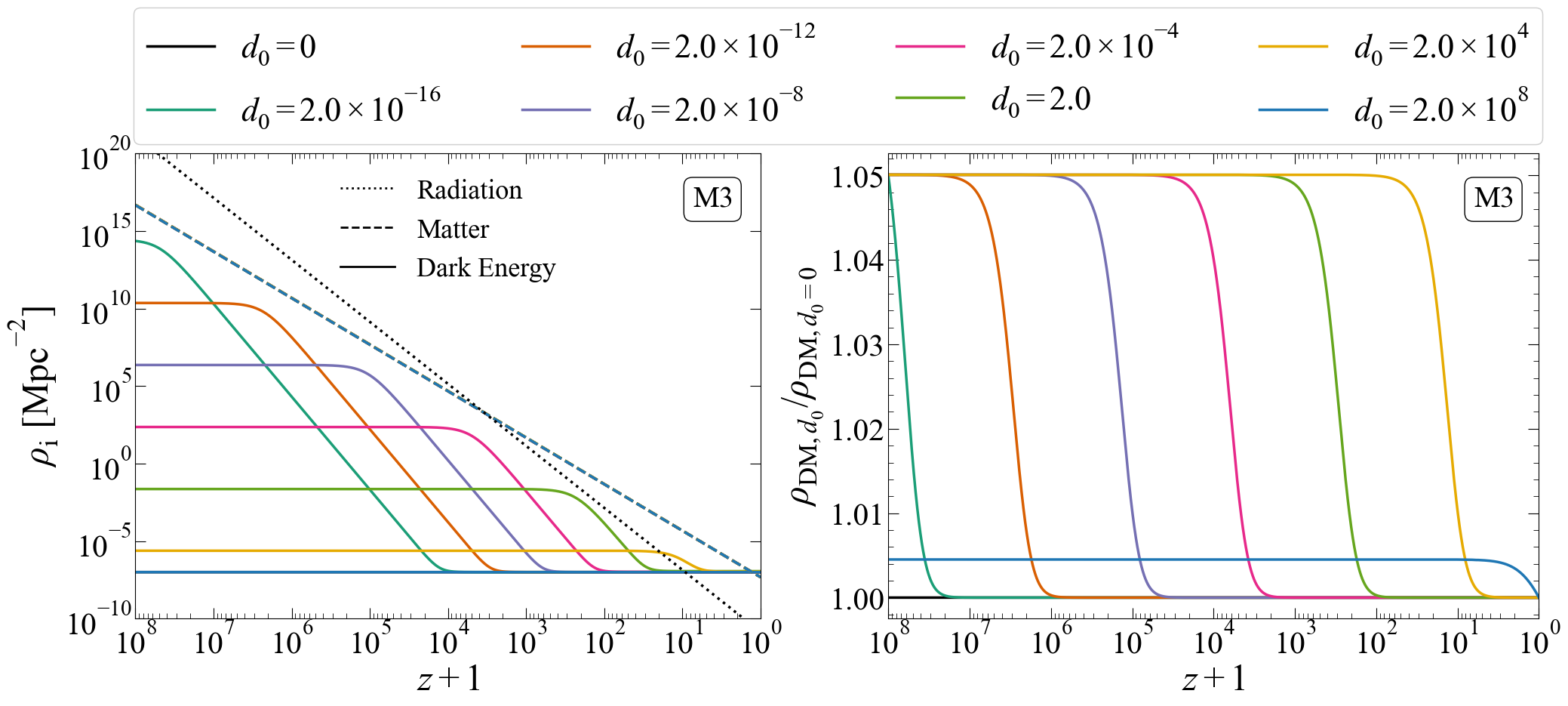}
\endminipage\hfill
\minipage{0.5\textwidth}
  \includegraphics[width=1\linewidth]{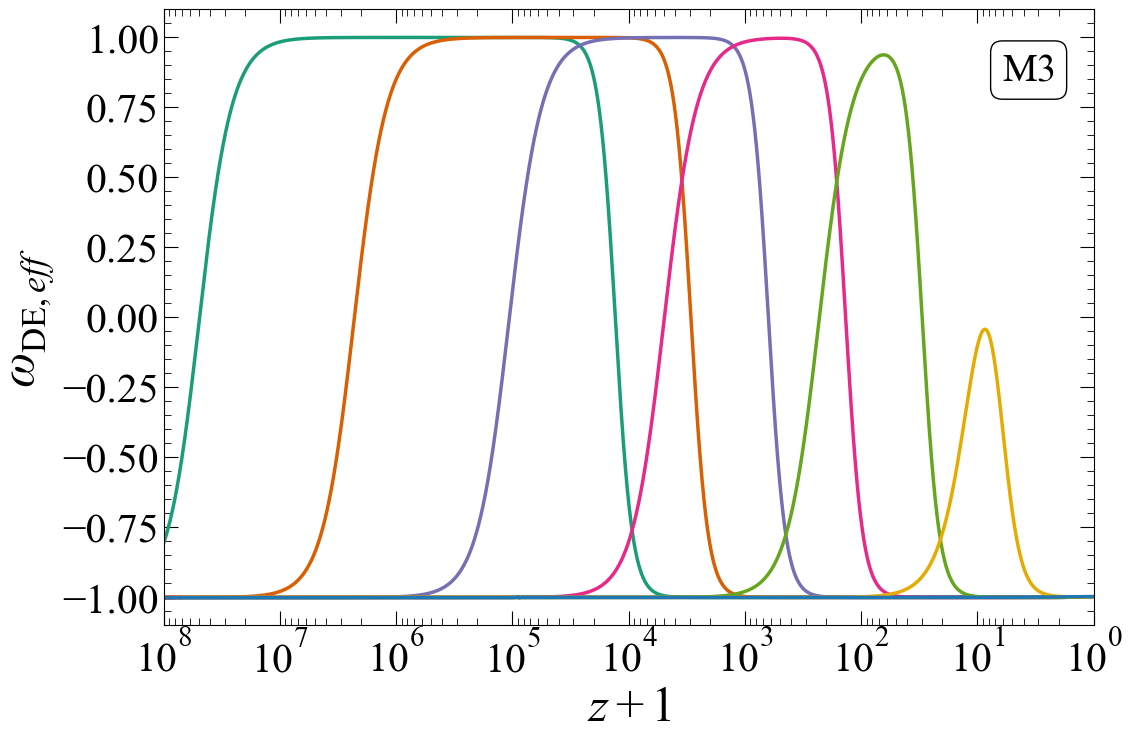}
\endminipage\hfill
\minipage{0.5\textwidth}
  \includegraphics[width=1\linewidth]{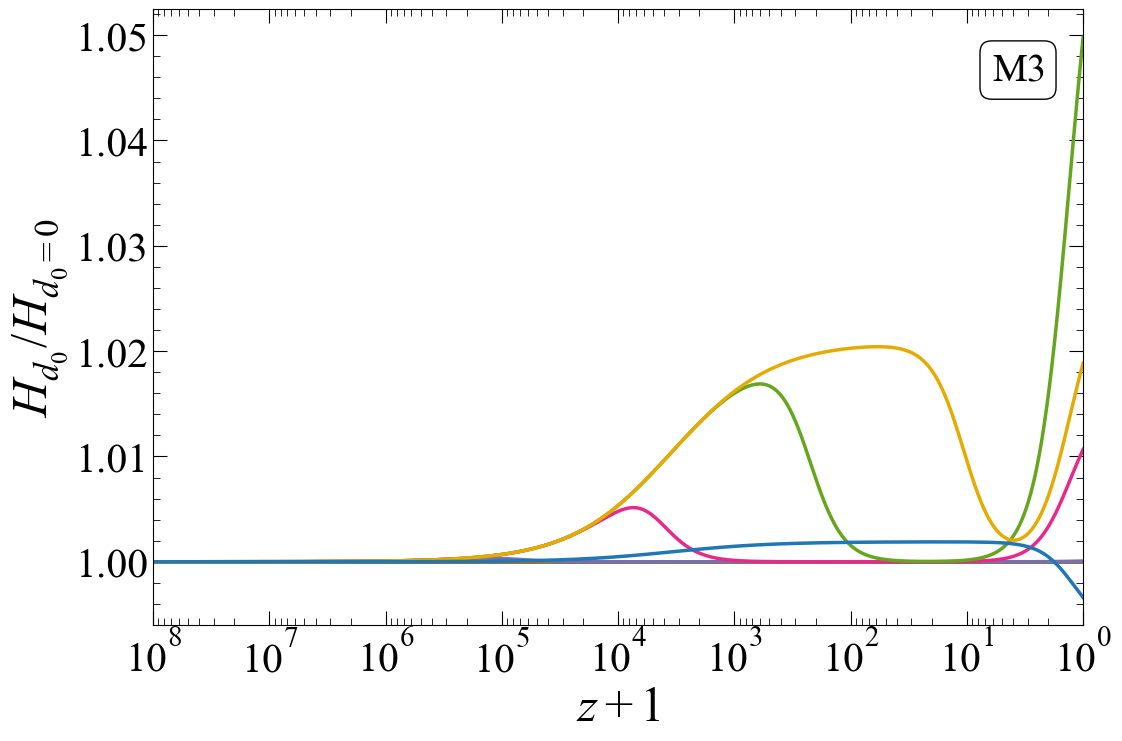}
\endminipage
\caption{Modified cosmological background evolution of the DE energy density (top left), DM energy density relative to the non-interacting case (top right), DE effective equation of state (bottom left) and the relative change in Hubble parameter compared to the non-interacting case (bottom right), for different coupling strengths of disformal DE-DM interactions (Model M3 in Table \ref{Table:iDEDM}).}
\label{Fig:M3:Bg}
\end{figure}

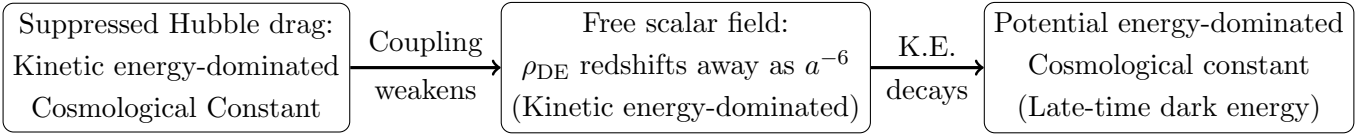
\begin{figure}[h]
    \centering
    \hspace*{-0.85cm}  
    \begin{tikzpicture}[
        node distance=1cm,
        box/.style={draw, rectangle, rounded corners, minimum width=3.5cm, minimum height=1cm, align=center}
    ]
        \node[box] (Step1) {Suppressed Hubble drag: \\[-1.25ex] Kinetic energy-dominated \\[-1.25ex] Cosmological Constant};
        \node[box, right=2cm of Step1] (Step2) {Free scalar field: \\[-1.25ex] $\rho_{\mathrm{DE}}$ redshifts away as $a^{-6}$ \\[-1.25ex] (Kinetic energy-dominated)};
        \node[box, right=1.5cm of Step2] (Step3) {Potential energy-dominated \\[-1.25ex] Cosmological constant \\[-1.25ex] (Late-time dark energy)};
        \draw[->, very thick] (Step1) -- node[above]{Coupling} node[below]{weakens} (Step2);
        \draw[->, very thick] (Step2) -- node[above]{K.E.} node[below]{decays} (Step3);
    \end{tikzpicture}
    \caption{Flow chart detailing the evolution of the DE field energy density $\rho_{\mathrm{DE}}$ from early till late times due to disformal interactions with the DM fluid (Models M3 and M4 in Table \ref{Table:iDEDM}).}
    \label{Schematic:Disformal}
\end{figure}

We now examine Model M3 from Table \ref{Table:iDEDM}, which assumes a pure disformal coupling where the functions $C$ and $D$ are constants --- $C=c_0$(=1, here) and $D=d_0$\,. For these interactions, the DE field $\phi$ is governed by the following equation of motion:
\bea
\Ddot{\phi} + \frac{3H\Dot{\phi} + V_{\phi}}{ \displaystyle 1 + \left[ \frac{\rho_{\mathrm{DM}}d_0}{c_0 - 2Xd_0} \right]} = 0\,.
\eea 
In Fig.~\ref{Fig:M3:Bg}, we have plotted the background evolution of the DE energy density (top left), DM energy density relative to the non-interacting case (top right), DE effective equation of state (bottom left) and the Hubble parameter relative to the non-interacting case (bottom right), for different coupling strengths of the disformal DE-DM interactions. As seen from Fig.~\ref{Fig:M3:Bg}, for a non-zero coupling constant $d_0$, the disformal coupling suppresses Hubble friction at early times. This leads to $\Ddot{\phi} \approx 0$ and a constant kinetic energy for the DE field, causing it to behave like a \emph{kinetic energy-dominated cosmological constant} during this epoch. As the universe expands, the DM energy density dilutes, and the coupling weakens. The DE kinetic energy remains constant until the effects of the coupling become subdominant to Hubble friction. Consequently, the DE field starts evolving like a free scalar field: its kinetic energy decays proportionally to $a^{-6}$ according to the standard Klein-Gordon equation. Eventually, the decaying kinetic energy becomes subdominant to its potential energy, and the DE field behaves like a \emph{potential energy-dominated cosmological constant} at late times. A schematic of this modified evolution of the DE energy density, $\rho_{\mathrm{DE}}$, is provided in Fig.~\ref{Schematic:Disformal}\,.

This early-time cosmological constant-like evolution of dark energy closely resembles that of \emph{Early Dark Energy} (EDE) models, which are strong candidates for resolving the Hubble tension \cite{2016-Karwal.Kamionkowski-Phys.Rev.D, 2018-Poulin.etal-Phys.Rev.Lett}. However, previously explored EDE models typically rely on finely-tuned potentials to maintain a Hubble-frozen state at early times, followed by a transition where the EDE field rolls down and oscillates about the minimum of its potential, causing its energy density to decay at a rate determined by the shape of its potential \cite{Smith:2019ihp, Agrawal:2019lmo}. 
Eventually, the field becomes completely subdominant at late times. In contrast, the early-time behavior in our model is a direct consequence of the disformal coupling, which suppresses Hubble friction and leads to a constant kinetic energy. Furthermore, the transition from a constant to decaying DE kinetic energy (as $a^{-6}$) is triggered by the dilution of dark matter. This unique evolution is evident from the DE effective equation of state transition: $\omega_{\mathrm{DE}, eff} = -1 \rightarrow 1 \rightarrow -1$ for different coupling strengths ($d_0$), as plotted in Fig.~\ref{Fig:M3:Bg}\,. Furthermore, our model uses a single scalar field for both Early Dark Energy and Late-time Dark Energy without requiring a fine-tuned potential.

The strength of the coupling, quantified by $d_0$\,, determines the transition redshift, $z_t$\,, from the constant kinetic energy phase ($\rho_{\mathrm{DE}} \approx \mathrm{K.E.}$) to the decaying phase ($\rho_{\mathrm{DE}} \propto a^{-6}$). A larger $d_0$ shifts this transition to later times (or smaller redshifts), as shown in Fig.~\ref{Fig:M3:Bg}\,. Similar to EDE models, this transition epoch contributes to the total energy budget, leading to a significant increase in the Hubble constant if the transition occurs around the redshift range $10^4 > z_t > 10^2$ or a reduction if it happens at late times ($z_t < 10$). This behavior is shown in Fig.~\ref{Fig:M3:Bg} (bottom right). In contrast to standard EDE models, the increase in the past expansion rate in our model results from the increased energy density of \emph{both} the DM fluid and the DE field. Since this behavior arises from disformal interactions between the two dark sector components, it is more appropriate to describe this cosmological evolution as a \emph{disformally interacting Early Dark Sector}.

Beyond $d_0$, the amplitude of the contribution of the DE field during the transition and its impact on $H_0$ are determined by the initial kinetic energy, $X_i$\,. Therefore, this class of interacting models introduces two additional parameters: $\{ d_0\,, X_i \}$.

\subsubsection{Impact on linear perturbations}

\begin{figure}[!htb]
\minipage{1\textwidth}
  \includegraphics[width=1\linewidth]{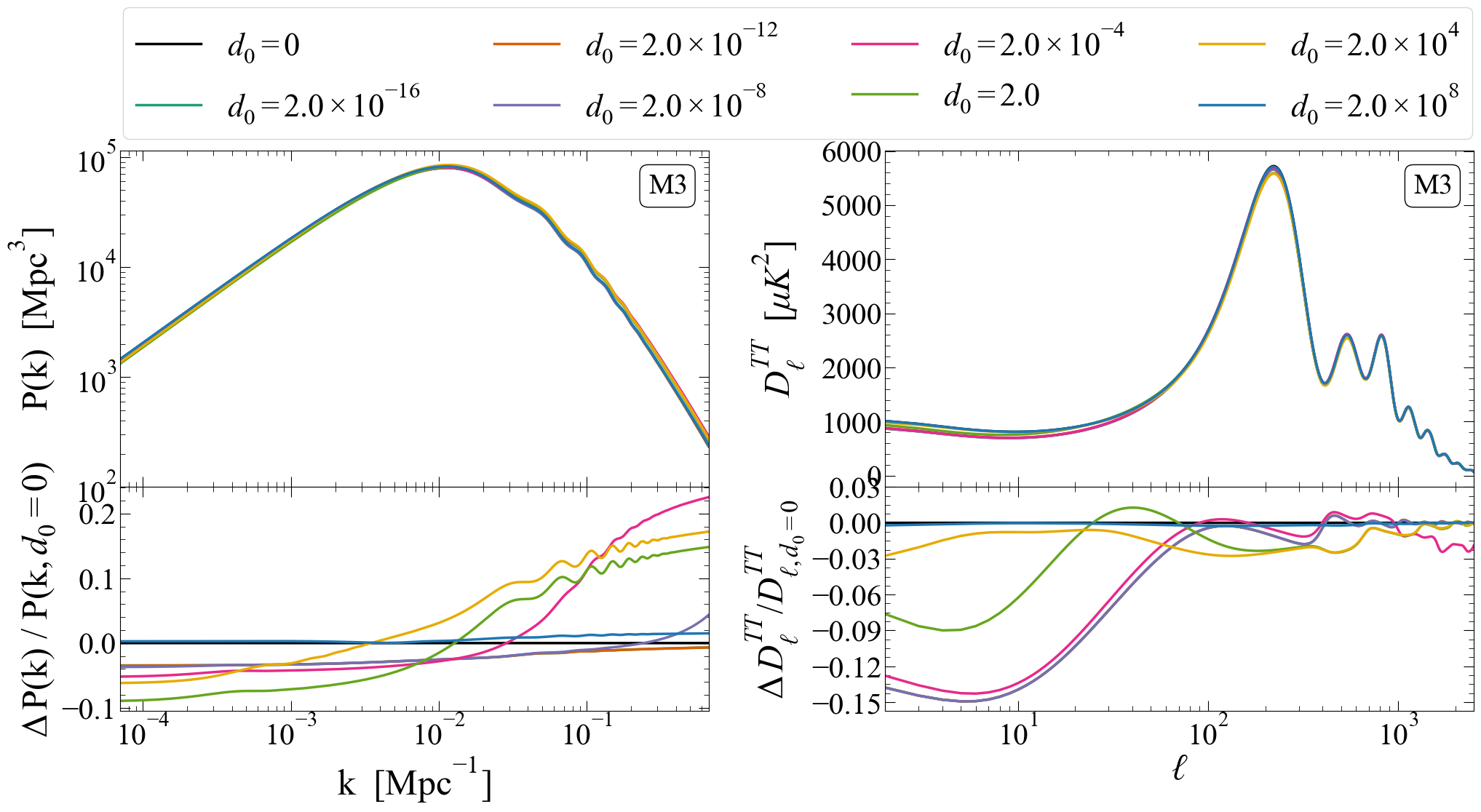}
\endminipage
\caption{Modified linear matter power spectrum, CMB temperature spectrum and relative deviation from their uncoupled counterparts due to disformal DE-DM interactions (Model M3 in Table \ref{Table:iDEDM}).}
\label{Fig:M3:mPk-DlTT}
\end{figure}

We now turn to the impact on linear perturbations, which are illustrated in Fig.~\ref{Fig:M3:mPk-DlTT}\,. We find that the interactions suppress the growth of matter density fluctuations for smaller values of $d_0$, leading to a reduced $\sigma_8$\,. Conversely, larger coupling strengths enhance structure growth on small scales, corresponding to a larger $\sigma_8$\,. Notably, modes that enter the horizon when $z \gg z_t$ undergo enhanced structure growth, while modes that remain super-horizon during this period undergo suppressed growth. The scale ($k$) of the transition from suppressed to enhanced structure growth in the matter power spectrum is strongly dependent on the transition redshift $z_t$, while the amplitude of this deviation from the non-interacting scenario depends on the initial kinetic energy configuration $X_i$ of the DE field.

On the other hand, the effect of the coupling remains significant over large angular scales (or low multipoles) but diminishes over small scales (higher multipoles) in the CMB power spectrum. Interestingly, the coupling leads to a power suppression in the TT spectrum compared to the non-interacting quintessence scenario. Specifically, we find a significant suppression of power on large angular scales, which gradually fades as $\ell \sim 100$\,. These modifications can be attributed to the effect of coupling on the contributions of the Sachs-Wolfe and Integrated Sachs-Wolfe (ISW) effects. 

The observed CMB exhibits a well-known tension with the standard $\Lambda$CDM model --- a suppression of power at large angular scales (low multipoles, $\ell$), particularly in the quadrupole moment ($\ell=2$). While a low quadrupole could be attributed to cosmic variance, a more detailed analysis by the Planck collaboration revealed a broader suppression of power. The Planck 2018 results demonstrated that the best-fit amplitude for the low-$\ell$ spectrum ($\ell < 30$) is approximately $10\%$ lower than that for the high-$\ell$ spectrum ($\ell \ge 30$)~\cite{2018-Planck-AA}.

Naturally, a number of theoretical models have been proposed to address the low-$\ell$ power suppression by modifying the physics of the early universe.
These models often propose a departure from the standard inflationary slow-roll conditions. For instance, some models suggest a pre-inflationary era or a slow-roll violation occurring approximately 60 e-folds before the end of inflation. Such a deviation from standard inflation can alter the primordial power spectrum on the largest scales, corresponding to the low-$\ell$ modes, leading to a suppression of power where the spectrum would otherwise be nearly scale-invariant \cite{Contaldi:2003zv,Shankaranarayanan:2004iq,White:2014aua}. In contrast, our disformal interaction model provides a completely different explanation. By modifying the late-time dynamics of the dark sector, it produces a suppression of power on large angular scales, offering a distinct alternative to early-universe scenarios.
A detailed analysis of how the disformal coupling affects individual contributions of the CMB spectra as well as parameter estimation for best fits and compatibility with datasets will be carried out in a future work.

This discussion of Model M3, with its constant coupling functions, represents the simplest form of a pure disformal coupling. Model M4, which incorporates a scalar field dependent disformal coupling ($D \equiv D(\phi)$), yields similar results and therefore does not require a separate, detailed discussion here. Similarly, Model M5, which assumes a kinetic term dependence ($D \equiv D(X)$), leads to a background evolution similar to Model M6 discussed in the Sec.~\ref{Subsec:PMC-Numerics}. For completeness, we discuss the numerical evolution of the background, matter and CMB power spectra for Models M4 and M5 in Appendix \ref{Appendix:M4,M5}.

\subsection{Approximate analytical derivation of DE density scaling for Models M1-M7}

We now look at the cosmological background evolution of the DE field under the influence of the interactions with dark matter. The following approximate analytical derivations determine the scaling of the dark energy (DE) density, $\rho_{\mathrm{DE}}$, during the radiation- and matter-dominated epochs. Note that we denote derivatives with respect to conformal time $\eta$ with a prime ($\phi'$).

\subsubsection{Model M1: Scalar Field-Dependent Conformal Coupling ($C(\phi), D=0$)}
We start with the DE-field-dependent conformal coupling. For this model, the DE field equation of motion and its corresponding energy density are:
\begin{eqnarray}
\Ddot{\phi} + 3H\Dot{\phi} + V_{\phi} + \frac{C_{\phi}}{2C}\rho_{\mathrm{DM}} = 0\,, \quad \quad \rho_{\mathrm{DE}} = \frac{\Dot{\phi}^2}{2} + V\,.
\end{eqnarray}
Assuming a standard exponential form $C(\phi) \propto e^{cf_1 \phi}$ (and neglecting the $V_\phi$ term at early times), the Klein-Gordon equation in conformal time becomes:
\begin{eqnarray}
\phi^{\prime\prime} + 2aH\phi^{\prime} + \frac{cf_1}{2}a^2\rho_{\mathrm{DM}} = 0\,.
\end{eqnarray}
\begin{enumerate}
\item {\bf Radiation Domination ($H \propto a^{-2}, \rho_{\mathrm{DM}} \propto a^{-3}$):}
We start by assuming $\phi^{\prime\prime} \simeq 0$ in the above equation during this epoch and justify this assumption with the following result:
\begin{eqnarray}
2aH\phi^{\prime} \propto - a^2\rho_{\mathrm{DM}} \implies \phi^{\prime} \propto - \frac{a^2\rho_{\mathrm{DM}}}{aH} \propto \frac{a^2 \cdot a^{-3}}{a \cdot a^{-2}} \propto a^0 \quad (\text{constant})\,.
\end{eqnarray}
Since $\phi'$ is constant, this validates our original assumption of $\phi^{\prime\prime}=0$. The DE density, then, scales as:
\begin{eqnarray}
\rho_{\mathrm{DE}} \simeq \frac{\phi^{\prime 2}}{2a^2} \propto a^{-2}\,.
\end{eqnarray}
\item {\bf Matter Domination ($H \propto a^{-3/2}, \rho_{\mathrm{DM}} \propto a^{-3}$):} Following a similar procedure, we define $\d T = a^{-3/2}\d t$, giving us the following form of the Klein-Gordon equation:
\begin{eqnarray}
\frac{\d^2\phi}{\d T^2} + \frac{3a^{3/2}H}{2}\frac{\d\phi}{\d T} + V_{\phi}a^3 + \frac{cf_1}{2}a^3\rho_{\mathrm{DM}} = 0\,.
\end{eqnarray}
Again, assuming $\d^2\phi/\d T^2 \simeq 0$ during the matter dominated epoch, we have:
\begin{equation}
\begin{aligned}
& \frac{\d\phi}{\d T} \simeq -\frac{cf_1}{3}\frac{a^{3/2}\rho_{\mathrm{DM}}}{H}\,.
\\ \implies
& \frac{\d\phi}{\d T} \propto a^{0} (\text{constant})\,,
\\ \implies
& \rho_{\mathrm{DE}} \simeq \frac{1}{2a^3}\left( \frac{\d\phi}{\d T} \right)^2 \propto a^{-3}\,.
\end{aligned}
\end{equation}
Since $\d\phi/\d T$ remains constant during matter dominated epoch, this justifies our assumption $\d^2\phi/\d T^2 \simeq 0$.
\item {\bf Late Times:} As the DE field kinetic energy decays during the early epochs, it is eventually dominated by its potential energy, leading to late-time acceleration.
\end{enumerate}
Therefore, we have:
\begin{eqnarray}
\rho_{\mathrm{DE}} \propto \begin{cases} a^{-2}~(\text{K.E. dominated}) & \text{during radiation domination} \\ a^{-3}~(\text{K.E. dominated}) & \text{during matter domination} \\ a^0~(\text{P.E. dominated}) & \text{at late times} \end{cases}
\end{eqnarray}

\subsubsection{Model M3: Pure \& Constant Disformal Coupling ($C=c_0, D=d_0$)}

As discussed before, the equations of motion for Model M3 possesses the characteristic of suppressing the Hubble friction term ($\propto 3H\Dot{\phi}$). The equation of motion governing the DE field $\phi$ under this coupling is given by:
\begin{equation}
\begin{aligned}
& \Ddot{\phi} + \frac{3H\Dot{\phi} + V_{\phi}}{ \displaystyle 1 + \left[ \frac{\rho_{\mathrm{DM}}d_0}{c_0 - 2Xd_0} \right]} = 0\,, \quad \quad \rho_{\mathrm{DE}} = \frac{\Dot{\phi}^2}{2} + V(\phi)\,.
\end{aligned}
\end{equation}
Fig.~\ref{Schematic:Disformal} provides a birds-eye view of the evolution of the DE field $\phi$ and its energy density $\rho_{\mathrm{DE}}$ for Models M3.
\begin{enumerate}
\item 
{\bf High redshift ($z \gg 1, \rho_{_{\mathrm{DM}}}$ is large):} In this regime, the disformal coupling term in the denominator (proportional to $\rho_{\mathrm{DM}}$) dominates the numerator terms, effectively leading to:
\begin{equation}
\begin{aligned}
& \Ddot{\phi} \simeq 0\,,
\\ \implies 
& \Dot{\phi} \simeq \mathrm{constant}\,.
\end{aligned}
\end{equation}
If the kinetic energy dominates initially, the DE density remains approximately constant:
\begin{eqnarray}
\rho_{\mathrm{DE}} \simeq \frac{\Dot{\phi}^2}{2} \simeq \mathrm{constant} \propto a^0\,.
\end{eqnarray}
This mimics the \emph{Early Dark Energy (EDE)}-like behavior.
\item {\bf Intermediate Redshifts:}  As $\rho_{\mathrm{DM}}$ decreases, the coupling influence diminishes, and the equation reverts to the standard Klein-Gordon form (for a sufficiently flat potential):
\begin{eqnarray}
\Ddot{\phi} + 3H\Dot{\phi} \simeq 0 \implies \Dot{\phi} \propto a^{-3}\,.
\end{eqnarray}
The kinetic energy density then decays rapidly:
\begin{eqnarray}
\rho_{\mathrm{DE}} \simeq \frac{\Dot{\phi}^2}{2} \propto a^{-6}\,.
\end{eqnarray}
\item {\bf Late Times:} The potential energy eventually dominates the decaying kinetic energy, leading to late-time acceleration.
\end{enumerate}
Therefore, we have:
\begin{equation}
\rho_{\mathrm{DE}} \propto \begin{cases}
a^0~(\mathrm{K.E. dominated}) & \text{at high redshifts (EDE-like behavior)} \\
a^{-6} & \text{at intermediate redshifts} \\
a^0~(\text{P.E. dominated}) & \text{at late times (cosmic acceleration)}
\end{cases}
\end{equation}
The value of the coupling strength $d_0/c_0$ determines the duration of the EDE-like epoch, with larger values leading to a more extended period of constant dark energy density.

\subsubsection{Model M4: Scalar-Field Dependent Disformal Coupling ($C = c_0, D \equiv D(\phi)$)}

Next, we look at Model M4 with a DE-field-dependent disformal coupling. The equation of motion governing the DE field and its energy density is given by:
\begin{eqnarray}
\displaystyle \Ddot{\phi} + \frac{3H\Dot{\phi} + V_{\phi} 
+ \displaystyle \frac{D_{\phi}X\rho_{\mathrm{DM}}}{c_0 - 2DX} }{1 + \displaystyle \frac{\rho_{\mathrm{DM}}D}{\left[ c_0 - 2DX \right] }} = 0\,, \quad \rho_{\mathrm{DE}} = \frac{\Dot{\phi}^2}{2} + V(\phi)\,.
\end{eqnarray}
Fig.~\ref{Schematic:Disformal} provides a birds-eye view of the evolution of the DE field $\phi$ and its energy density $\rho_{\mathrm{DE}}$ for Models M4.
\begin{enumerate}
\item 
{\bf High redshift ($z \gg 1, \rho_{_{\mathrm{DM}}}$ is large):} In this regime, the disformal coupling term, proportional to $\rho_{\mathrm{DM}}$, in the numerator and denominator dominate, effectively leading to:
\begin{eqnarray}
\Ddot{\phi} + \frac{D_{\phi}}{2D}\Dot{\phi}^2 \simeq 0\,.
\end{eqnarray}
Assuming a simple exponential + power-law form, $D(\phi) = d_0\,\phi^{df_0}\,e^{df_1\,\phi}$, we have:
\begin{equation}
\begin{aligned}
& \Ddot{\phi} + \left( \frac{df_1}{2} + \frac{df_0}{2\phi} \right)\Dot{\phi}^2 \simeq 0\,,
\\ \implies
& \Dot{\phi} \propto \phi^{-df_0/2}\,e^{-df_1\phi/2}\,,
\\ \implies
& \rho_{\mathrm{DE}} \simeq \frac{\Dot{\phi}^2}{2} \propto \phi^{-df_0}\,e^{-df_1\phi}\,.
\end{aligned}
\end{equation}
Since there is little change in the scalar field value $\phi$ at high redshifts, the dark energy density remains approximately constant, mimicking an early dark energy-like behavior as Model M3.
\item
{\bf Intermediate and Late Times:} As the universe expands and $\rho_{\mathrm{DM}}$ decreases, the influence of the disformal coupling diminishes, allowing the scalar field to evolve more freely under the influence of the Hubble friction.
Therefore, the behavior of the dark energy density at intermediate and late times is similar to that in Model M3.
\end{enumerate}
Therefore, we have:
\begin{equation}
\rho_{\mathrm{DE}} \propto \begin{cases}
a^0~(\mathrm{K.E. dominated}) & \text{at high redshifts (EDE-like behavior)} \\
a^{-6} & \text{at intermediate redshifts} \\
a^0~(\text{P.E. dominated}) & \text{at late times (cosmic acceleration)}
\end{cases}
\end{equation}
The values of the coupling strength parameters $d_0$, $df_0$ and $df_1$ determine the duration of the EDE-like epoch, 
with larger values leading to a more extended period of constant dark energy density.

\subsubsection{Models M2, M5, M6: Kinetic-Dependent Couplings}

These models, including the conformal kinetic coupling (M2), disformal kinetic coupling (M5) and the pure-momentum coupling (M6), all admit a similar scaling solution at early times. Taking Model M2 as the representative case with $C(X) \propto e^{ck_1X}$ and assuming $V_\phi \simeq 0$ at early times:
\begin{eqnarray}
\Ddot{\phi} + \frac{3H\Dot{\phi}}{1 - \rho_{\mathrm{DM}}(ck_1 + ck_1^2X)/2} = 0\,, \quad \rho_{\mathrm{DE}} = X(1 - ck_1\rho_{\mathrm{DM}}) + V\,.
\end{eqnarray}

Fig.~\ref{Schematic:M6} provides a birds-eye view of the evolution of the DE field $\phi$ and its energy density $\rho_{\mathrm{DE}}$ for Models M2, M5 and M6.
\begin{enumerate}
\item {\bf High Redshift ($z \gg 1$, $\rho_{\mathrm{DM}}$ is large):}
To avoid negative energy states, we assume $ck_1 < 0$. The large $\rho_{\mathrm{DM}}$ term in the denominator leads to $\Ddot{\phi} \simeq 0$ (constant kinetic energy, $X$). The DE density is dominated by the interaction term:
\begin{eqnarray}
\rho_{\mathrm{DE}} \simeq -ck_1\rho_{\mathrm{DM}}X \propto a^{-3}\,.
\end{eqnarray}
Therefore, the DE density tracks the dark matter density, exhibiting a scaling solution.
\item {\bf Intermediate and Late Times:}
As the coupling weakens, the field reverts to the standard kinetic decay and the DE density is dominated by the decaying kinetic term leading to $\Dot{\phi} \propto a^{-3}$, $\rho_{\mathrm{DE}} \propto a^{-6}$, followed by potential domination ($a^0$).
\end{enumerate}
Therefore, for Models M2, M3 and M6, we have:
\begin{eqnarray}
\rho_{\mathrm{DE}} \propto \begin{cases} a^{-3}~(\text{K.E. dominated, scaling}) & \text{at high redshifts} \\ a^{-6} & \text{at intermediate redshifts} \\ a^0~(\text{P.E. dominated}) & \text{at late times} \end{cases}
\end{eqnarray}

\subsubsection{Model M7: Pure Momentum Coupling Special Case}

Model M7 is a special case where the scalar field follows the standard Klein-Gordon equation ($\Ddot{\phi} + 3H\Dot{\phi} + V_{\phi} = 0$), but the DE density is explicitly coupled to DM: 
\begin{eqnarray}
\rho_{\mathrm{DE}} = X + V + 2\,m_1\,\rho_{\mathrm{DM}}\,.
\end{eqnarray}
\begin{enumerate}
\item {\bf High Redshift:}
The $\rho_{\mathrm{DE}}$ is dominated by the DM-dependent term:
\begin{eqnarray}
\rho_{\mathrm{DE}} \simeq 2\,m_1\,\rho_{\mathrm{DM}} \propto a^{-3}\,.
\end{eqnarray}
\item {\bf Late Times:}
Hubble friction quickly drives the kinetic term $X$ to zero, and the DE density transitions directly to potential domination.
\end{enumerate}
\begin{eqnarray}
\rho_{\mathrm{DE}} \propto \begin{cases} a^{-3}~(\text{DM scaling}) & \text{at high redshifts} \\ a^0~(\text{P.E. dominated}) & \text{at late times} \end{cases}
\end{eqnarray}
\subsection{Implications of the coupling for the Hubble constant ($H_0$)}

\begin{figure}[!htb]
\minipage{0.5\textwidth}
  \includegraphics[width=1\linewidth]{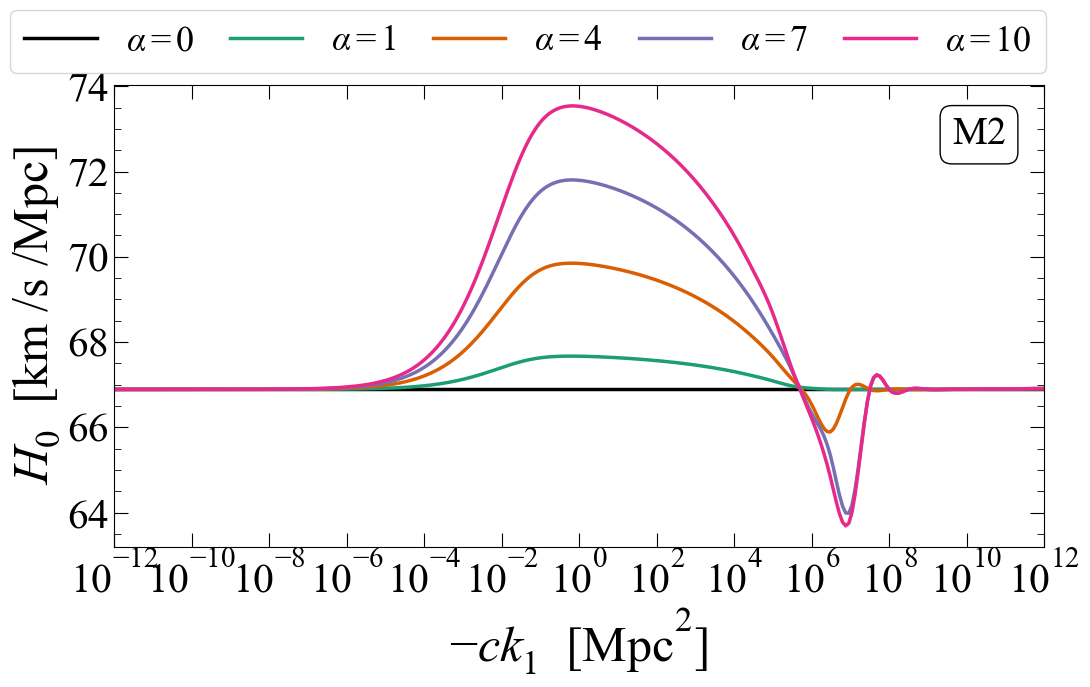}
\endminipage\hfill
\minipage{0.5\textwidth}
  \includegraphics[width=1\linewidth]{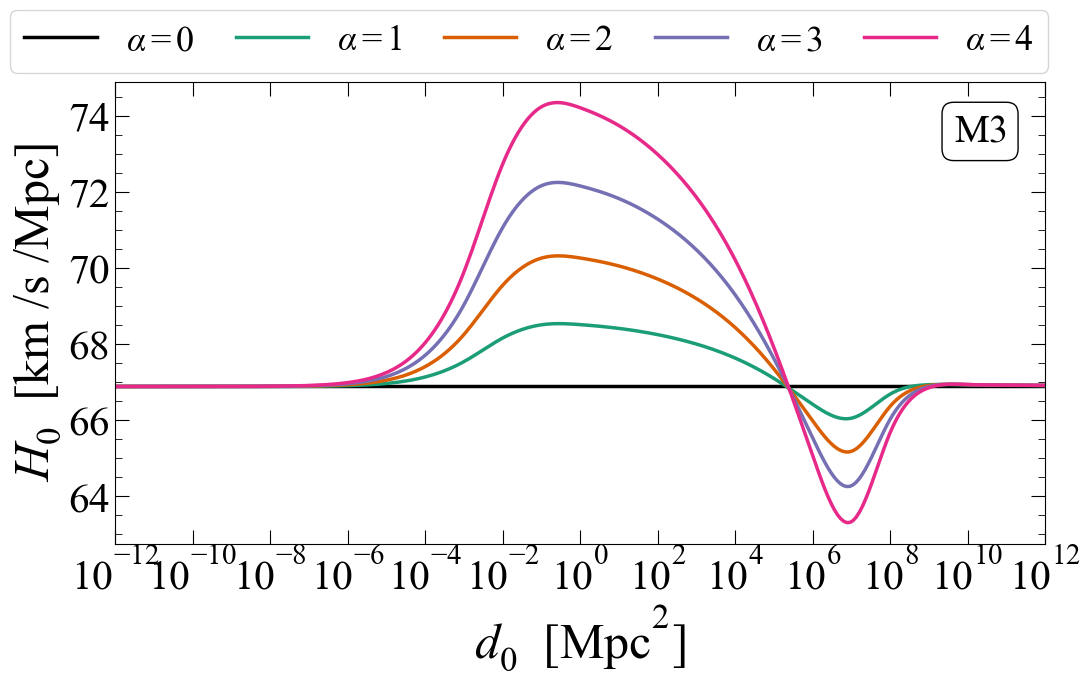}
\endminipage\hfill
\minipage{0.5\textwidth}
  \includegraphics[width=1\linewidth]{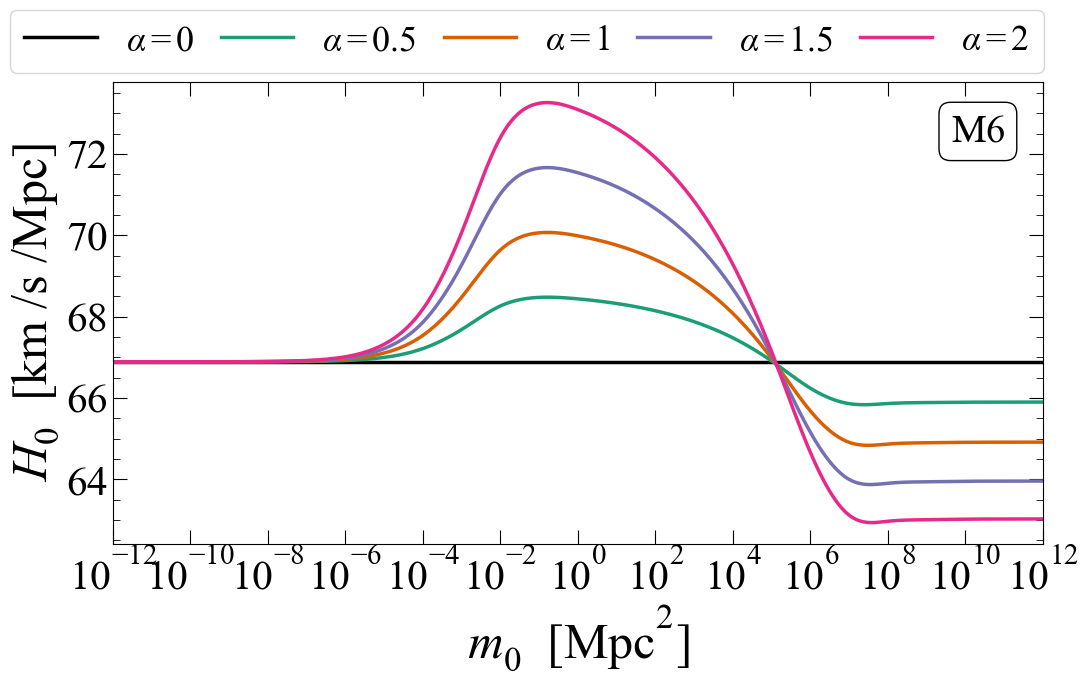}
\endminipage\hfill
\minipage{0.5\textwidth}
  \includegraphics[width=1\linewidth]{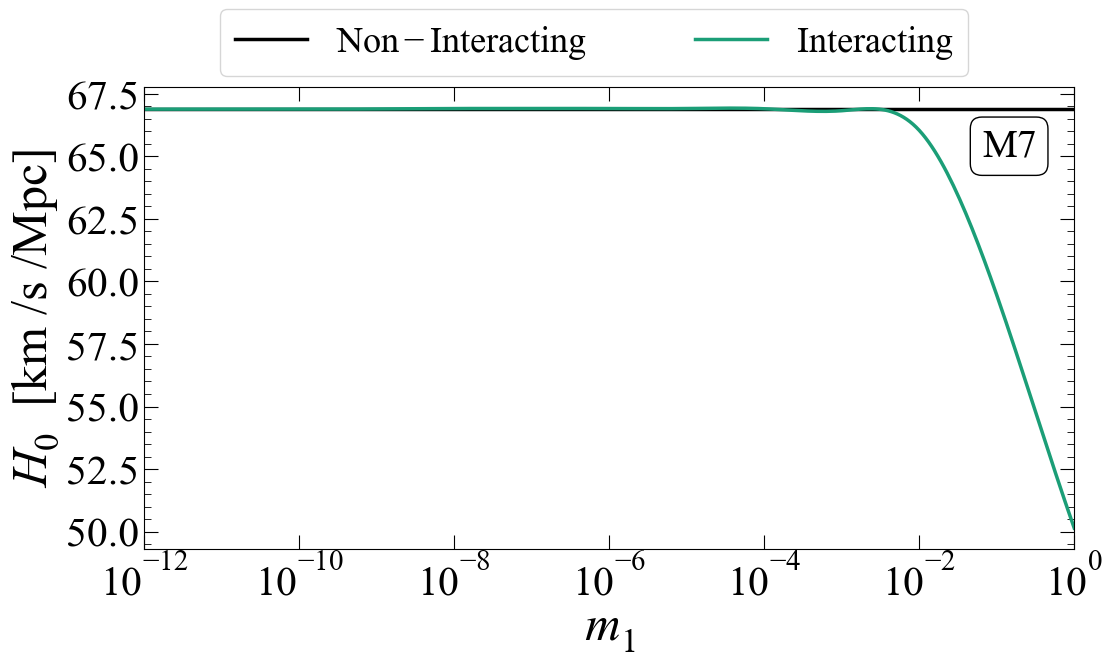}
\endminipage
\caption{Impact of the various DE-DM interactions on the Hubble constant $H_0$.
Black solid line represents the non-interacting scenario (corresponding to $\alpha=0$). Note that we have fixed the cosmological parameters to the Planck 2018 best-fit values \cite{2018-Planck-AA}.}
\label{Fig:H0}
\end{figure}

As discussed, the interactions can lead to significant shifts in the value of the Hubble constant by altering the expansion history. We now perform a quantitative analysis of the implication of the DE-DM interactions on the Hubble tension. To do this, we consider Model M3 as the representative case. Following our discussion in Sec.~\ref{SubSec:Disformal-Numerics} for this model, the shift in $H_0$ is quantified by two parameters:
\begin{enumerate}
\item  {\bf Coupling Strength ($d_0$):} This parameter controls the \emph{transition redshift ($z_t$)} at which the dark energy (DE) density shifts from the constant, EDE-like plateau phase to the rapidly decaying phase ($a^{-6}$). A larger $d_0$ corresponds to a later $z_t$.
\item {\bf Initial Kinetic Energy ($X_i$):} This determines the \emph{amplitude} of the EDE-like plateau. A larger $X_i$ (for a fixed $d_0$) leads to a larger contribution of early dark energy.
\end{enumerate}
Next, we define a dimensionless parameter $\alpha$ related to the initial conditions as:
\begin{eqnarray}
\left(\frac{\mathrm{d}\phi}{\mathrm{d}N}\right)_{N=N_i}^2 = \alpha \times \left(d_{0}\right)^{-1} \times 10^{-22} \quad \text{or} \quad X_{\mathrm{i}} \simeq \alpha \times \left(d_{0}\right)^{-1} \times 2.1\times 10^{9}\,.
\end{eqnarray}
Increasing $\alpha$ thus increases the initial DE kinetic energy relative to the coupling strength. We use similar definitions to analyze Models M2 and M6.
Figure \ref{Fig:H0} illustrates the impact of varying $\alpha$ and the respective coupling strengths on $H_0$ for Models M2, M3, M6, and M7. We find that:
\begin{enumerate}
\item {\bf $H_0$ Increase:} For Models M2, M3, and M6, increasing the parameter $\alpha$ (i.e., increasing the EDE contribution) results in a \emph{quantifiable increase in the value of $H_0$}.

\item {\bf Optimal Parameter Range:} The maximal increase in $H_0$ occurs for coupling strength ranges $d_0 \sim 10^{-2}-10^{0}$ (or corresponding $ck_1, m_0$ values). This range corresponds to a transition redshift of approximately $10^{4} \lesssim z_t \lesssim 10^{2}$.

\item {\bf $H_0$ Decrease:} Very large coupling strengths lead to late-time transition redshifts ($z_t \lesssim 10^2$). This significantly alters the angular diameter distance to the last scattering surface, resulting in a \emph{decreased value of $H_0$}, consistent with results found in the EDE literature for late-time intervention.

\item {\bf Distinction from traditional EDE:} Unlike standard EDE models, the shift in $H_0$ in our interacting framework is a combined consequence of the EDE-like behavior *and* the enhanced energy density of dark matter at high redshifts due to the coupling. Therefore, the required \emph{early dark energy contribution is less than $10\%$} to achieve the necessary shift in $H_0$ to alleviate the tension.

\item {\bf Model M7:} Model M7, a special case of pure-momentum coupling, consistently results in a \emph{lower value of $H_0$} compared to the non-interacting $\Lambda$CDM scenario, thereby worsening the Hubble tension.
\end{enumerate}

In this regard, we find that Model M4 gives similar results as M3. Implications of Model M1 on the Hubble tension were discussed in Ref.~\cite{2022-Johnson.etal-JCAP}.
\begin{table}[htbp]
\vspace*{-0.50cm}
\hspace*{-0.75cm}
    \centering
\begin{tabular}{||c|l|l|l||} 
\hline \hline  
  Model 
& Form of coupling 
& Effect on $\rho_{\mathrm{DE}}$ and $H_0$
& Effect on P(k), $\sigma_8$ and $C_{\ell}^{\mathrm{TT}}$
\\
\hline \hline 
M1 & $\begin{array}{l}
        C(\phi) = c_0\,\phi^{cf_0}\,e^{cf_1\,\phi} \\
        \displaystyle D = 0
      \end{array}$
   & $\begin{array}{cc}
        \rho_{\mathrm{DE}} \propto a^{-2} \rightarrow a^{-3} \rightarrow a^0 \\
        \mbox{Value of}~H_0~\mbox{increases}~(\uparrow)
     \end{array}$
   & $\begin{array}{l}
        \text{Large scales: P(k) decreases}~(\downarrow)  \\[-0.5em]
        \text{Small scales: P(k) increases}~(\uparrow) \implies \sigma_8 \uparrow \\
        \ell\lesssim30: C_{\ell}^{\mathrm{TT}} \uparrow \quad \text{(Moderate change}\footnotemark[5]  \\[-0.5em]
        \ell\gtrsim30: C_{\ell}^{\mathrm{TT}} \downarrow \quad \text{~over low and high $\ell$)}
      \end{array}$
\\
\hline
M3 & $\begin{array}{l}
       C = c_0 \\
       \displaystyle D = d_0
       \\[0.75em]
      \end{array}$ 
   & \multirow{2}{*}{\parbox[c][2\totalheight][l]{4.75cm}{\raggedright \vspace{-0.3cm} EDE-like evolution: \\[-0.5em] $\rho_{\mathrm{DE}} \propto a^0 \xrightarrow{z_t} a^{-6} \rightarrow a^0 $ \\[-0.5em] (KECC\footnotemark[1] $\rightarrow$ FF\footnotemark[2] $\rightarrow$ PECC\footnotemark[3]) \\ Transition epoch $z_t$\footnotemark[4]: \\[-0.5em] $z_t\lesssim10: H_0\downarrow$ \\[-0.5em] $z_t\gg10: H_0\uparrow$ }}
   & \multirow{2}{*}{\parbox[c][2\totalheight][l]{7.6cm}{\raggedright \vspace{-0.3cm} 
   Large scales: P(k) $\downarrow$ \\[-0.5em] Small scales: Low $d_0$: \,P(k)$\downarrow\implies\sigma_8\downarrow$ \\[-0.5em] \hspace{2.25cm}High $d_0$: P(k)$\uparrow\implies\sigma_8\uparrow$ 
   \\ 
   $\ell\lesssim70$: $C_{\ell}^{\mathrm{TT}} \downarrow$ \text{(Large suppression\footnotemark[5] at low $\ell$)}
   \\[-0.5em] 
   $\ell\gtrsim70$: $C_{\ell}^{\mathrm{TT}} \downarrow$ \text{(Small suppression\footnotemark[5] at high $\ell$)}
   \\[-0.5em]
   }}
\\
\cline{0-1}
M4 & $\begin{array}{l}
       C = c_0 \\
       \displaystyle D(\phi) = d_0\,\phi^{df_0}\,e^{df_1\,\phi} 
       \\[0.75em]
      \end{array}$ 
   &
   &
\\
\hline
M2 & $\begin{array}{l}
       C(X) = c_0\,e^{ck_1\,X} \\
       \displaystyle D = 0
      \end{array}$
   & \multirow{2}{*}{\parbox[c][2\totalheight][l]{4.3cm}{\raggedright \vspace{2.5cm} $\rho_{\mathrm{DE}} \propto a^{-3} \xrightarrow{z_t} a^{-6} \rightarrow a^0$ \\[-0.5em] (DM $\rightarrow$ FF\footnotemark[2] $\rightarrow$ PECC\footnotemark[3]) \\ Transition epoch $z_t$\footnotemark[4]: \\[-0.5em] $z_t\lesssim10: H_0\downarrow$ \\[-0.5em] $z_t\gg10: H_0\uparrow$}}
   & $\begin{array}{l}
        \text{All scales: P(k)} \uparrow \implies \sigma_8 \uparrow \\
        \ell\lesssim80-100: C_{\ell}^{\mathrm{TT}} \uparrow \\[-0.5em]
        (\text{Large enhancement\footnotemark[5] at low $\ell$})
      \end{array}$
\\
\cline{0-1} \cline{4-4}
M5 & $\begin{array}{l}
       C = c_0 \\
       \displaystyle D(X) = d_0\,e^{dk_1\,X}
      \end{array}$ 
   &
   & $\begin{array}{l}
        \text{Large scales: P(k)} \downarrow \\[-0.5em]
        \text{Small scales: Low}~d_0: \text{P(k)} \downarrow \implies \sigma_8 \downarrow \\[-0.5em]
        \text{\hspace{2.25cm}High}~d_0: \text{P(k)} \uparrow \implies \sigma_8 \uparrow \\
        \text{Low}~d_0-\ell\lesssim70:C_{\ell}^{\mathrm{TT}}\downarrow \text{(Large suppression\footnotemark[5])} \\[-0.5em]
        \hspace{1.65cm}\ell\gtrsim70:C_{\ell}^{\mathrm{TT}}\downarrow \text{(Small suppression\footnotemark[5])} \\[-0.5em]
        \text{High}~d_0-\ell<10~\text{and}~\ell>100: C_{\ell}^{\mathrm{TT}}\downarrow \\[-0.5em]
        \hspace{1.65cm} 10 \lesssim \ell \lesssim 100: C_{\ell}^{\mathrm{TT}}\uparrow
      \end{array}$
\\
\cline{0-1}\cline{4-4}
M6 & $\begin{array}{l}
       MT(X)= \\
       m_0\,X^{mk_0}\,e^{mk_1\,X}
      \end{array}$
   &  
   & $\begin{array}{l}
        \text{Large scales: P(k)} \downarrow \\[-0.5em]
        \text{Small scales: P(k)} \uparrow \implies \sigma_8 \uparrow \\
        \ell<70: C_{\ell}^{\mathrm{TT}} \uparrow (\text{Large increase\footnotemark[5] at low $\ell$}) \\[-0.5em]
        \ell\gtrsim70: C_{\ell}^{\mathrm{TT}} \downarrow (\text{Small suppression\footnotemark[5] at high $\ell$})
      \end{array}$
\\
\cline{0-3}
M7 & $MT(X)=\displaystyle \frac{m_1}{X}$ 
   & $\begin{array}{l}
       \rho_{\mathrm{DE}} \propto a^{-3} \rightarrow a^0 \\[-0.5em]
       \hspace{1cm} (\text{DM} \rightarrow \text{PECC}\footnotemark[3]) \\
       \multicolumn{1}{c}{H_0 \downarrow}
      \end{array}$
   & $\begin{array}{l}
        \text{All scales: P(k)} \uparrow \implies \sigma_8 \uparrow  \\
        \ell<70: C_{\ell}^{\mathrm{TT}} \uparrow (\text{Large increase\footnotemark[5] at low $\ell$}) \\[-0.5em]
        \ell\gtrsim70: C_{\ell}^{\mathrm{TT}} \downarrow (\text{Small suppression\footnotemark[5] at high $\ell$})
   \end{array}$
\\
\hline \hline
\end{tabular}
\footnotetext[1]{KECC denotes Kinetic-Energy-dominated Cosmological Constant}
\footnotetext[2]{FF denotes Free Field: Kinetic-energy–driven scalar field, governed by the free Klein-Gordan equation}
\footnotetext[3]{PECC denotes Potential-Energy-dominated Cosmological Constant}
\footnotetext[4]{$z_t$ depends on the coupling strengths $\{ck_1, d_0, df_0, df_1, dk_1, m_0, mk_0, mk_1\}$: Coupling $\uparrow \implies z_t \downarrow$}
\footnotetext[5]{For reference, \emph{Large} change: $\gtrsim$10\%; \emph{Moderate} change: $\sim$ 5-6\%; \emph{Small} change: $\lesssim$ 1-2\% deviation in $C_{\ell}^{\mathrm{TT}}$}
    \caption{Table summarizes the key characteristics of each model at both the background FLRW and linear perturbation levels, highlighting the unique features and cosmological implications of the different coupling types, including modifications relative to the non-interacting scenario.}
    \label{Table:Summary}
\end{table}
\subsection{Summary of results}
\label{Subsec:Summary}

To provide a comprehensive overview of the key results, we summarize the characteristics and main findings for all models discussed in this work in Table \ref{Table:Summary}\,. This table consolidates the key characteristics of each model at both the background FLRW and linear perturbation levels, highlighting the unique features and cosmological implications of the different coupling types. Specifically, we compare their effects on the background evolution of dark energy, the Hubble expansion rate, the matter power spectrum, and the CMB temperature power spectrum. This allows for a direct comparison of the distinct phenomenologies arising from purely conformal, disformal, and pure-momentum couplings, providing a clear roadmap to the potential of each model in addressing current cosmological tensions.

\section{Conclusions and Discussions}
\label{Sec:Conclusions}

In this work, we have provided a first-principles, field-theoretic framework for DE-DM interactions using a general disformal coupling. By moving beyond the simple fluid description, our approach offers a rich set of scenarios like conformal, disformal and pure-momentum coupling models with distinct cosmological behaviors. Our analysis of these models at both the background and linear perturbation levels reveals compelling connections to the major tensions faced by the standard $\Lambda$CDM model.

Our most significant finding is related to \emph{Models M3 and M4}, which feature a pure disformal coupling. These models naturally lead to an \emph{interacting Early Dark Sector} where the dark energy field's evolution exhibits a unique three-stage behavior: an early-time cosmological constant-like phase driven by a constant kinetic energy, a subsequent decay phase at a rate proportional to $a^{-6}$, and a late-time potential-driven cosmological constant-like phase accounting for late dark energy. This evolution, purely driven by the disformal coupling and the dynamics of dark matter, offers a physically motivated alternative to the existing \emph{EDE models}, which often require finely-tuned scalar field potentials. The ability of these models to increase the CMB-inferred Hubble constant, $H_0$\,, by a transition occurring around the epoch of recombination, offers a promising pathway to alleviating the \emph{Hubble tension}.

Furthermore, these models exhibit distinct signatures in the linear matter and CMB power spectra. Interacting Early Dark Sector models with a high transition redshift ($z_t \gtrsim 10^5$) suppress structure growth, while those with a lower transition redshift lead to enhanced small-scale structure growth. On the other hand, they produce a crucial power suppression on large angular scales in the CMB temperature spectrum. This suppression is consistent with the observed \emph{low-$\ell$ CMB anomaly} from Planck, demonstrating that our framework offers a potential resolution for this persistent observational puzzle. We contrast these results with other models in our framework, such as the pure-momentum coupling Models M6 and M7, which predict enhanced power at low-$\ell$, making it disfavored by current data.

The unique signatures of the disformally interacting Early Dark Sector provide a clear avenue for future observational tests. Upcoming high-precision surveys from telescopes like the James Webb Space Telescope (JWST), along with ground-based projects such as the Dark Energy Spectroscopic Instrument (DESI), the Euclid mission, and the Vera C. Rubin Observatory's LSST, will provide unprecedented data. These observations will be able to probe the subtle modifications to the expansion history and structure growth, offering a critical test for our disformal coupling framework and its potential to resolve cosmic tensions.

To date, these observations have primarily relied on electromagnetic signals, which have provided critical insights but are subject to limitations in probing certain aspects of cosmic evolution. A promising new avenue to distinguish our disformally coupled dark energy models from $\Lambda$CDM lies in the detection of \emph{gravitational wave (GW) memory}~\cite{Chakraborty:2024ars,Chakraborty:2025qcu}. This phenomenon, a permanent spacetime displacement caused by passing gravitational waves, is sensitive to the underlying cosmological model. In Refs.~\cite{Chakraborty:2024ars,Chakraborty:2025qcu}, the authors have shown that the integrated GW memory from high-redshift sources carries a distinctive signature, with amplification factors potentially reaching 100 in certain cosmological backgrounds.  This places it within the detection capabilities of next-generation observatories like Cosmic Explorer and the Einstein Telescope. Because our disformal coupling framework uniquely modifies the expansion history and the effective dark energy equation of state—especially at early times—it imparts a distinct signature on GW memory, offering a powerful, independent probe to test our models against competing dark energy scenarios and potentially resolve the current cosmological tensions.

\begin{acknowledgments}
The work is supported by the SERB-Core Research
Grant (Project SERB/CRG/2022/002348).
\end{acknowledgments}

\appendix

\section{Disformal transformation of DHOST to quintessence theory}
\label{Appendix:DT-DHOST-Quintessence}

In this Appendix, we discuss how one can arrive at a quintessence scalar field description in the Einstein frame from some theory in the Jordan frame. To do this, let us start with the following form of the Lagrangian in the Jordan frame:
\begin{equation} \label{AEq:A1}
\begin{aligned}
\Tilde{S} = \int \d^4x\sqrt{\Tilde{g}}
&
\left[ \frac{M^{2}_{\mathrm{Pl}}}{2}\Tilde{f}(\phi, \Tilde{X})\Tilde{R} + \Tilde{G}_{2}(\phi, \Tilde{X}) + \Tilde{A}_{1}(\phi, \Tilde{X})\Tilde{\mathcal{L}}^{\phi}_1 + \Tilde{A}_{2}(\phi, \Tilde{X})\Tilde{\mathcal{L}}^{\phi}_2
\right. \\
& \left.
+ \Tilde{A}_{3}(\phi, \Tilde{X})\Tilde{\mathcal{L}}^{\phi}_3 + \Tilde{A}_{4}(\phi, \Tilde{X})\Tilde{\mathcal{L}}^{\phi}_4 + \Tilde{A}_{5}(\phi, \Tilde{X})\Tilde{\mathcal{L}}^{\phi}_5 \right] \,.
\end{aligned}
\end{equation}
where we have \cite{BenAchour:2016cay}:
\begin{equation}
\begin{aligned}
\mathcal{L}^{\phi}_1 & \equiv \phi_{\mu\nu}\phi^{\mu\nu}\,, \quad 
\mathcal{L}^{\phi}_2 \equiv \left( \Box\phi \right)^2\,, \quad 
\mathcal{L}^{\phi}_3 \equiv \Box\phi\phi_{\mu\nu}\phi^{\mu}\phi^{\nu}\,, \\
\mathcal{L}^{\phi}_4 & \equiv  \phi_{\mu\nu}\phi^{\nu}\phi^{\mu\sigma}\phi_{\sigma}\,, \quad
\mathcal{L}^{\phi}_5 \equiv \left( \phi_{\mu\nu}\phi^{\mu}\phi^{\nu} \right)^2\,.
\end{aligned}
\end{equation}
The action \eqref{AEq:A1} corresponds to a quadratic DHOST description of the scalar field $\phi$. Next, we perform a generalized disformal transformation:
\begin{equation} \label{AEq:GDT}
\Tilde{g}_{\mu\nu} = C(\phi, X)g_{\mu\nu} + D(\phi, X)\phi_{\mu}\phi_{\nu}\, ,
\end{equation}
This transforms the action \eqref{AEq:A1} to the following form:
\bea
S = \int \mathrm{d}^{4}x\sqrt{-g}
& &
\left[ \frac{M^{2}_{\mathrm{Pl}}}{2}f(\phi, X)R + G_{2}(\phi, X) + A_{1}(\phi, X)\mathcal{L}^{\phi}_1 + A_{2}(\phi, X)\mathcal{L}^{\phi}_2 \right. \\
& & \left.
+ A_{3}(\phi, X)\mathcal{L}^{\phi}_3 + A_{4}(\phi, X)\mathcal{L}^{\phi}_4 + A_{5}(\phi, X)\mathcal{L}^{\phi}_5 \right] \,,
\nonumber
\eea
where the functions $f, G_2, A_1-A_5$ are related to the disformal and original functions, $C, D, \Tilde{f}, \Tilde{G}_2$, and $\Tilde{A}_1-\Tilde{A_5}$ in the following manner \cite{BenAchour:2016cay}:
\begin{subequations}
\begin{align}
f & = \sqrt{C(C - 2DX)}\,\Tilde{f} \,, \\
G_2 & = \sqrt{C^3(C - 2DX)}\,\Tilde{G}_2 \,, \\
A_1 & = - h + \sqrt{C^{3}(C - 2DX)}\,\mathcal{T}_{11}\Tilde{A}_1 \,, \\
A_2 & = h + \sqrt{C^{3}(C - 2DX)}\,\mathcal{T}_{22}\Tilde{A}_2 \,, \\
A_3 & = -h_X + C^{3/2}\sqrt{C-2DX}\left[ \Tilde{f}\gamma_3 + \Tilde{X}_X\Tilde{f}_{\Tilde{X}}\lambda_3 + \mathcal{T}_{13}\Tilde{A}_1 + \mathcal{T}_{23}\Tilde{A}_2 + \mathcal{T}_{33}\Tilde{A}_3 \right] \,, \\
A_4 & = h_X + C^{3/2}\sqrt{C-2DX}\left[ \Tilde{f}\gamma_4 + \Tilde{X}_X\Tilde{f}_{\Tilde{X}}\lambda_4 + \mathcal{T}_{14}\Tilde{A}_1 + \mathcal{T}_{44}\Tilde{A}_4 \right] \,, \\
A_5 & = \sqrt{C^3(C-2DX)}\left[ \Tilde{f}\gamma_5 + \Tilde{X}_X\Tilde{f}_{\Tilde{X}}\lambda_5 + \sum^{i=1}_{i=5}\mathcal{T}_{i5}\Tilde{A}_i  \right] \,, \\
h & = -\sqrt{\frac{CD^2}{C - 2DX}}\,\Tilde{f} \,.
\end{align}
\end{subequations}
The form of the functions $\gamma_i$ and $\mathcal{T}_{ij}$ is complex and is not provided here. We refer the readers to Ref.~\cite{BenAchour:2016cay} for a more detailed discussion and the exact form of these functions.

\noindent Imposing the condition $f = 1$ takes us to the Einstein frame which gives us the condition:
\begin{eqnarray}
\Tilde{f} = \frac{1}{\sqrt{C(C - 2DX)}}\,.
\end{eqnarray}
Furthermore, we can have $A_1 = 0$ and $A_2 = 0$ which gives us:
\begin{subequations} \label{AEq-DHOST-terms-JF}
\begin{align}
\Tilde{A}_1 & = \frac{h}{\sqrt{C^{3}(C - 2DX)}\,\mathcal{T}_{11}}\,, \\
\Tilde{A}_2 & = \frac{-h}{\sqrt{C^{3}(C - 2DX)}\,\mathcal{T}_{22}}\,, \\
\Tilde{A}_3 & = \frac{1}{\mathcal{T}_{33}}\left[ \frac{h_X}{\sqrt{C^{3}(C - 2DX)}} - \Tilde{f}\gamma_3 - \Tilde{X}_X\Tilde{f}_{\Tilde{X}}\lambda_3 - \mathcal{T}_{13}\Tilde{A}_1 - \mathcal{T}_{23}\Tilde{A}_2 \right] \,, \\
\Tilde{A}_4 & = \frac{1}{\mathcal{T}_{44}}\left[ \frac{h_X}{\sqrt{C^{3}(C - 2DX)}} - \Tilde{f}\gamma_4 - \Tilde{X}_X\Tilde{f}_{\Tilde{X}}\lambda_4 - \mathcal{T}_{14}\Tilde{A}_1 \right] \,, \\
\Tilde{A}_5 & = -\frac{1}{\mathcal{T}_{55}}\left[ \Tilde{f}\gamma_5 + \Tilde{X}_X\Tilde{f}_{\Tilde{X}}\lambda_5 + \sum^{i=1}_{i=4}\mathcal{T}_{i5}\Tilde{A}_i \right]\,.
\end{align}
\end{subequations}
Since we have $\Tilde{f}$, $h$, $\gamma_i$ and $\mathcal{T}_{ij}$ in terms of the disformal factors $C$ and $D$, imposing $A_i = 0$ in the Einstein frame gives us specific functional forms of $\Tilde{A}_i$ (from the Jordan frame) in terms of the disformal factors. This means that, in principle, we can have a $k-$essence DE field description in the Einstein frame (which includes quintessence) with arbitrary form of the disformal coupling functions $C$ and $D$, provided we start with a DHOST theory given by Eqs.~\eqref{AEq-DHOST-terms-JF} in the Jordan frame. Moreover, one can verify that this theory in the Jordan frame belongs to the DHOST subclass type Ia theories which are disformally related to the Horndeski (and therefore quintessence) theories \cite{BenAchour:2016cay}. Therefore, the actions in the two frames of reference remain ghost-free and without any instabilities.

\section{Modeling the dark sector interactions}
\label{Appendix:Summary-Previous-Works}

In this Appendix, we briefly discuss the construction of the field-theoretic frameworks describing DE-DM interactions developed by the current authors in previous works \cite{2021-Johnson.Shankaranarayanan-Phys.Rev.D, 2022-Johnson.etal-JCAP, Bansal:PRD:2025}.
\subsection{Dark sector interactions from $f(R)$ theories}
\label{Appendix:iDEDM-from-f(R)}

To demonstrate the construction of an interacting dark sector model with a strong theoretical foundation, we start with the following toy model \cite{2021-Johnson.Shankaranarayanan-Phys.Rev.D}. Consider the following action:
\begin{eqnarray}
S = \int \d^4x\sqrt{-\Tilde{g}}\left[ \frac{M_{\mathrm{Pl}}^2}{2}\Tilde{f}(R, \upchi) + \Tilde{L}_{\chi}(\Tilde{g}_{\mu\nu}, \upchi)\right] + \int \d^4x\sqrt{-g}\mathcal{L}_{_{\mathrm{SM}}} \,.
\end{eqnarray}
Here, the scalar field $\upchi$ represents the dark matter field and $\mathcal{L}_{_{\mathrm{SM}}}$ represents the standard model particles moving along the Einstein frame ($g_{\mu\nu}$) geodesics. We see that the scalar field $\upchi$ is non-minimally coupled to the geometric part of the action in the Jordan frame ($\Tilde{g}_{\mu\nu}$). To remove this non-minimal coupling, we perform a conformal transformation and move to the Einstein frame. In this case, we use the following conformal transformation:
\begin{equation}
\label{eq:conftrans}
{g}_{\mu \nu}=\Omega^{2} \tilde{g}_{\mu \nu}\,,
\quad \mbox{where} \quad
\Omega^{2}= F(\tilde{R},\tilde{\upchi}) \equiv \frac{\partial f(\tilde{R}, \tilde{\upchi})}{\partial \tilde{R}}\,.
\end{equation}
After this transformation and a field redefinition, the action in the Einstein frame takes the following form:
\begin{equation}
 \label{eq:Scde}
S = \int \d^4x \sqrt{-g}\left[\dfrac{M_{\mathrm{Pl}}^2}{2}R-\dfrac{1}{2}g^{\mu \nu}\nabla_{\mu}\phi \nabla_{\nu}\phi - U(\phi)-\dfrac{1}{2} e^{2\alpha(\phi)}g^{\mu \nu}\nabla_{\mu}\upchi \nabla_{\nu}\upchi -e^{4 \alpha(\phi)} V(\upchi) + \mathcal{L}_{_{\mathrm{SM}}} \right] \,.
\end{equation}
Here, we have used the mapping:
\begin{eqnarray}
& & \phi \equiv M_{\mathrm{Pl}} \sqrt{\frac{3}{2}}\mathrm{ln}F\,, \\ 
& & U \equiv \frac{M_{\mathrm{Pl}}^2}{2}\left(\frac{F\tilde{R}-f}{F^{2}}\right)\,. \nonumber
\end{eqnarray}
Therefore, the conformal transformation and the field redefinitions result in a coupling between the dark energy ($\phi$) and dark matter ($\chi$) fields via. the coupling function $\alpha(\phi)$ in the Einstein frame.
Now, we can derive the energy-momentum tensor and the interaction term for the two scalar fields as the following:
\begin{eqnarray}
\label{eq:emtensor}
T^{(\chi)}_{\mu \nu} &=&  e^{2\alpha(\phi)}\left(\nabla_{\mu}\upchi \nabla_{\nu}\upchi - \dfrac{1}{2}g_{\mu \nu} \nabla^{\sigma} \upchi \nabla_{\sigma} \upchi - e^{2 \alpha(\phi)}g_{\mu \nu}V(\upchi)\right)\,, \\
T^{(\phi)}_{\mu \nu} &=&\nabla_{\mu}\phi \nabla_{\nu}\phi-\dfrac{1}{2}g_{\mu \nu}\nabla^{\sigma}\phi \nabla_{\sigma}\phi-g_{\mu \nu}U(\phi)\,, \\
\label{eq:interaction}
Q_{\nu}^{\rm (F)}  &=& \nabla^{\mu} T_{\mu\nu}^{(\chi)} =  -e^{2\alpha(\phi)} \alpha_{,\phi}(\phi) \nabla_{\nu} \phi \left[ \nabla^{\sigma} \upchi \nabla_{\sigma} \upchi + 4 e^{2\alpha(\phi)} V(\upchi) \right]\,.
\end{eqnarray}
Given that, the energy momentum tensor of a perfect fluid can be expressed as:
\bea
T^{(\mathrm{DM})}_{\mu\nu}=p_{\mathrm{DM}} g_{\mu \nu} + (\rho_{\mathrm{DM}} + p_{\mathrm{DM}}) u_{\mu} u_{\nu} \,.
\eea
One can map the scalar field description of the dark matter to the fluid description by defining its energy density $\rho_{\mathrm{DM}}$, pressure $p_{\mathrm{DM}}$, and the four-velocity $u^{(\mathrm{DM})}_{\mu}$ as
\begin{subequations}
\label{eq:dmrhop4v}
\begin{align}
p_{\mathrm{DM}} & = -\dfrac{1}{2}e^{2 \alpha(\phi)}\left[g^{\mu \nu} \nabla_{\mu} \upchi \nabla_{\nu} \upchi + e^{2\alpha(\phi)}V(\upchi) \right]\,, \\ 
\rho_{\mathrm{DM}} & = -\dfrac{1}{2}e^{2 \alpha(\phi)}\left[g^{\mu \nu} \nabla_{\mu} \upchi \nabla_{\nu} \upchi - e^{2\alpha(\phi)}V(\upchi) \right]\,, \\
u^{(\mathrm{DM})}_{\mu} & = -\left[-g^{\gamma\delta} \nabla_{\gamma}\upchi \nabla_{\delta}\upchi \right]^{-\frac{1}{2}} \nabla_{\mu} \upchi\,.
\end{align}
\end{subequations}
After making this identification, one can write down the interaction term as:
\begin{equation}
\label{eq:interaction02}
Q_{\nu} = \nabla_{\mu} T^{(\mathrm{DM})\mu}_{\nu} =  -e^{2\alpha(\phi)} \alpha_{,\phi}(\phi) \nabla_{\nu} \phi \left[ \nabla^{\sigma} \upchi \nabla_{\sigma} \upchi + 4 e^{2\alpha(\phi)} V(\upchi) \right] =  -\alpha_{,\phi}(\phi) \nabla_{\nu}\phi (\rho_{\mathrm{DM}}  - 3 p_{\mathrm{DM}})\,.
\end{equation}
Identifying $T^{(\mathrm{DM})}=T^{(\mathrm{DM}) \mu}_{\mu} = -(\rho_{\mathrm{DM}} - 3 p_{\mathrm{DM}})$, we get:
\begin{equation}
\label{eq:traceinter}
Q_{\nu}^{\rm (F)}  = T^{(\mathrm{DM})} \nabla_{\nu}\alpha(\phi)\,.
\end{equation}
Now, background and perturbed equations for the interacting dark sector can be obtained by substituting the perturbed FLRW line element to these equations.

\subsection{Dark sector interactions from Horndeski theories}

Next, we construct an interacting DE-DM framework starting with a quadratic-Horndeski Lagrangian \cite{Bansal:PRD:2025}. To do this, we consider the following action:
\bea \label{Eq:Horndeski action}
S = \int \mathrm{d}^{4}x\sqrt{-\Tilde{g}}
& &
\left[\frac{M^{2}_{\mathrm{Pl}}}{2}\Tilde{G}_{4}(\phi, \Tilde{X})\Tilde{R} + \Tilde{G}_{2}(\phi, \Tilde{X}) - \Tilde{G}_{3}(\phi, \Tilde{X})\Tilde{\Box}\phi
\right. \\ 
& & \left. 
+ \frac{M^{2}_{\mathrm{Pl}}}{2}\Tilde{G}_{4\Tilde{X}}[(\Tilde{\Box}\phi)^{2} - \Tilde{\nabla}_{\mu}\Tilde{\nabla}_{\nu}\phi\Tilde{\nabla}^{\mu}\Tilde{\nabla}^{\nu}\phi] + P_{1}(\upchi, \Tilde{Y})\right] \, .
\nonumber
\eea
where $\Tilde{X} = -\Tilde{g}^{\mu\nu}\Tilde{\nabla}_{\mu}\phi\Tilde{\nabla}_{\nu}\phi\,/2$, $\Tilde{\Box}\phi = \Tilde{g}^{\mu\nu}\Tilde{\nabla}_{\mu}\Tilde{\nabla}_{\nu}\phi$ and $\Tilde{Y} = -\Tilde{g}^{\mu\nu}\Tilde{\nabla}_{\mu}\upchi\Tilde{\nabla}_{\nu}\upchi\, /2$. $\Tilde{G}_{2}, \Tilde{G}_{3}$, and $\Tilde{G}_{4}$ are arbitrary functions of $\phi$ and $\Tilde{X}$. To remove the non-minimal coupling between gravitation and the field $\phi$, we perform an extended conformal transformation \cite{Bekenstein:1992pj}:
\bea \label{Eq:Extended conformal transformation}
\Tilde{g}_{\mu\nu} = \Tilde{G}^{-1}_{4}\left( \phi, \tilde{X} \right)g_{\mu\nu} = \mathrm{e}^{2\alpha(\phi, X)} \, g_{\mu\nu}  \, ,
\eea
Under this transformation and function redefinitions (refer to \cite{Bansal:PRD:2025} for a detailed derivation), we get the following action in the Einstein frame:
\bea \label{Eq:Field-kinetic coupled:Final action}
\begin{aligned}
S = \int \mathrm{d}^{4}x\sqrt{-g}
&
\left[ \frac{M^{2}_{\mathrm{Pl}}}{2}R + G_{2}(\phi, X) - G_{3}(\phi, X)\Box\phi + A_{1}(\phi, X)\phi_{\mu\nu}\phi^{\mu\nu} + A_{2}(\phi, X)(\Box\phi)^{2} \right. \\
& \left.
+ A_{3}(\phi, X)(\phi_{\mu\nu}\phi^{\mu}\phi^{\nu})^{2} + A_{4}(\phi, X)\Box\phi\phi_{\mu\nu}\phi^{\mu}\phi^{\nu} + A_{5}(\phi, X)\phi_{\mu\nu}\phi^{\nu}\phi^{\mu\sigma}\phi_{\sigma}\right. \\
& \left. 
 + \mathrm{e}^{4\alpha(\phi, X)}P_{1}\left(\chi, \mathrm{e}^{-2\alpha(\phi, X)}Y \right)\right] \, ,
\end{aligned}
\eea
where the functions $A_{1}(\phi, X) - A_{5}(\phi, X)$ are:
\begin{subequations} \label{Eq:Field-kinetic coupled:DHOST terms}
\begin{align}
&
A_{1}(\phi, X) = - A_{2}(\phi, X)
= + M_{\mathrm{Pl}}^{2}\left( \frac{\alpha_{X}}{1 - 2X\alpha_{X}} \right) \,, 
\\ &
A_{3}(\phi, X) 
= - 2M^{2}_{\mathrm{Pl}}\left( \frac{(\alpha_{X})^3}{1 - 2X\alpha_{X}} \right) \,,
\\ &
A_{4}(\phi, X)
= + 2M^{2}_{\mathrm{Pl}}\left(\frac{(\alpha_{X})^2}{1 - 2X\alpha_{X}} \right) \,, \\ &
A_{5}(\phi, X) 
= + M^{2}_{\mathrm{Pl}}(\alpha_{X})^2\left(\frac{1 + 2X\alpha_{X}}{1 - 2X\alpha_{X}}\right) \,.
\end{align}
\end{subequations}
Theories of this form fall under a special class of theories referred to as \emph{Degenerate higher-order scalar-tensor theories} or DHOST theories \cite{Langlois:2018dxi}. Specifically, these theories belong to the class Ia which do not suffer from ghosts or instabilities \cite{Langlois:2015skt}.

Following a similar procedure as we did in Sec.~\ref{Appendix:iDEDM-from-f(R)}, we map the dark matter field to a perfect fluid with the following relations for $\rho_{\mathrm{DM}}$, $p_{\mathrm{DM}}$ and $u^{\mathrm{(DM)}}_{\mu}$:
\begin{subequations}
\begin{align}
\rho_{\mathrm{DM}} & = \mathrm{e}^{2\alpha(\phi, X)}\left[ 2YP_{1Z} - \mathrm{e}^{2\alpha(\phi, X)}P_{1} \right] \,,
\\ p_{\mathrm{DM}} & = \mathrm{e}^{4\alpha(\phi, X)}P_{1} \,,
\\ u^{\mathrm{DM}}_{\mu} & = -[-g^{\alpha\beta}\chi_{\alpha}\chi_{\beta}]^{-\frac{1}{2}}\chi_{\mu} \,.
\end{align}
\end{subequations}
Note that these relations generalize the results of Eq.~\eqref{eq:dmrhop4v} to include non-canonical descriptions of the DM field $\upchi$\,. Under this mapping, the stress-energy tensor describing the DE field and the interaction strength are given by the following:
\bea
T_{\mu\nu}^{(\mathrm{DE})} & & = T_{\mu\nu}^{(II,\phi)} + (3p_{\mathrm{DM}} - \rho_{\mathrm{DM}}) \, \alpha_X\phi_{\mu}\phi_{\nu} \,, \\
Q_{\nu}^{(II)} & & = (3p_{\mathrm{DM}} - \rho_{\mathrm{DM}})\nabla_{\nu}\alpha(\phi, X) = T^{\mathrm{(DM)}}\nabla_{\nu}\alpha(\phi, X)\,.
\label{AEq-IS-Horndeski}
\eea
Interestingly, we note that, while we started with two different descriptions of the theory in the Jordan frame, i.e., $f(R)$ theories and quadratic-Horndeski theories, the form of interaction strength and the DM field-to-fluid mapping in the two theories are similar. Eq.~\eqref{AEq-IS-Horndeski} generalizes the form of interactions derived in Eq.~\eqref{eq:traceinter} to include an additional dependence of the coupling function $\alpha$ on the DE field's kinetic term $X$.

\section{Models M1 and M2: Purely conformal DE-DM interactions}
\label{Appendix:Conformal}

\begin{figure}[!htb]
\minipage[b]{0.5\textwidth}
  \includegraphics[width=1\linewidth]{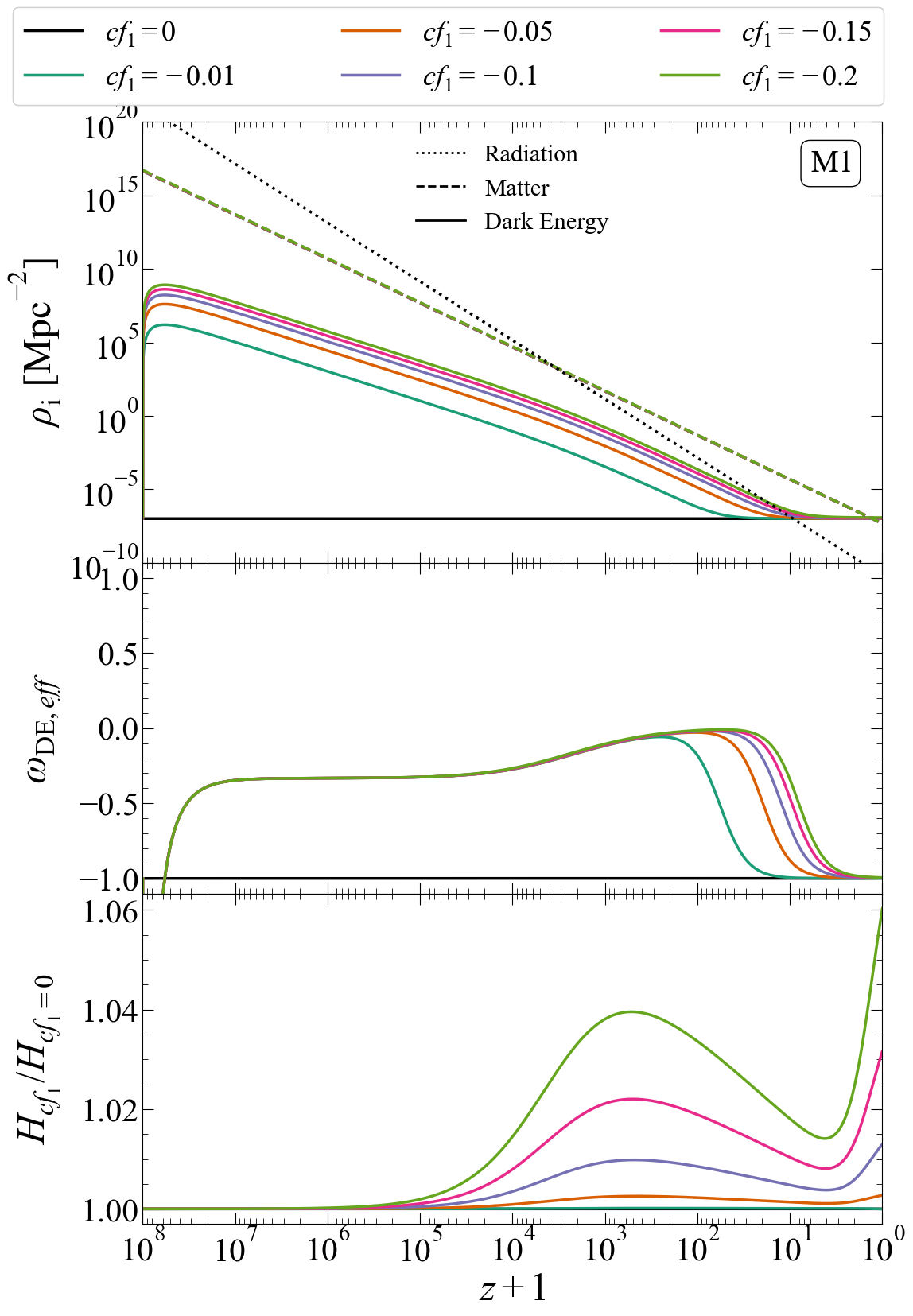}
\endminipage\hfill
\minipage[b]{0.5\textwidth}
  \includegraphics[width=1\linewidth]{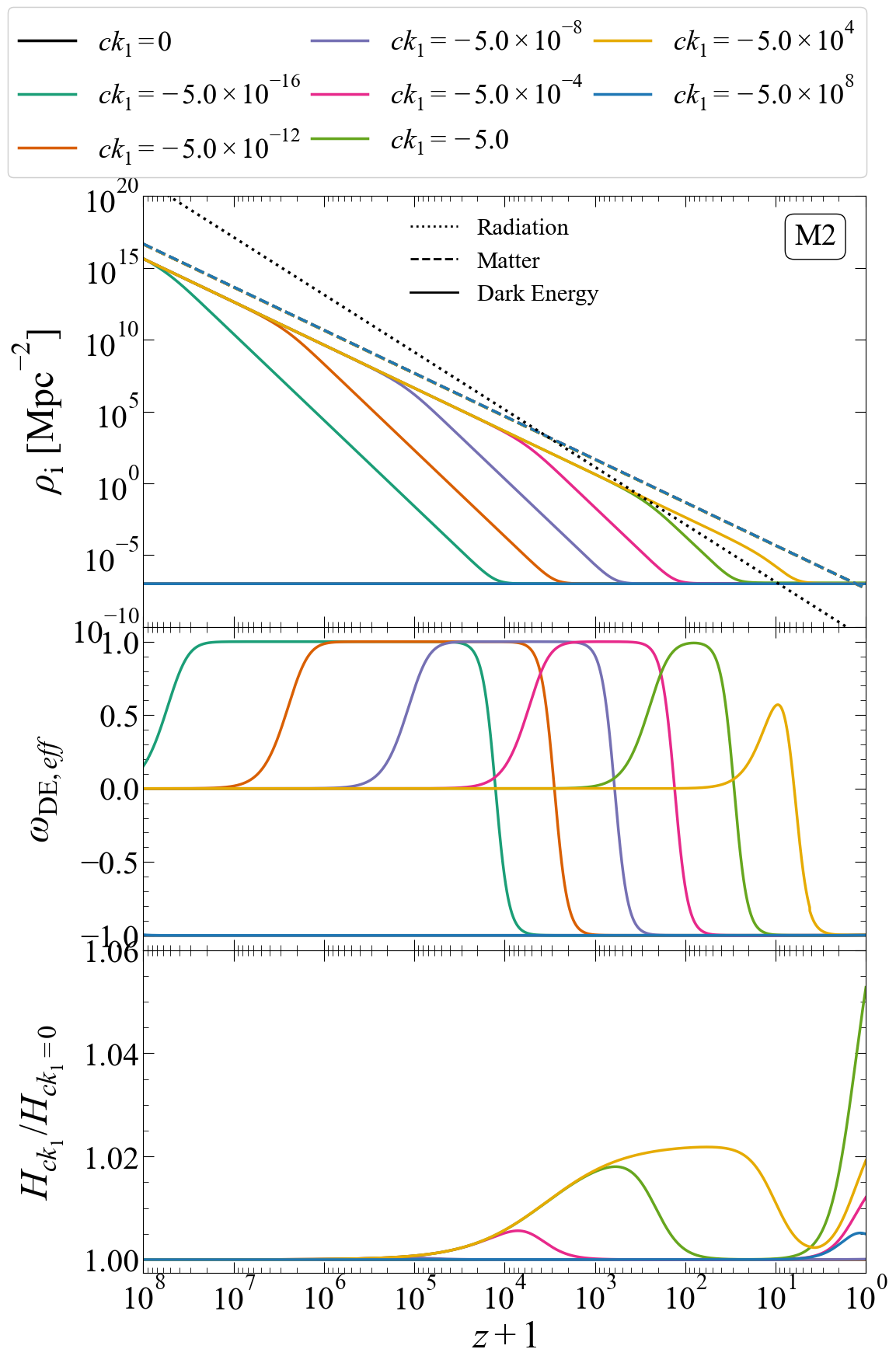}
\endminipage
\caption{Modified background evolution of the DE energy density (top), its effective equation of state (center) and the deviation of the Hubble parameter from the non-interacting case (bottom) due to purely conformal DE-DM interactions. \emph{Left}: Scalar field coupling corresponding to Model M1, \emph{Right}: Kinetic coupling corresponding to Model M2 from Tables \ref{Table:iDEDM} and \ref{Table:Summary}\,.}
\label{Fig:M1,M2:Bg}
\end{figure}
\begin{figure}[!htb]
\minipage{1\textwidth}
  \includegraphics[width=1\linewidth]{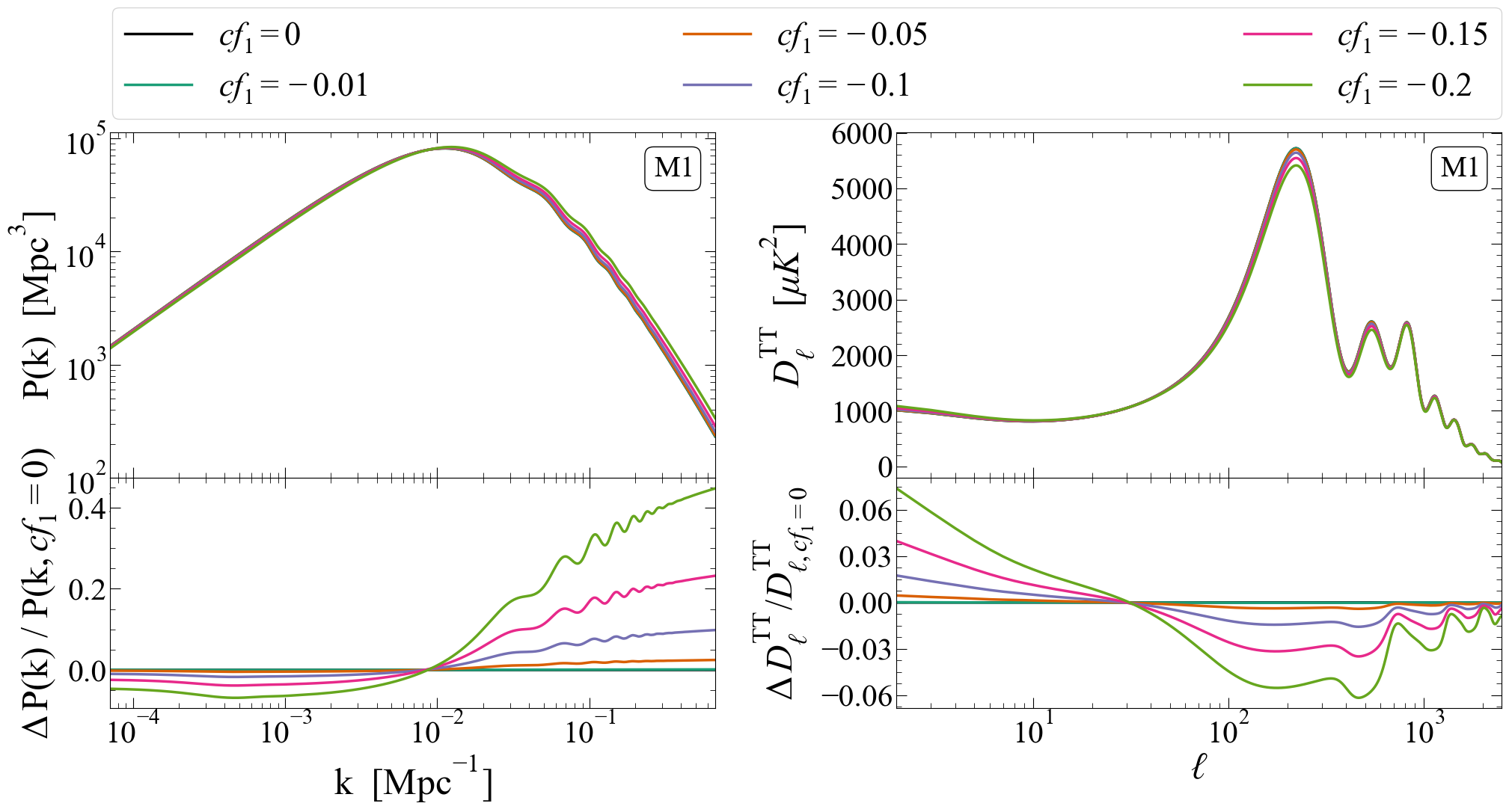}
\endminipage
\caption{Linear matter power spectrum (left), CMB temperature spectrum (right) and relative deviation from the uncoupled scenario due to scalar field dependent conformal DE-DM interactions (Model M1 in Tables \ref{Table:iDEDM} and \ref{Table:Summary}).}
\label{Fig:M1:mPk-DlTT}
\end{figure}

\begin{figure}[!htb]
\minipage{1\textwidth}
  \includegraphics[width=1\linewidth]{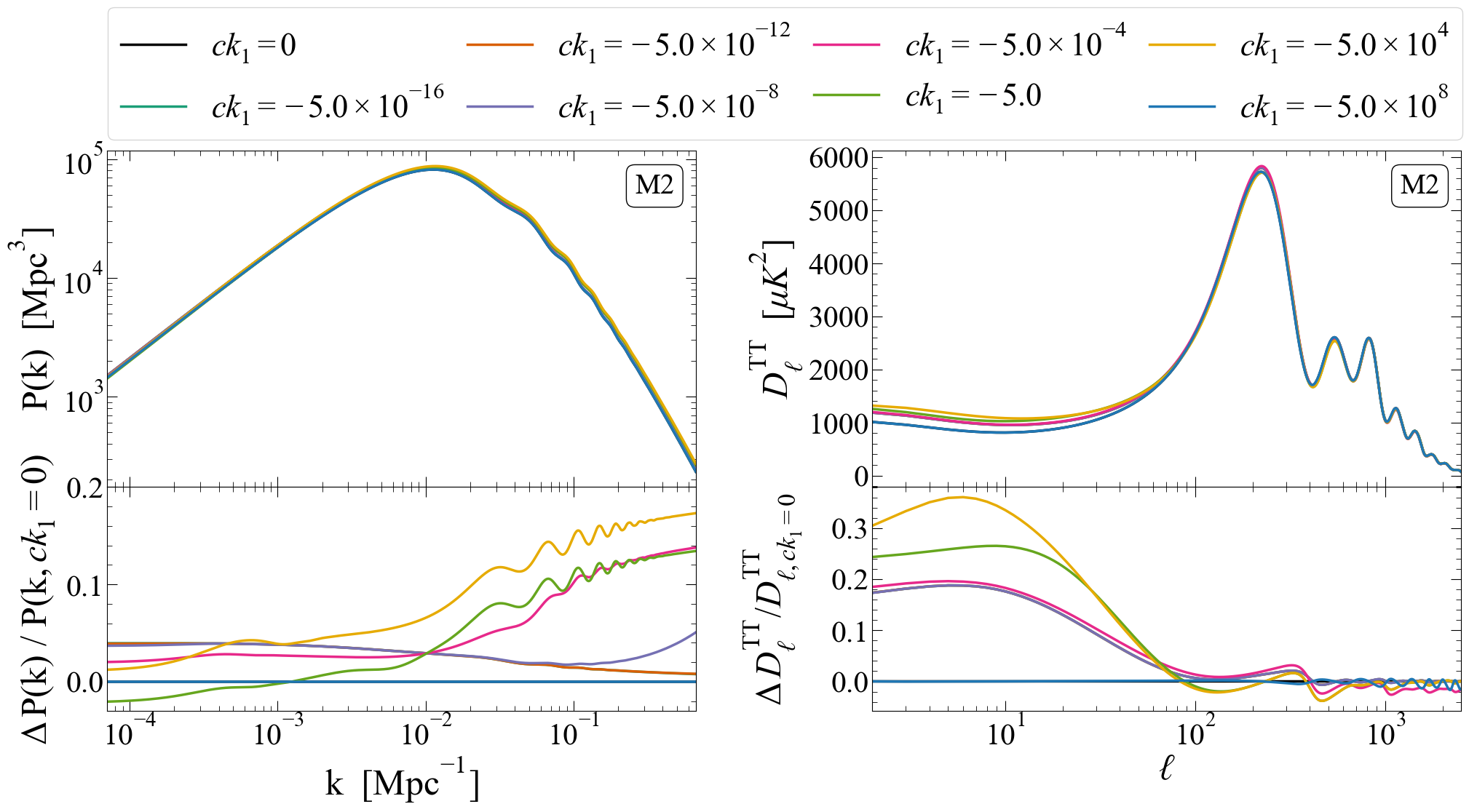}
\endminipage
\caption{Linear matter power spectrum (left), CMB temperature spectrum (right) and relative deviation from the uncoupled scenario due to kinetic dependent conformal DE-DM interactions (Model M2 in Tables \ref{Table:iDEDM} and \ref{Table:Summary}).}
\label{Fig:M2:mPk-DlTT}
\end{figure}

In this appendix, we examine purely conformal interactions between DE  and DM, specifically Models M1 and M2 from Table \ref{Table:iDEDM}\,. The cosmological background evolution for these models is shown in Fig.~\ref{Fig:M1,M2:Bg}\,, while the linear matter and CMB temperature power spectra are shown in Figs.~\ref{Fig:M1:mPk-DlTT}, \ref{Fig:M2:mPk-DlTT}\,. For numerical analysis, we assume an exponential dependence of the couplings $C(\phi)$ and $C(X)$ on the field $\phi$ and its kinetic term $X$, respectively.

\subsection{Model M1: Scalar Field-Dependent Conformal Coupling}
\label{Appendix:M1}

Model M1 corresponds to a conformal scalar field-dependent coupling ($C\equiv C(\phi), D=0$), a class of models that have been extensively studied in the literature \cite{Amendola:1999er, Boehmer:2008av, Beyer:2010mt, CarrilloGonzalez:2017cll, 2021-Johnson.Shankaranarayanan-Phys.Rev.D, 2022-Johnson.etal-JCAP, Bansal:PRD:2025}. We find that for a sufficiently large coupling strength (on the order of $10^{-1}$), the effective equation of state of the DE field evolves as $\omega_{\mathrm{DE,eff}} \approx -1/3 \to 0 \to -1$\,.

Unlike the disformal coupling models discussed in Sec.~\ref{SubSec:Disformal-Numerics}\,, this model does not generate interactions localized in time. In other words, the DE-DM coupling is non-trivial from the onset of matter domination and persists until the present time. This is evident from the evolution of the Hubble parameter relative to the uncoupled scenario, as plotted in the bottom-left panel of Fig.~\ref{Fig:M1,M2:Bg}\,. These couplings generally increase the value of the Hubble constant and enhance the growth of matter density perturbations on small scales, leading to a larger $\sigma_8$ value compared to the uncoupled case. In the CMB TT spectrum, we find a power enhancement on large angular scales (low multipoles) and a suppression on small angular scales (high multipoles). Notably, the deviation in the CMB TT power at the acoustic peaks is relatively large compared to that observed in the disformal interaction models.

\subsection{Model M2: Kinetic Energy-Dependent Conformal Coupling}
\label{Appendix:M2}

Model M2 corresponds to a purely kinetic energy-dependent conformal coupling ($C\equiv C(X), D=0$). The cosmological evolution in this model shares similarities with the pure-momentum coupling Model M6, discussed in Sec.~\ref{Subsec:PMC-Numerics}\,.

At high redshifts, a non-zero kinetic coupling suppresses Hubble friction, leading to a constant kinetic energy for the DE field. As the DM energy density decays and the coupling weakens, the DE kinetic energy begins to dilute at a rate of $a^{-6}$\,. In this model, the DE energy density takes the form:
\bea
\rho_{\mathrm{DE}} = X + V - \rho_{\mathrm{DM}}\,X\,ck_1\,.
\eea
For the DE field to avoid negative energy density states, the coupling strength $ck_1$ must be negative. Therefore, dark energy evolves like a DM fluid at early times, a free scalar field with a dominating but decaying kinetic energy at intermediate times, and a potential-driven cosmological constant at late times. The value of the constant $ck_1$ determines the localized, non-trivial impact on the Hubble parameter, as shown in the bottom-right panel of Fig.~\ref{Fig:M1,M2:Bg}\,. Unlike Model M6, the presence of background energy transfer means the DM fluid does not simply decay proportionally to $a^{-3}$\,.

Similar to the pure-momentum coupling models (M6 and M7), the conformal kinetic coupling leads to a power enhancement over low multipoles and power suppression over larger multipoles in the CMB TT spectrum. Additionally, it leads to an enhanced growth of matter density fluctuations. However, for lower values of the coupling strength, the linear-level energy coupling in this model results in enhanced growth compared to the pure-momentum coupling scenario, which did not show any visible effect on the evolution of linear perturbations.

\section{Models M4 and M5: Field and kinetic dependent disformal couplings}
\label{Appendix:M4,M5}

\begin{figure}[!htb]
\minipage{0.5\textwidth}
  \includegraphics[width=1\linewidth]{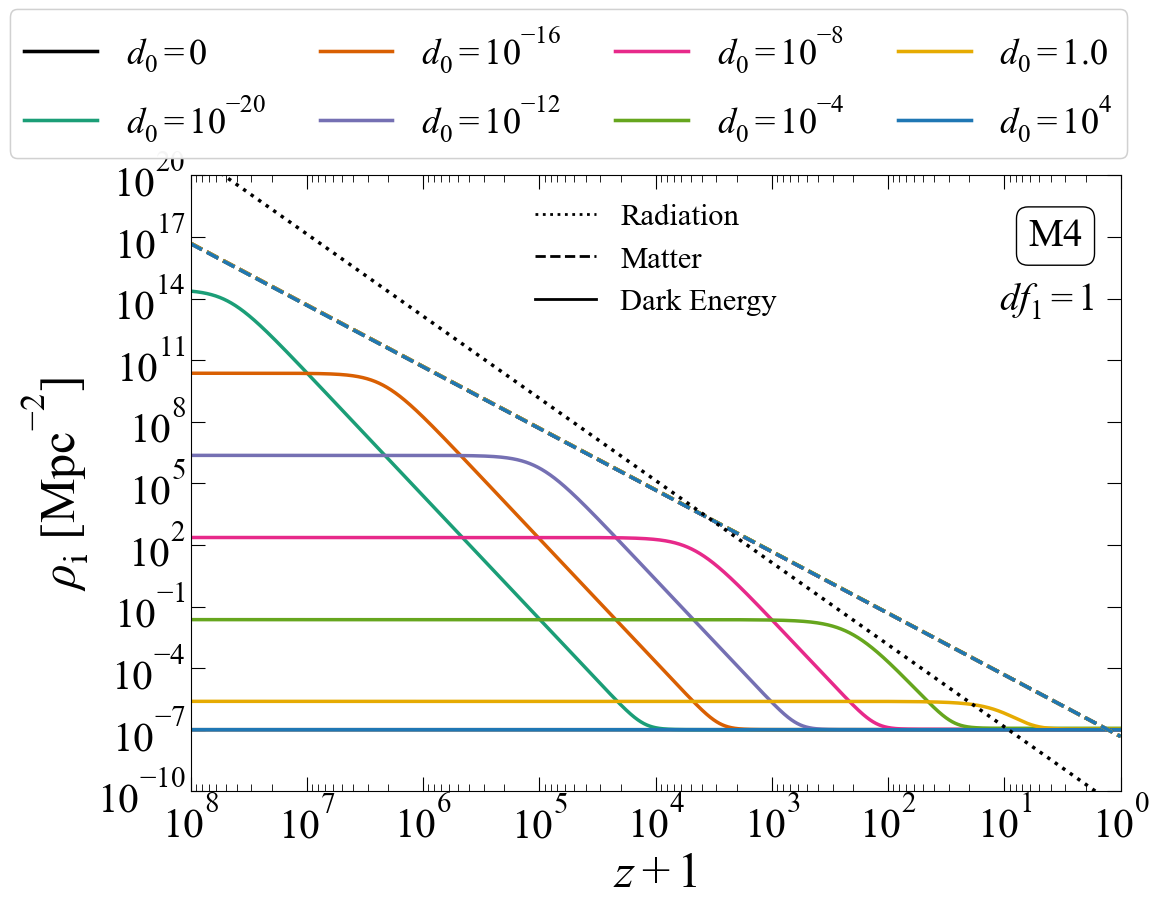}
\endminipage\hfill
\minipage{0.5\textwidth}
  \includegraphics[width=1\linewidth]{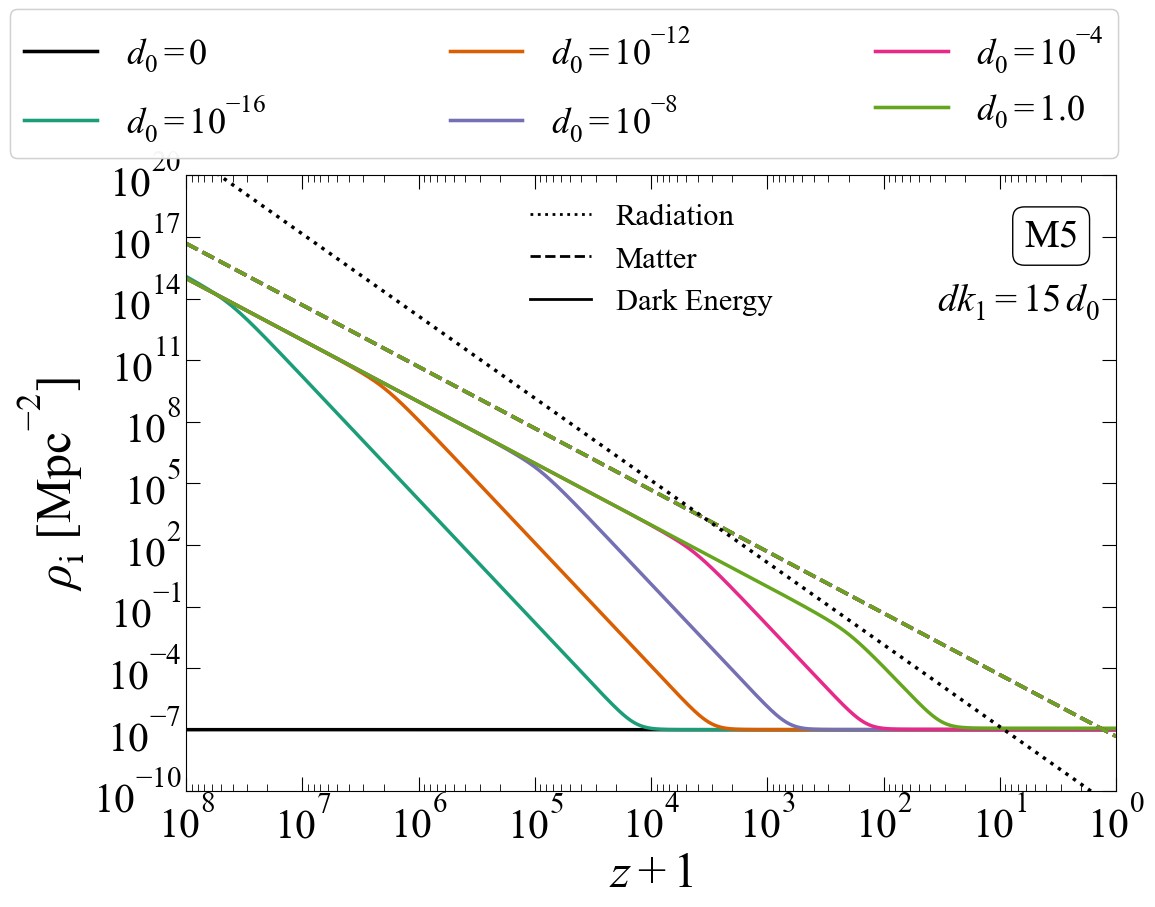}
\endminipage\hfill
\minipage{0.5\textwidth}
  \includegraphics[width=1\linewidth]{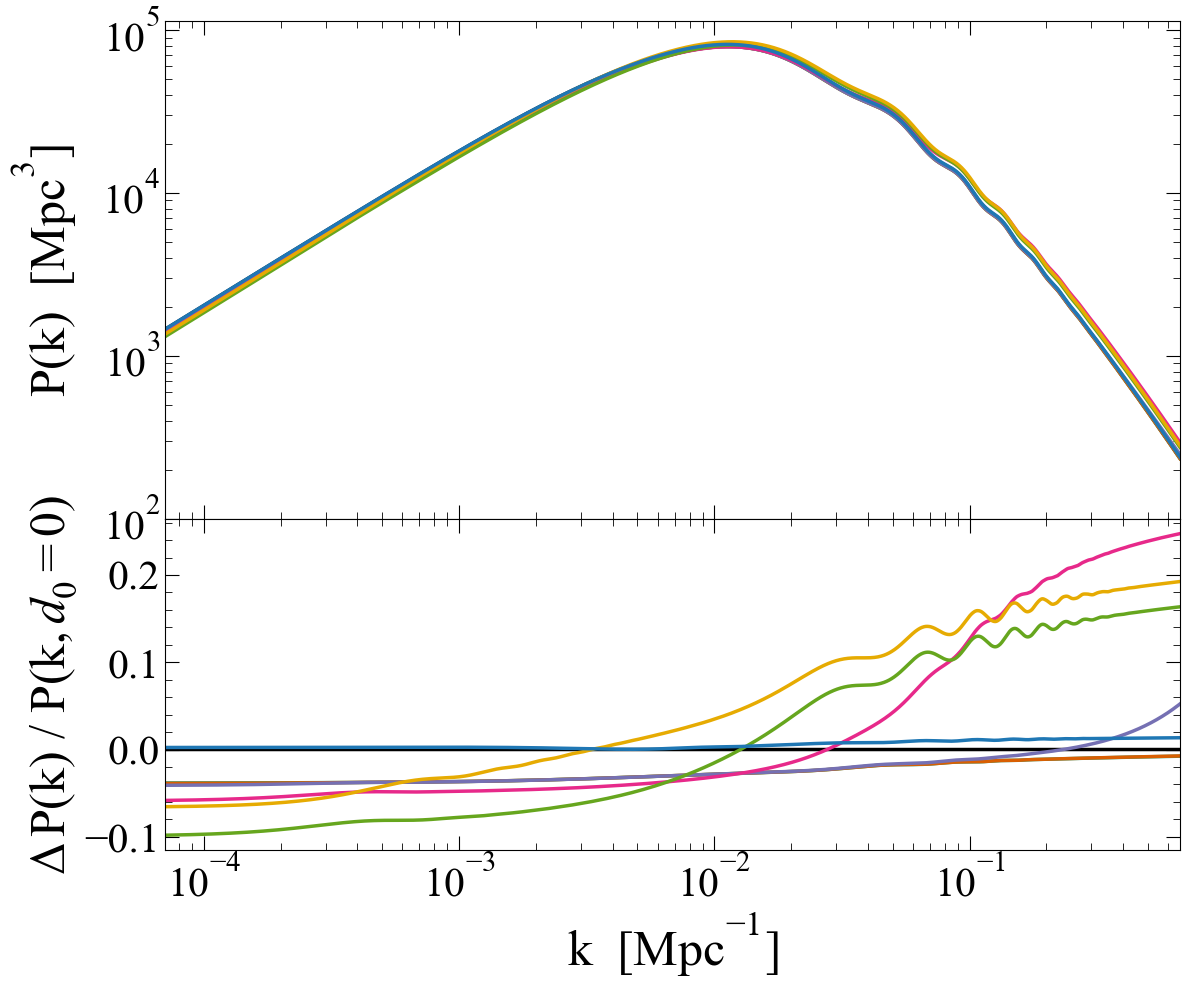}
\endminipage\hfill
\minipage{0.5\textwidth}
  \includegraphics[width=1\linewidth]{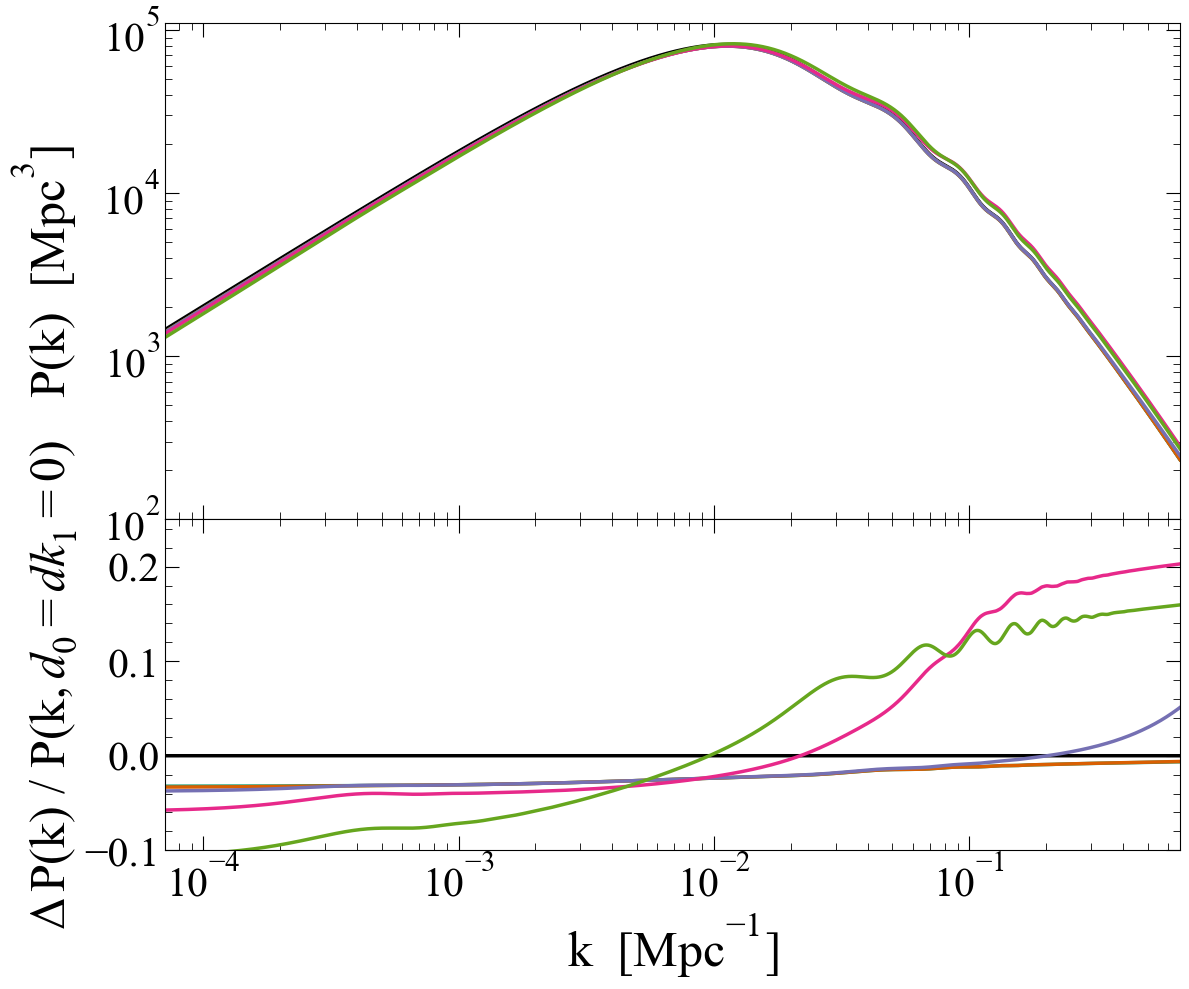}
\endminipage\hfill
\minipage{0.5\textwidth}
  \includegraphics[width=1\linewidth]{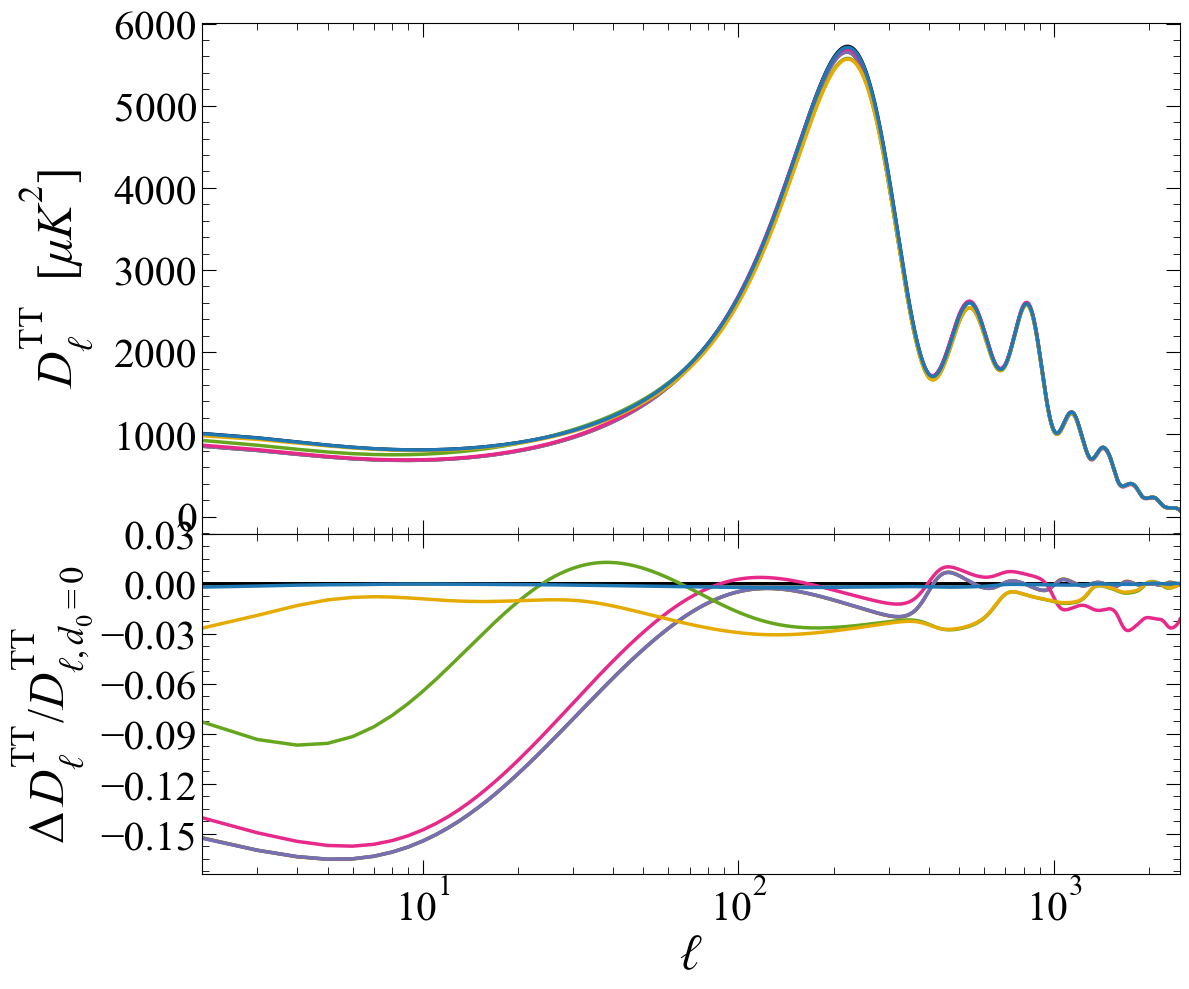}
\endminipage\hfill
\minipage{0.5\textwidth}
  \includegraphics[width=1\linewidth]{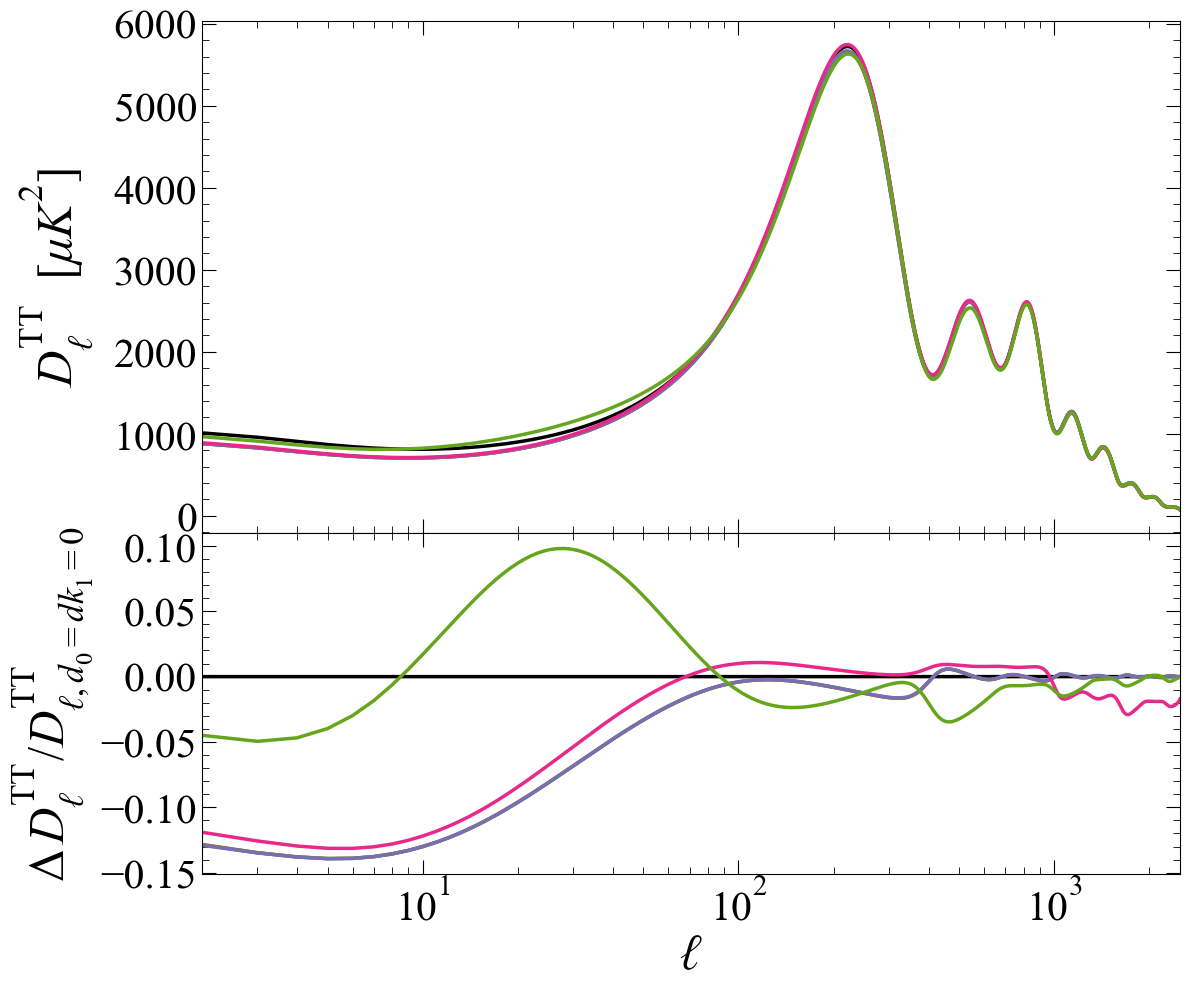}
\endminipage
\caption{Evolution of the energy densities of different components (top), linear matter power spectrum (center) and CMB temperature spectrum (bottom) in the field-dependent (Model M4) and kinetic energy-dependent (Model M5) disformal DE-DM interactions from Tables \ref{Table:iDEDM} and \ref{Table:Summary}\,.}
\label{Fig:M4,M5}
\end{figure}
In this appendix, we discuss the characteristic modifications introduced by pure disformal couplings that depend on the scalar field ($D \equiv D(\phi)$) and the kinetic energy ($D \equiv D(X)$), corresponding to Models M4 and M5. For our numerical analysis, we have assumed an exponential dependence on the scalar field $\phi$ for Model M4 and on the kinetic term $X$ for Model M5, respectively:
\bea
& &
\mathrm{M4}: C=1,~D(\phi) = d_0\,e^{df_1\,\phi}\,,
\\
& &
\mathrm{M5}: C=1,~D(X) = d_0\,e^{dk_1\,X}\,.
\eea
For both models, we set $X_i\,d_0 \approx 0.02$, where $X_i$ denotes the initial kinetic energy of the DE field. The background evolution of the energy densities, the linear matter power spectrum, and the CMB TT spectrum for the two models are shown in Fig.~\ref{Fig:M4,M5}.

\subsection{Model M4: Scalar Field-Dependent Disformal Coupling}

For Model M4, we have set $df_1=1.0$ and varied $d_0$. As evident from the modified scalar field equation (in Table \ref{Table:iDEDM}) and the left-panel plots in Fig.~\ref{Fig:M4,M5}, the cosmological evolution is functionally identical to that of Model M3. The key difference lies in the value of $d_0$ required to produce a similar evolution, which is shifted by several orders of magnitude (four orders for $df_1=1.0$) compared to Model M3 due to the exponential coupling. The resulting EDE-like evolution of the DE scalar field and the modifications to the matter and CMB power spectra are identical to the results obtained for M3.

\subsection{Model M5: Kinetic Energy-Dependent Disformal Coupling}

For Model M5, we have set $dk_1=15\,d_0$\,. Similar to Models M3 and M4, the kinetic energy of the DE field remains constant at early times. The disformal kinetic coupling in M5 directly contributes to the DE energy density, which takes the form:
\bea
\rho_{\mathrm{DE}} = X + V + 2\rho_{\mathrm{DM}}\,X^2\,dk_1\,d_0\,e^{dk_1\,X}\,.
\eea
To avoid negative energy density states for the DE field, we assume $d_0 > 0$ and $dk_1 > 0$\,. As in the pure-momentum coupling Model M6 and the conformal Model M2, the DE field evolves in three distinct stages: like a dark matter fluid at early times, a free-scalar field at intermediate times, and a potential-driven cosmological constant at late times. The linear matter and CMB power spectra are illustrated in the right-panel plots of Fig.~\ref{Fig:M4,M5}\,.

Similar to Models M3 and M4, we find a suppressed structure growth for smaller values of $d_0$ and an enhanced growth on small scales for larger values of $d_0$. However, we observe that the CMB spectrum in M5 is slightly enhanced compared to Models M3 and M4.

\section{Comparing the numerical accuracy of our code with \texttt{CLASS}}
\label{Appendix:CLASS-Comparison}

\begin{figure}[!htb]
\minipage[h]{0.5\textwidth}
  \includegraphics[width=1\linewidth]{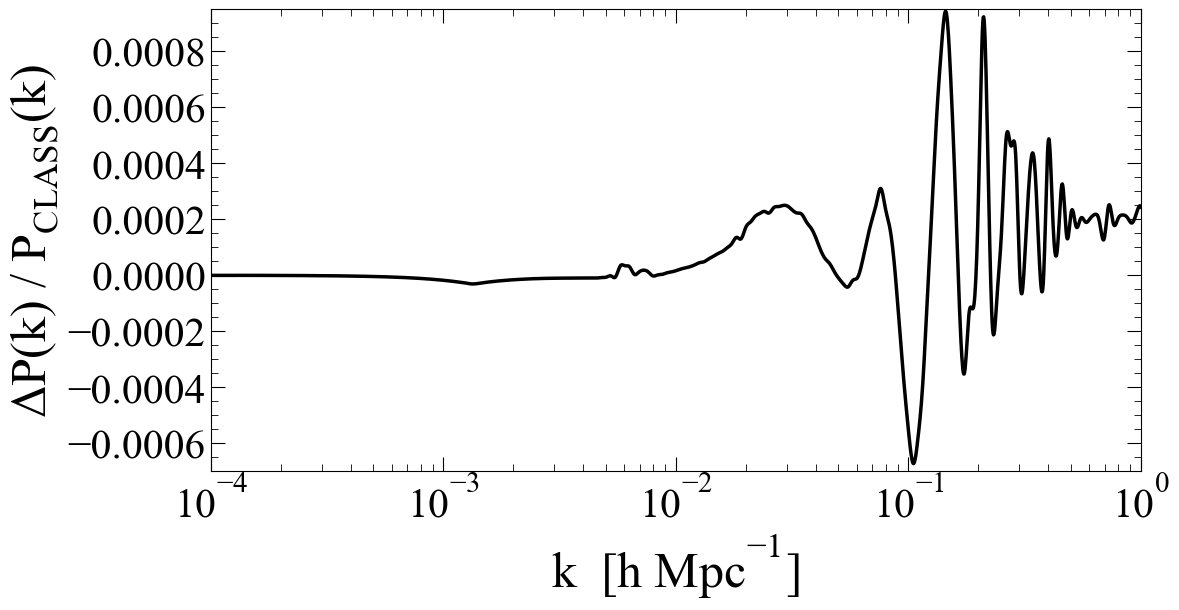}
\endminipage\hfill
\minipage[h]{0.5\textwidth}
  \includegraphics[width=1\linewidth]{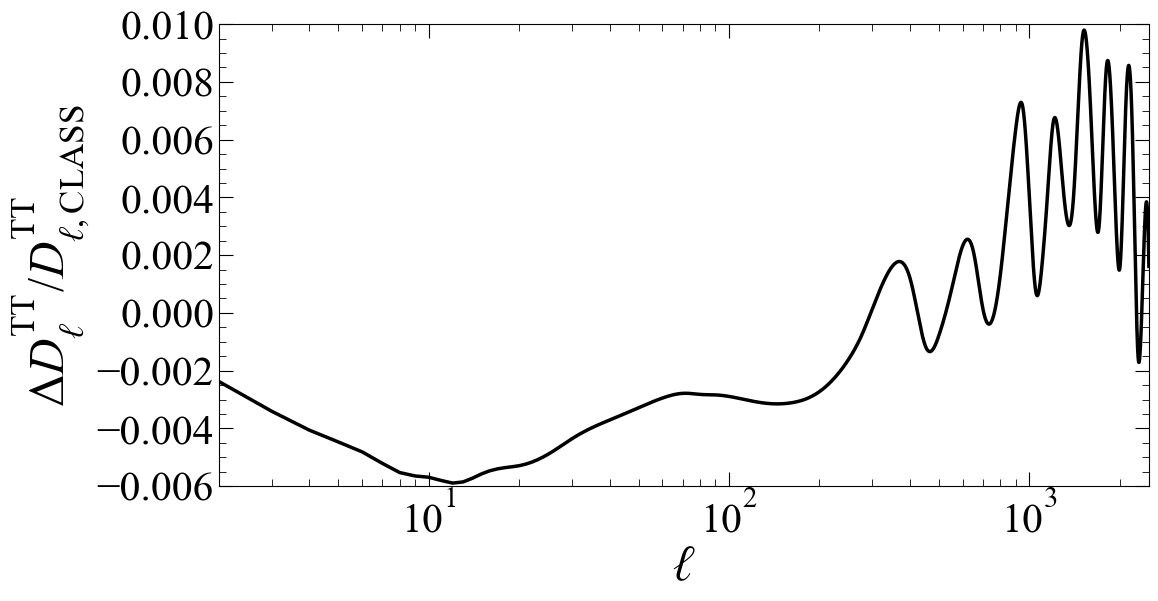}
\endminipage
\caption{Deviation in our constructed linear matter power spectrum and the CMB temperature power spectrum for the $\Lambda$CDM model with the results computed using the Boltzmann code \texttt{CLASS}.}
\label{Fig:CLASS-Comparison}
\end{figure}

In this appendix, we provide an overview of the custom Python code developed for the numerical analysis of our disformal coupling framework. We also compare its accuracy for the $\Lambda$CDM model with the results from the well-established Boltzmann code, \texttt{CLASS} \cite{Lesgourgues:2011re}.

For the background evolution, our code solves the coupled Friedmann equations and the equations of motion for the dark matter fluid and the dark energy scalar field using forward shooting. The dark energy scalar field's potential normalization, $V_0$\,, is dynamically calculated to match the background parameters from the Planck 2018 constraints \cite{2018-Planck-AA}. For recombination history, we solve the Saha and Peebles equations, using a corrected three-level atom approach to improve accuracy. We acknowledge that this method does not reach the high precision of state-of-the-art recombination codes like H\textsc{\small{y}}R\textsc{\small{ec}}~\cite{HyRec-2011} or C\textsc{\small{osmo}}R\textsc{\small{ec}}~\cite{CosmoRec-2010}, which account for a detailed multi-level atom and radiative-transfer effects. We model hydrogen reionization as well as first and second helium reionization using the tanh model \cite{Lewis:2008wr}. For linear perturbations, our code works in the conformal Newtonian gauge, employing the tight-coupling approximation for photons and baryons at early times and the full Boltzmann system thereafter. The Boltzmann hierarchy for temperature, polarization and neutrino multipoles is stopped at $l_{\mathrm{max}} = 46$ and we have used the cut-off method discussed in \cite{Ma:1995ey} to prevent any unphysical reflection of power from $l_{\mathrm{max}}$ to lower multipoles. For computing the CMB spectra, we have used the line-of-sight integration approach.

For our comparison, we calibrated the cosmological parameters to the Planck 2018 constraints, specifically using the following values:
\bea
h &=& 0.6732117, \quad \omega_{\mathrm{b}} = 0.0223828, \quad 
\omega_{\mathrm{cdm}} = 0.1201075, \quad N_{\mathrm{eff}} = 3.046, \nonumber \\
T_{0,\mathrm{CMB}} &=& 2.7255~\mathrm{K}, \quad Y_{\mathrm{He}} = 0.2454006, \quad
\tau_{\mathrm{reio}} = 0.0543084, \nonumber \\
n_{\mathrm{s}} &=& 0.9660499, \quad A_{\mathrm{s}} = 2.100549\times 10^{-9}.
\eea
Fig.~\ref{Fig:CLASS-Comparison} shows the deviation of our code's linear matter power spectrum and CMB temperature power spectrum for the $\Lambda$CDM model from the results obtained using \texttt{CLASS}. Our matter power spectrum agrees with \texttt{CLASS} within $0.1\%$, the $\sigma_8$ value agrees within $0.01\%$, and the CMB TT spectrum agrees within $1\%$. The primary deviation is located in the small-scale regime around the acoustic peaks.

While the Planck collaboration's analysis requires a higher level of theoretical precision, our code's accuracy is sufficient for the present work. We have focused on analyzing the \emph{relative modifications} to the power spectra introduced by our general disformal coupling framework, rather than on providing absolute predictions. For future directions, including MCMC analysis for parameter estimation and compatibility tests with datasets, we plan to utilize publicly available, state-of-the-art Boltzmann solvers.

\input{Reference.bbl}

\end{document}

%% file: Reference.bbl
%